\documentclass[11pt,a4paper]{article}
\usepackage{amsfonts,amssymb,epsfig,amsmath,mathtools}
\usepackage[utf8]{inputenc}
\usepackage{textcomp,setspace}
\usepackage{tabularx,bm,booktabs,multirow}
\usepackage[inline]{enumitem}
\usepackage{cite}
\usepackage{hyperref,caption,subcaption}
\usepackage{xcolor,tikz,graphicx}
\usetikzlibrary{shapes.geometric,positioning}
\usetikzlibrary{cd}
\renewcommand{\thefootnote}{\fnsymbol{footnote}}

\allowdisplaybreaks

\begin{document}

\makeatletter \@addtoreset{equation}{section} \makeatother
\renewcommand{\theequation}{\thesection.\arabic{equation}}
\renewcommand{\thefootnote}{\alph{footnote}}

\begin{titlepage}
\begin{center}

\hfill {\tt KIAS-P18001}\\

\vspace{2cm}

{\LARGE\bf On elliptic genera of 6d string theories}

\vspace{2cm}

{\Large Joonho Kim$^1$, Kimyeong Lee$^1$, Jaemo Park$^2$}

\vspace{0.5cm}

\textit{$^1$School of Physics, Korea Institute for Advanced Study, Seoul 02455, Korea.}\\

\vspace{0.3cm}

\textit{$^2$Department of Physics, POSTECH, Pohang 37673, Korea}\\

\vspace{1cm}

E-mails: {\tt joonhokim@kias.re.kr, klee@kias.re.kr, jaemo@postech.ac.kr}

\end{center}

\vspace{2.5cm}

\begin{abstract}
\begin{spacing}{1.2}
\normalsize

We study the elliptic genera of 6d strings based on their modular properties.
They are weak Jacobi forms of weight $0$, whose indices are determined from the 2d chiral anomalies. We propose the ansatz for the elliptic genera
which reflects the analytic structure of instanton partition functions. Given a finite amount of initial BPS data,
 we completely determine the elliptic genera of 6d strings in various 6d SCFTs.
We also apply our ansatz to study $\mathcal{N}=(2,0)$ and $(1,1)$ little strings as well as $\mathcal{N}=(1,0)$ heterotic little strings, 
for which T-duality of little string theories supplies a sufficient number of initial BPS data.
The anomaly polynomials of 6d little strings are worked out, which is needed for the elliptic genera bootstrap.
In some little string theories, the elliptic genera must have the extra contributions from the Coulomb branch, which correspond to the additional zero modes
for the full strings. The modified ansatze for such elliptic genera are  also discussed.

\end{spacing}

\end{abstract}

\end{titlepage}

\setcounter{tocdepth}{2}
\tableofcontents

% \pagebreak
\section{Introduction}
\label{sec:intro}

Non-critical strings play an important role in understanding the physics of 6d superconformal field theories (SCFTs) and little string theories (LSTs) \cite{Witten:1995zh,Strominger:1995ac}. In this paper, we study the supersymmetric partition functions on Omega-deformed $\mathbf{R}^4 \times T^2$ for various 6d SCFTs and LSTs. They are Witten indices which capture the bound states of winding and momentum modes, coming from multiple numbers of 6d BPS strings on $T^2$. In 6d gauge theories, these observables are 6d uplifts of the instanton partition functions \cite{Nekrasov:2002qd,Nekrasov:2003rj}, which were first introduced to derive the Seiberg-Witten prepotentials of 4d $\mathcal{N}=2$ gauge theories \cite{Seiberg:1994rs}.

The $\mathbf{R}^4 \times T^2$ partition function is a tensor branch observable. Recall that 6d superconformal and little string theories are equipped with $\mathcal{N}=(1,0)$ tensor multiplets, consisting of a 2-form potential $B_2$ whose field strength $H_3$ is subject to the self-duality condition $H_3 = \star H_3$, a real scalar $\varphi$, and a superpartner fermion $\lambda^A$. It is the VEV of the scalar $\varphi$ which parametrizes the tensor branch moduli space of vacua and determines a tension of the 6d string, the source of the tensor multiplet.  The 6d string acquires a non-zero tension at a generic point of the tensor branch, such that the string number fugacity $\mathfrak{n} \equiv \exp{(-\text{vol}(T^2)\cdot \langle\varphi \rangle)}$ becomes a sensible expansion parameter of the $\mathbf{R}^4 \times T^2$ partition function. One can write the partition function as the weighted sum over the 6d string elliptic genera with different numbers of strings. More precisely, the $\mathbf{R}^4 \times T^2$ partition function is given as
\begin{align}
	\label{eq:6d-decom}
	\textstyle Z_{6d} = \mathcal{I}_{0}\cdot \left(1 + \sum_{k} 	 \mathfrak{n}^{k}  \cdot \mathcal{I}_{k}\right).
\end{align}
The overall factor $\mathcal{I}_0$ is the Witten index for pure momentum states decoupled from winding modes. The coefficient $\mathcal{I}_{k}$ captures the BPS spectrum of an infinite tower of momentum modes and $k$ winding modes, corresponding to the elliptic genus of $k$ strings. It turns {out} to be strongly constrained by the modular and symmetry properties.

The 6d string elliptic genus $\mathcal{I}_k$ depends on the complex modulus $\tau$ of the $T^2$ and various chemical potentials for the $U(1)$ charges in the maximal tori of the 6d symmetry group. We collectively denote all chemical potentials by $z$. The elliptic genus $\mathcal{I}_k$ is a weak Jacobi form of weight $0$ and index $\mathfrak{i}(z)$, transforming under the modular transformation $\tau \rightarrow \tfrac{a\tau + b}{c\tau+d}$, $z \rightarrow \tfrac{z}{c\tau+d}$ with $(\begin{smallmatrix}a&b\\c&d\end{smallmatrix}) \in \text{SL}(2,\mathbf{Z})$ as follows:
\begin{align}
\label{eq:elliptic-genera-modular-property}
\mathcal{I}  (\tau, z)\ \longrightarrow\ \mathcal{I} \left(\frac{a\tau + b}{c\tau+d}, \frac{z}{c\tau+d}\right) = \varepsilon(a,b,c,d)\, \exp{\left(\frac{-\pi i c \cdot \mathfrak{i} (z)}{c\tau + d} \right)}\ \mathcal{I} (\tau, z)
\end{align}	
where the index $\mathfrak{i}(z)$ is completely determined by
 the worldsheet chiral anomaly of 6d strings \cite{Benini:2013nda,Benini:2013xpa}.
Combined with a separate observation on the pole structure of the elliptic genus, induced from the zero modes that parametrize the moduli space of 6d strings, \eqref{eq:elliptic-genera-modular-property} nearly solves the elliptic genus $\mathcal{I}_k$ in an appropriate ring of weak Jacobi forms up to finite numerical coefficients \cite{Haghighat:2014pva,Cai:2014vka,Huang:2015sta,Huang:2015ada,Haghighat:2015ega,DelZotto:2016pvm,Gu:2017ccq}. The problem of finding the 6d string elliptic genus has been reduced to determining the coefficients through comparison with an initially given set of the BPS data. In this way, the elliptic genera were successfully \emph{bootstrapped} out for the instanton strings in minimal SCFTs \cite{DelZotto:2016pvm} and also for the chains of E- and M-strings \cite{Gu:2017ccq}.

In this paper, we apply this approach to broader classes of 6d SCFTs. 
Specifically we are interested in various self-dual string theories, which are defined as IR limit of 2d gauge theories. 
The initial BPS data are obtained from the gauge theory side.
Obviously this is just one convenient way of obtaining the BPS data and the bootstrapping procedure can equally be applied to the cases
where the gauge theory description is not available. Also we make a technical improvement over \cite{DelZotto:2016pvm}. When the 6d string
theories have global symmetry, we can consider the elliptic genus with the corresponding chemical potentials. The elliptic genus of the 6d string theories should be described by suitable Weyl-invariant Jacobi forms. We explicitly work out such Weyl-invariant Jacobi forms wherever needed.

In addition, we also focus on circle compactified LSTs, bootstrapping their $\mathbf{R}^4 \times T^2$ partition functions. A characteristic feature of the LST is T-duality that identifies two apparently distinct LSTs on $S^1$, at different circle radii $R' = \alpha'/R$, by exchanging the winding and momentum modes. As the supersymmetric partition function is protected and insensitive to the circle radius, T-duality implies the equivalence of the $\mathbf{R}^4 \times T^2$ partition functions for a dual pair of LSTs. This has been confirmed for several examples, such as $(2,0)$ and $(1,1)$ LSTs of A-type \cite{Kim:2015gha} and their orbifold variations \cite{Hohenegger:2016eqy,Kim:2017xan} which are engineered from type IIA and IIB NS5-branes on transverse $\mathbf{R}^4$ and $\mathbf{R}^4/\Gamma_\text{AD}$ backgrounds. Assuming the general equivalence of the BPS spectra for all T-dual pairs of circle compactified LSTs, a sufficient amount of the initial BPS data will be given such that the $\mathbf{R}^4 \times T^2$ partition function can be constructed through the iterated bootstrap of the 6d string elliptic genus. For example, it can reproduce the $\mathbf{R}^4 \times T^2$  partition functions of A-type $(2,0)$ and $(1,1)$ LSTs which were first obtained in \cite{Kim:2015gha} using the worldsheet gauge theories of little strings.
Since the bootstrap approach to the LST partition functions does not use the gauge theory description of little strings,
it is also applicable to any general LSTs whose T-duality relations have been established.
In this work, we will consider D-type $(2,0)$ and $(1,1)$ LSTs as well as $SO(32)$ and $E_8 \times E_8$ heterotic LSTs,
which arise as the worldvolume theory of type II and heterotic NS5-branes in the decoupling limit $g_s \rightarrow 0$ \cite{Seiberg:1997zk}.

For the \emph{full} strings which completely wrap the transverse circle to the NS5-branes,
the bootstrap computation shows that the conjectured form of the pole structure,
which is generally expected for the 6d BPS partition function \cite{Gopakumar:1998jq}, does not always hold in their elliptic genera.
It is because the 2d superconformal field theories of little strings has the target space with a tubelike region,
where strings escape from NS5-branes \cite{Witten:1995gx,Witten:1997yu,Diaconescu:1997gu}.
This is reflected in the elliptic genera as the additional poles which indicate the presence of the extra bosonic zero modes parametrizing the run-away motions \cite{Seiberg:1999xz,Aharony:1999dw}. 
Based on the modified ansatze which include the additional zero modes, one can bootstrap the elliptic genera of the full winding modes. 
We also remark that the ADHM gauge theories for $\mathcal{N}=(1,1)$ $SO(2n)$ instantons and $\mathcal{N}=(1,0)$ $Sp(n)$ instantons with 1 antisymmetric and 16 fundamental hypermultiplets analogously develop the extra poles in their elliptic genera.
To obtain the proper 6d spectrum, one still has to separately remove the extra states' contribution from the partition function.
See also \cite{Hwang:2016gfw,Hwang:2014uwa} for removal of the extra contributions in the instanton partition functions of 5d SYMs, obtained from their suitable $S^1$ reductions.
On the contrary, the \emph{fractional} strings which partially wrap the transverse circle must end on a pair of NS5-branes,
not escaping to the bulk. Using the BPS data coming from T-duality relation between circle compactified LSTs,
we find the elliptic genera of various string chains in D-type $(2,0)$ LSTs and $E_8\times E_8$ heterotic LSTs.
These fractional string chains include what appear in their relative 6d SCFTs, i.e., D-type $(2,0)$ SCFTs and E-string SCFTs \cite{Gadde:2015tra,Kim:2015fxa,Gu:2017ccq}, while many of them are unique to LSTs.

The rest of this paper is organized as follows. Section~\ref{sec:modular-anomaly} reviews the modular bootstrap of the 6d string elliptic genera \cite{DelZotto:2016pvm}, refining the conjectured form of the elliptic genera. {Along with it, we clarify the relation between the 2d chiral
 anomaly of the 6d string theories and modular properties of their elliptic genera.} In Section~\ref{sec:anomaly-polynomial},
we study the anomaly polynomial of little strings in maximally supersymmetric LSTs and heterotic LSTs. In Section~\ref{sec:examples}, we construct
the $\mathbf{R}^4 \times T^2$ partition functions of LSTs by the iterated bootstrap of the elliptic genera, based on the T-duality relations. Section~\ref{sec:conclusion} concludes with brief discussions.

\noindent{\textbf{Note added:} As this work is being finished, the paper \cite{DelZotto:2017mee} appears on arXiv which partially overlaps with the current work. }

\section{Elliptic genera of 6d strings}
\label{sec:modular-anomaly}

In this section, we will study the strings of 6d SCFTs and LSTs on $\mathbf{R}^4 \times T^2$ in the tensor branch.
They are the BPS string configurations which preserve at least 2d $\mathcal{N}=(0,4)$ supersymmetry.
They have  non-zero tension proportional to the VEV of a tensor multiplet scalar.
Wrapping the $T^2$, they preserve $SO(4)_T = SU(2)_l \times SU(2)_r$ symmetry that rotates the $\mathbf{R}^4$ space.
The 6d R-symmetry $SU(2)_R$, the 6d gauge symmetry $G$, the 6d flavor symmetry $F$ are also visible
in the  $(0,4)$ SCFT of the strings.

The elliptic genus of the 6d strings is the supersymmetric partition function on $T^2 = S^1_t \times S^1_x$ with the periodic boundary condition, defined as
\begin{align}
  \label{eq:elliptic-genera}
  \mathcal{I}_k = \text{Tr}_{RR}\,\left[ (-1)^F
e^{2\pi i (\tau H_L - \bar{\tau} H_R)}\, e^{2\pi i \epsilon_+(J_{r} + J_{R})} e^{2\pi i \epsilon_- J_{l}}   e^{2\pi i z\cdot J_z} \right].
\end{align}
The complex structure $\tau$ of the torus $T^2$ is conjugate to the left-moving Hamiltonian $H_L = \frac{H + P}{2}$.
With $(0,4)$ supersymmetry, the right-moving Hamiltonian $H_R = \frac{H -P}{2}$ can be written in terms of the supercharges $Q^{\dot{\alpha}A}$
where $\alpha$, $\dot{\alpha}$, $A$ respectively denote the doublet indices of $SU(2)_l$, $SU(2)_r$, $SU(2)_R$.
For $Q \equiv Q^{\dot{1}2}$ and $Q^\dagger \equiv -Q^{\dot{2}1}$, $H_R \sim \{Q, Q^\dagger\}$ such that the elliptic genus {is} independent of $\bar{\tau}$,
if one introduces the chemical potentials to generate the mass gap, lifting all the zero modes.
The Cartan generators of $SU(2)_{l}$, $SU(2)_{r}$, $SU(2)_{R}$ are denoted by $J_{l}$, $J_r$,  $J_R$. Only two of three combinations $J_l$ and $(J_r + J_R)$ commute with the supercharges $Q$ and $Q^\dagger$. We introduce their conjugate chemical potentials as $2\pi\epsilon_-$ and $2\pi \epsilon_+$, respectively. They uplift the zero modes for the center-of-mass motion of the strings on $\mathbf{R}^4 \subset \mathcal{M}_6$ \cite{Nekrasov:2002qd,Nekrasov:2003rj}.
We will collectively denote by $J_z$ and $2\pi z$ the Cartan generators and the chemical potentials introduced for the 6d gauge symmetry $G$ and the 6d flavor symmetry $F$.

\subsection{High temperature free energy}
\label{subsec:freeenergy}

Let us study the high temperature free energy of the elliptic genus $\mathcal{I}_k$ to derive its modular property.
The elliptic genus $\mathcal{I}_k$ is the supersymmetric partition function on the Euclidean torus $T^2 = S^1_t \times S^1_x$,
which has the periodicity $(t,x) \sim (t,x+2\pi)\sim (t+2\pi \,\text{Im}\tau, x+2\pi \,\text{Re}\tau)$.
The torus metric is given by
\begin{align}
\label{eq:metric}
ds^2  = (1 + \mu^2) \left(dt + \frac{\mu\,dx}{1 + \mu^2} \right)^2 + \frac{dx^2}{1 + \mu^2}
% \left(d\alpha + \frac{\tau_1}{\sqrt{\tau_1^2 + \tau_2^2}}d\sigma\right)^2 + \left(\frac{\tau_2}{\sqrt{\tau_1^2 + \tau_2^2}} d\sigma\right)^2
\end{align}
with $\tau = \frac{\beta}{2\pi}(\mu + i)$.
Insertion of $e^{2\pi i z \cdot J_z}$ introduces the $U(1)$ background gauge field as
\begin{align}
\mathcal{A}_z = \frac{2\pi  z}{\beta }  dt \cdot J_z,
\end{align}
where the normalization of the generator $J_z$ is captured by $d_z \equiv \text{tr}\, (J_z J_z)$.

We reduce the elliptic genus along $S^1_t$ to reach the high temperature limit $\beta \ll 1$.
For the Kaluza-Klein reduction, we recast the metric \eqref{eq:metric} and the background gauge field $\mathcal{A}_z$ into
\begin{align}
ds^2= e^{2\phi}(dt + a )^2 + g dx^2\,,\qquad
\mathcal{A}_z = \Phi_z (dt + a) +  A_z  dx\,.
\end{align}
This identifies the dilaton $e^{2\phi}$, the graviphoton $a$, the 1d gauge field $A_z$, the 1d scalar field $\Phi_{z}$ as
\begin{align}
\label{eq:1d-field-classical-value}
e^{2\phi} =(1 + \mu^2)\,,\qquad a = \frac{\mu\,dx}{1 + \mu^2}\,,\qquad
A_{z} = -\frac{2\pi z}{\beta} \cdot\frac{\mu\, dx}{1 + \mu^2}\cdot J_z\,,\qquad
\Phi_{z} = \frac{2\pi z}{\beta}\cdot J_z.
\end{align}
% where $\sigma \sim \sigma +1 $ and $\alpha \sim \alpha + |\tau|$ parametrize the twisted torus, i.e.,
% \begin{align}
% t = \frac{2\pi\tau_2}{\sqrt{\tau_1^2 + \tau_2^2}} \alpha, \quad x = 2\pi\sigma + \frac{2\pi\tau_1}{\sqrt{\tau_1^2 + \tau_2^2}} \alpha.
% \end{align}
% The chemical potentials, such as $2\pi \epsilon_+$ and $2\pi \epsilon_-$, are introduced as the 2d $U(1)$ background gauge fields.
% We collectively denote the chemical potentials by $2\pi z$ and the Cartan generators conjugate to $2\pi z$ by $J_z$. We also define
% The background gauge fields $\mathcal{A}_z$ are given by
% \begin{align}
% \mathcal{A}_z = \frac{2\pi  z}{\beta }  dt \cdot J_z  \equiv \Phi_z \left({dt}+ \frac{\mu\,dx}{1 + \mu^2} \right) + A_z dx,
% % \mathcal{A}_v = \frac{2\pi v}{\sqrt{\tau_1^2 + \tau_2^2}} d\alpha  \equiv \Phi_v \left({d\alpha}+ \frac{\tau_1}{\sqrt{\tau_1^2 + \tau_2^2}}d\sigma\right) + A_v d\sigma
% \end{align}
% which can reduced to the 1d background gauge and scalar fields as follows.
We now apply the analysis of \cite{DiPietro:2014bca,Golkar:2015oxw,Kim:2017zyo} to the reduced 1d system on $S^1_x$.
After the $S^1_t$ reduction, there are the massless degrees of freedom whose determinants appear in the 1d effective action as non-local terms.
These non-local terms are real-valued since all the background fields in the Euclidean quantum mechanics have been chosen to be real.
On the other hand, the imaginary part of the effective action can be obtained from local terms, such as the Euclidean Chern-Simons term,
which can be fixed by the 2d chiral anomaly \cite{Banerjee:2012iz,DiPietro:2014bca,Golkar:2015oxw,Kim:2017zyo}.
Let us split the imaginary terms into the gauge invariant and non-invariant ones.
The gauge invariant action  generally takes the form of $i \int a f(\Phi, \phi)$ and $i \int d\phi\ g(\Phi, \phi)$.
It must produce the anomalous factor $\exp{\left(\frac{\pi i}{6} (c_R - c_L)\right)}$ under the transformation $a \rightarrow a + \frac{\beta\, dx}{2\pi}$,
which corresponds to the 2d global diffeomorphism $(t,x) \rightarrow (t + \frac{\beta x}{2\pi}, x)$.
Matching the global anomaly fixes the gauge invariant action $\mathcal{S}^{(1)}$  to be \cite{Golkar:2015oxw}
\begin{align}
 \mathcal{S}^{(1)} = \frac{\pi i}{6}\frac{\left(c_R - c_L \text{\footnotesize{ (mod 12)}}\right)}{\beta}\int_{S^1_x} a + \mathcal{O}(\beta^0)
\end{align}
where $(c_R - c_L)$ is the 2d gravitational anomaly. Similarly, the gauge non-invariant action $\mathcal{S}^{(2)}$ must match
the 2d chiral anomaly under the $U(1)$ gauge transformation, i.e., $\delta_\epsilon \Phi_z = 0$ and $\delta_{\epsilon} A_z = d\epsilon_z$.
Recall that the 2d chiral anomaly $\Delta$ is encoded in the 4-form anomaly polynomial $I_4$ by the descent formalism, such that
\begin{align}
\Delta = \sum_z \frac{ i n_z}{4\pi }\ \int_{T^2} \text{tr}\,(d\epsilon_z \,\mathcal{A}_z)   \quad \longleftrightarrow \quad I_4 =  \sum_z \frac{n_z}{4}\,\text{tr}\mathcal{F}^2_z
\end{align}
where the sum is taken over all background $U(1)$ gauge fields. Dimensionally reducing it on $S^1_t$,
\begin{align}
\label{eq:1d-effective-action-anomaly}
\Delta_\text{1d} = \sum_z \frac{i \beta\, n_z}{4\pi}\int_{S^1_x} \text{tr}\,(d\epsilon_z \, \Phi_z).
\end{align}
This must be reproduced by the gauge non-invariant action $\mathcal{S}^{(2)}$ under the $U(1)$ gauge transformation, implying that $\mathcal{S}^{(2)}$ has to be
\begin{align}
\label{eq:non-inv-cs-action}
\mathcal{S}^{(2)} = \sum_z \frac{i \beta\, n_z}{4\pi}\int_{S^1_x}  \text{tr}\left( A_z\, \Phi_z \right) + \mathcal{O}(\beta^0) .
% = |\tau| \sum_{i,j} i\left(c_1^{(i,j)} - c_2^{(i,j)}\right) \int  \Phi_{(i)}  \Phi_{(j)} a
\end{align}

We evaluate the imaginary part of the high temperature free energy $f_h$ by inserting the background values \eqref{eq:1d-field-classical-value}
into the effective action $\mathcal{S}^{(1)} + \mathcal{S}^{(2)}$. It is given by
\begin{align}
\label{eq:free-energy-im}
\text{Im}\,f_h(\tau) &= \frac{2\pi^2 }{\beta} \frac{\mu}{1+\mu^2}  \bigg( \frac{c_R - c_L \text{\footnotesize{ (mod 12)}}}{6} -\sum_z  d_z n_z z^2\bigg)  \nonumber\\ &= \text{Im}\, \Bigg[\frac{\pi i}{\tau} \bigg( \frac{c_R - c_L \text{\footnotesize{ (mod 12)}}}{6} - \sum_z d_z n_z z^2\bigg)\Bigg].
\end{align}
On the other hand, the Casimir energy $E_0$ of the elliptic genus, defined by $\mathcal{I}_k = q^{E_0} \left(1 + \sum_{n=1}^\infty  q^n j_{k,n} \right)$ with $q\equiv e^{2\pi i \tau}$,
makes a dominant contribution to the low temperature free energy $f_l$, i.e.,
\begin{align}
 f_l(\tau) = 2\pi i\tau E_0 + \mathcal{O}(\tau \,e^{2\pi i \tau}).
\end{align}
Since the modular transformation $\tau \rightarrow -\frac{1}{\tau}$ and $z \rightarrow \frac{z}{\tau}$ inverts the temperature, in the $\tau \rightarrow i0^+$ limit,
the free energies $f_l(-\frac{1}{\tau})$ with $f_h(\tau)$ must be identified  up to an anomalous factor $\mathfrak{i}(z)$. We find that
the Casimir energy $E_0$ and the anomalous factor $\mathfrak{i}(z)$ are given by
\begin{align}
\label{eq:index-polynomial-free}
\quad E_0 = -\frac{c_R - c_L }{12} \text{\footnotesize{ (mod 1)}}, \quad \mathfrak{i} (z) = -\sum_z d_z n_z  \cdot z^2.
\end{align}
We conclude that the elliptic genus $\mathcal{I}_k$ behaves under the modular transformation
$\tau \rightarrow \tfrac{a\tau + b}{c\tau+d}$, $z \rightarrow \tfrac{z}{c\tau+d}$ with $(\begin{smallmatrix}a&b\\c&d\end{smallmatrix}) \in \text{SL}(2,\mathbf{Z})$
as a weak Jacobi form of weight $0$ and index $-\sum_z d_z n_z  \cdot z^2$, i.e.,  \cite{Benini:2013nda,Benini:2013xpa}
\begin{align}
\label{eq:modular-property}
 \mathcal{I}_k \left(\frac{a\tau + b}{c\tau+d}, \frac{z}{c\tau+d}\right) = \varepsilon(a,b,c,d)\, \exp{\left(\frac{\pi i c }{c\tau + d} \cdot \sum_z d_z n_z   z^2 \right)}\ \mathcal{I}_k (\tau, z)
\end{align}	
where $\varepsilon(a,b,c,d)$ is a phase factor.

{The above derivation clearly shows that the 2d chiral anomaly of the 6d string theory determines the index of weak Jacobi forms of the corresponding elliptic genus. Note that the above argument is quite general so that we need not the gauge theory description of the 6d string theory.}

\subsection{Analytic properties}
\label{subsec:ansatz}

We expect the elliptic genus of the 6d strings to have the following structure: \cite{DelZotto:2016pvm}
\begin{align}
\label{eq:ansatz}
\mathcal{I}_k(\tau, z) = \eta(\tau)^{n_0}  \cdot \frac{\mathcal{N}(\tau, z)}{\mathcal{D} (\tau, z)}.
\end{align}
The overall factor $\eta(\tau)^{n_0}$ has been introduced to absorb the Casimir energy $E_0$ given in \eqref{eq:index-polynomial-free}, while
the numerator $\mathcal{N}(\tau, z)$ and denominator $\mathcal{D} (\tau, z)$ are Jacobi forms whose $q$-expansion starts at $q^0$ order.
This means that the exponent $n_0$ of the Dedekind eta function is given by
\begin{align}
n_0 = 24 E_0 = -2 (c_R - c_L) \text{\footnotesize{ (mod 24)}}.
\end{align}
We will always assume that the Casimir energy $E_0$ of the elliptic genus is non-positive and $|E_0| \leq 1$,
which hold true for all 6d theories studied throughout the paper.

\subsubsection{Pole structure}

The elliptic genus \eqref{eq:elliptic-genera} develops various poles at certain values of chemical potentials,
which lift the bosonic zero modes that parameterize the moduli space of the 6d strings. We will predict the location of poles by
inspecting these zero modes, making a conjecture on the denominator $\mathcal{D}(\tau,z)$ in \eqref{eq:ansatz}.

As the 6d strings wrapping on $T^2$ can freely move along the $\mathbf{R}^4$ plane, there exist the zero modes for their center-of-mass motion.
Had there not been the chemical potentials $\epsilon_1 \equiv \epsilon_+ + \epsilon_-$ and $\epsilon_2 \equiv \epsilon_+ - \epsilon_-$,
the elliptic genus would have suffered from these infrared divergences. 
Since the center-of-mass zero modes have been lifted by $\epsilon_1$ and $\epsilon_2$,
the elliptic genus must have two poles at $\epsilon_1 = 0$ and $\epsilon_2=0$.
Precisely speaking, we expect the 6d \emph{single} particle index on $\mathbf{R}^4 \times T^2$, defined by
\begin{align}
f_{\rm 6d}(\tau, z) \equiv \text{PE}^{-1}\,[ \textstyle Z_{6d} (\tau, z)] \quad \text{with} \quad
\text{PE} \left[ f(\tau, z) \right] \equiv \exp \left(\sum_{p=1}^\infty \frac{1}{p} \cdot f (p\tau, pz)	\right),
\end{align}
to have a simple pole at $\epsilon_1 = 0$ and $\epsilon_2=0$ \cite{Gopakumar:1998jq}. This causes the $k$ string elliptic genus $\mathcal{I}_k$
to have the following factor in the denominator $\mathcal{D}(\tau,z)$ \cite{DelZotto:2016pvm,Gu:2017ccq}.
\begin{align}
\label{eq:den-center-of-mass-k}
\tilde{\mathcal{D}}^{\rm com}_k (\tau, z) = \prod_{m=1}^{k} \frac{\theta_1 \left(m \epsilon_1\right)}{\eta^3} \frac{\theta_1 \left(m \epsilon_2\right)}{\eta^3}.
\end{align}
For the chain of $\{k_1,k_2,\cdots,k_n\}$ strings, the above factor is generalized as follows.
\begin{align}
\label{eq:den-center-of-mass}
\tilde{\mathcal{D}}^{\rm com}_{\{k_i\}} (\tau, z) = \prod_{i=1}^n  \tilde{\mathcal{D}}^{\rm com}_{k_i} (\tau, z) =
\prod_{i=1}^n\prod_{m=1}^{k_i} \frac{\theta_1 \left(m \epsilon_1\right)}{\eta^3} \frac{\theta_1 \left(m \epsilon_2\right)}{\eta^3}.
\end{align}
We notice that the $q$-expansion of \eqref{eq:den-center-of-mass-k} and \eqref{eq:den-center-of-mass} starts from $q^0$ as required in \eqref{eq:ansatz}.

The 6d strings are also the Yang-Mills instanton solitons in the 6d gauge theories.
As the translational zero modes along $\mathbf{R}^4$ have already been taken care of, here we focus on the bosonic zero modes that span
the \emph{reduced} instanton moduli space. We first consider the elliptic genus of $k$ $SU(2)$ instanton strings.
Taking the $q \rightarrow 0$ limit, it  is reduced to the Witten index of $k$ instantons in the 5d $SU(2)$ gauge theory
whose denominators are known from the 5d partition functions \cite{Hollowood:2003cv}. The poles are located at
\begin{align}
 \left.\begin{aligned}
        a\epsilon_1 + b \epsilon_2 + \alpha (\mathbf{a})  &= 0\quad \\
        a\epsilon_1 + b \epsilon_2 - \alpha (\mathbf{a})  &= 0\quad
       \end{aligned}
 \right\}
 \ \text{for positive integers $(a,b)$ such that $a b \leq k$}
\end{align}
where $\alpha (\mathbf{a})$ is the gauge holonomy for a positive root $\alpha  \in \Delta_+ = \{e_1 - e_2\}$, e.g.,
$\alpha (\mathbf{a}) = a_1 - a_2$. The entire denominator $\mathcal{D}(\tau,z)$ in the elliptic genus of $k$ $SU(2)$ instanton strings is a product of $\eqref{eq:den-center-of-mass-k}$ and
\begin{align}
\label{eq:SU2-ansatz}
\tilde{\mathcal{D}}^{SU(2)}_{k}  (\tau, z)= & \prod_{\substack{ ab\leq k \\  a,b > 0}} \frac{\theta_1 (a\epsilon_1 + b \epsilon_2 + \alpha (\mathbf{a})  )}{\eta^3}\frac{\theta_1 ( a\epsilon_1 + b \epsilon_2 - \alpha (\mathbf{a})  )}{\eta^{3}}
\end{align}
whose $q$-expansion starts at the $q^0$ order as required in \eqref{eq:ansatz}.

We recall that non-Abelian $G$ instantons can be constructed by embedding $SU(2)$ BPST instantons into $G$ \cite{Belavin:1975fg,Bernard:1977nr}.
For embedding $SU(2)$, we choose 3 generators of $G$ satisfying the $SU(2)$ algebra.
All possible choices of embedding are labeled by positive roots of $G$.
Denoting 3 generators by $T^{a}_{\alpha}$ with $a=1,2,3$, for a given positive root $\alpha \in \Delta_+$, the trace between them takes the form of
\begin{align}
\text{tr}\left(T^a_{\alpha} T^b_{\alpha}\right) = c_{\alpha} \delta^{ab}.
\end{align}
The constant $c_\alpha$ is normalized to be $1$ for every long root $\alpha$. Under such normalization,
 the constant $c_\alpha$ for a short root $\alpha$ becomes
\begin{align}
c_{\alpha} &= 2\quad \text{ if }\ G=Sp(N),\,SO(2N+1),\,F_4\\
c_{\alpha} &= 3\quad \text{ if }\ G=G_2.
\end{align}
Starting from the $SU(2)$ BPST solution carrying an instanton charge $k_{SU(2)}$, one can construct the $G$ instanton solution by embedding it to $\alpha \in \Delta^+$ of $G$. It carries an instanton charge $k_G = c_{\alpha}\, k_{SU(2)}$.
So the short root embedding can only produce the $G$ instanton solutions with  $k_G \geq c_\alpha$.
Such embedding structure must be reflected in the denominator of the elliptic genus of $G$ instanton strings.
In fact, the denominator $\mathcal{D}(\tau,z)$ of the $k$ string elliptic genus $\mathcal{I}_k$  is a product of $\eqref{eq:den-center-of-mass-k}$ and
\begin{align}
\label{eq:inst-ansatz}
\tilde{\mathcal{D}}^{G}_k  (\tau, z) =
\prod_{\alpha \in \Delta_l} \tilde{\mathcal{D}}^{SU(2)}_{k,\alpha}  (\tau, z) \cdot \prod_{\alpha \in \Delta_s}
\tilde{\mathcal{D}}^{SU(2)}_{\lfloor{k/c_\alpha}\rfloor,\alpha}  (\tau, z),
\end{align}
where $\tilde{\mathcal{D}}^{SU(2)}_{k,\alpha}  (\tau, z)$ is the $SU(2)$ denominator \eqref{eq:SU2-ansatz} after replacing $e_1 - e_2$ with a given root $\alpha$ of $G$.
More generally, for the chain of $\{k_1,\cdots, k_n\}$ strings in the $G = G_1  \otimes \cdots \otimes G_n$ quiver gauge theory, the denominator $\mathcal{D}(\tau,z)$ is generalized as a product of \eqref{eq:den-center-of-mass} and
\begin{align}
\tilde{\mathcal{D}}^{G}_{ \{k_i\}}  (\tau, z) =
\prod_{i=1}^n \tilde{\mathcal{D}}^{G_i}_{k_i}  (\tau, z)
\end{align}
where $\tilde{\mathcal{D}}^{\varnothing}_k  (\tau, z)$ is understood as $1$.
We checked  $\tilde{\mathcal{D}}^{\rm com}_k (\tau, z)\cdot \tilde{\mathcal{D}}^{G}_k  (\tau, z)$ in the $q\rightarrow 0$ limit  agrees with 
the denominator of the Witten index for 5d $k$ $G$ instantons in the following cases:
\begin{enumerate*}[label=(\arabic*)]
% \item $SU(2)$ $k=7$,
\item $G=SO(6)$, $k=3$,
\item $G=E_{6,7,8}$, $k=1$,
\item $G=SO(5)$, $k=3$,
\item $G=Sp(2)$, $k=4$,
\item $G=G_2$, $k=3$
\end{enumerate*}
\cite{Hwang:2014uwa,Cremonesi:2014xha,Hayashi:2017jze,Kim:2017}. In summary, we propose\footnote{As discussed in Section~\ref{sec:examples}, this ansatz should be modified when the ellipic genus has additional contributions from the Coulomb branches.}
\begin{align}
\label{eq:den-ansatz}
\mathcal{D} (\tau, z) = \tilde{\mathcal{D}}^{\rm com}_{\{k_i\}} (\tau, z)\cdot \tilde{\mathcal{D}}^{G}_{ \{k_i\}}  (\tau, z).
\end{align}

\subsubsection{Weyl invariant Jacobi forms}

The 6d string elliptic genus $\mathcal{I}_k$ is strongly constrained by the modular property \eqref{eq:modular-property} and the Weyl invariance of the global symmetry of 6d strings.
As the denominator \eqref{eq:den-ansatz} itself is a weak Jacobi form of certain weight $w_d$ and index $\mathfrak{i}_d$,
 % the modularity \eqref{eq:modular-property} requires
 the numerator $\mathcal{N}(\tau, z)$ has to be a weak Jacobi form of weight $(w_d - \frac{n_0}{2})$
and index $\mathfrak{i} + \mathfrak{i}_d$ to match the modularity \eqref{eq:modular-property} of the entire elliptic genus $\mathcal{I}_k$.
Similarly,
% And also,
as the denominator \eqref{eq:den-ansatz}  is invariant under the Weyl reflections of $SU(2)_l$, $SU(2)_d \subset SU(2)_r \times SU(2)_R$,
the 6d gauge group $G$, the 6d flavor group $F$, the numerator should also manifest the Weyl invariance.

% $\epsilon_+ \rightarrow -\epsilon_+$, $\epsilon_- \rightarrow -\epsilon_-$, $\mathbf{a} \rightarrow w_G(\mathbf{a})$ for all $w_G \in \text{Weyl}[G]$,

One way to guarantee the Weyl invariance is to express the numerator $\mathcal{N}(\tau, z)$
as the Weyl invariant Jacobi forms of $SU(2)_l$, $SU(2)_d$, $G$, and $F$ \cite{DelZotto:2016pvm,Gu:2017ccq}.
For a simple Lie algebra $R$, the Weyl invariant Jacobi forms of $R$ depend on the complex structure $\tau$ of $T^2$
and the chemical potentials $\mathbf{m} \equiv (m_1, m_2, \cdots,m_{|R|})$ conjugate to the Cartan generators of $R$.
They are characterized by two integers $w$ and $m>0$ and have the following properties \cite{MR781735,MR1163219,Sakai:2011xg}:
\begin{itemize}[itemsep=0pt,topsep=0pt]
	\item Weyl invariance
	\vspace{-0.5\baselineskip}
	\begin{align}
	\label{eq:weyl-inv-c1}
	\varphi_{w,m}(\tau, w_R(\mathbf{m})) = \varphi_{w,m}(\tau, \mathbf{m}) \qquad \text{for all }w_R\in \text{Weyl}[R]
	\end{align}
	\item Modular property
	\vspace{-0.5\baselineskip}
	\begin{align}
	\varphi_{w,m}\left(\frac{a\tau + b}{c \tau + d}, \frac{\mathbf{m}}{c\tau+d}\right) = \left(c\tau + d\right)^w \exp{\left( \frac{ \pi i m c  }{c\tau + d} \ \mathbf{m} \cdot \mathbf{m}\right)}
	\ \varphi_{w,m}(\tau, \mathbf{m})
	\end{align}
	\item Quasi-periodicity
	\vspace{-0.5\baselineskip}
	\begin{align}
	\varphi_{w,m}(\tau, \mathbf{m} + \mathbf{a} + \tau \mathbf{b}) = e^{-\pi i m   (\tau \mathbf{b}\cdot \mathbf{b} + 2 \mathbf{m} \cdot \mathbf{b})}\ \varphi_{w,m}(\tau, \mathbf{m})
	\end{align}
	\item Fourier expansion
	\vspace{-0.5\baselineskip}
	\begin{align}
	\label{eq:weyl-inv-c4}
	\varphi_{w,m}\left(\tau, \mathbf{m}\right) = \sum_{n=0}^\infty \sum_{\bm{\mu}} c(n, \bm{\mu})\cdot e^{2\pi i (n\tau+\bm{\mu} \cdot \mathbf{m})}
	\end{align}
\end{itemize}
The weight and index of $\varphi_{w,m}\left(\tau, \mathbf{m}\right)$ are $w$ and $-\frac{m}{2}(\mathbf{m} \cdot \mathbf{m})$, respectively.
We also note that a Weyl invariant Jacobi form $\varphi_{w,m}\left(\tau, \mathbf{m}\right)$ of $R$ can be constructed as a linear combination of level-$m$ theta functions of the affine Lie algebra $\hat{R}$,
defined as follows \cite{Bertola:1999,Sakai:2017ihc}.
\begin{align}
\label{eq:theta-def}
\Theta_{\bm{\lambda},m} (\tau, \mathbf{m}) \equiv \sum_{\bm{\alpha} \in \mathbf{adj}_R} \exp{\left(\pi i \tau (\bm{\alpha}+\bm{\lambda}/m)^2 + 2\pi i (m \bm{\alpha}+\bm{\lambda}) \cdot \mathbf{m} \right)}
\end{align}
It implies that the number of independent Weyl invariant Jacobi forms of index  $-\frac{m}{2}(\mathbf{m} \cdot \mathbf{m})$ is the same as
the number of level-$m$ fundamental representations of the affine Lie algebra $\hat{R}$ \cite{MR1163219,Sakai:2017ihc}.
Furthermore, the algebra of Weyl invariant Jacobi forms of $R$ over the algebra of modular forms $\mathbb{C}[E_4,E_6]$
with an integer-valued $m$ is freely generated by the following $\text{rank}(\hat{R})$ generators \cite{MR1163219}
\begin{align}
\label{eq:list-gen}
 \varphi_{-w_j,m_j} \quad \text{ for }j \in \{0,1,\cdots,\,\text{rank}(R)\}
\end{align}
except the case of $R =  E_8$. Here $\{w_j\}$ and $\{m_j\}$ collect the order of independent Casimirs
and the level of fundamental representations of $\hat{R}$, respectively.

The explicit forms of the generators \eqref{eq:list-gen}
are written in many literatures such as \cite{MR1163219,Bertola:1999,Sakai:2011xg,Sakai:2017ihc}.
For $R = A_n$ and $B_n$, all the $(n+1)$ generators
\begin{align}
SU(n+1):& \quad \varphi_{0,1}, \varphi_{-2,1}, \varphi_{-3,1}, \cdots, \varphi_{-n-1,1}\\
SO(2n+1):& \quad \varphi_{0,1}, \varphi_{-2,1}, \varphi_{-4,1}, \cdots, \varphi_{-2n,1}
\end{align}
can be constructed from the generating functions found in \cite{Bertola:1999}. Among the $C_n$ generators,
\begin{align}
Sp(n):& \quad \varphi_{0,1}, \varphi_{-2,1}, \varphi_{-4,1}, \varphi_{-6,2},\varphi_{-8,2}, \cdots, \varphi_{-2n,2},
\end{align}
all index-$2$ generators are identical to the $B_n$ generators, i.e., $\varphi_{-2l,2}^{C_n} = \varphi_{-2l,1}^{B_n}$ \cite{Bertola:1999}.
Also for $R=D_n$,
\begin{align}
SO(2n):& \quad \varphi_{0,1}, \varphi_{-2,1}, \varphi_{-4,1}, \varphi_{-n,1},\varphi_{-6,2}, \varphi_{-8,2}, \cdots, \varphi_{-2n+2,2},
\end{align}
all index-$2$ generators are  identical to the $B_n$ generators, i.e., $\varphi_{-2l,2}^{D_n} = \varphi_{-2l,1}^{B_n}$ \cite{Bertola:1999}.
The remaining index-$1$ generators for $C_n$ and $D_n$ can be constructed from the level-$1$ fundamental theta functions.
For example, the $D_n$ generators $\varphi_{0,1}$, $\varphi_{-2,1}$, $\varphi_{-4,1}$, $\varphi_{-n,1}$ can be written as follows.
\begin{align}
\label{eq:dn-gen}
\varphi_{-n,1} &= \frac{\prod_{i=1}^{n} \theta_1(a_i)}{\eta^{3n}}, \quad \varphi_{-4,1} = \frac{1}{\eta^{12}} \left(\frac{\prod_{i=1}^{n} \theta_3(a_i)}{\theta_3(0)^{n-4}} - \frac{\prod_{i=1}^{n} \theta_4(a_i)}{\theta_4(0)^{n-4}} -\frac{\prod_{i=1}^{n} \theta_2(a_i)}{\theta_2(0)^{n-4}}\right)\\
\varphi_{-2,1} &= \frac{\left(\theta_3(0)^4 + \theta_4(0)^4\right)}{\eta^{12}}\cdot \left(\frac{\prod_{i=1}^{n} \theta_3(a_i)}{\theta_3(0)^{n-4}} - \frac{\prod_{i=1}^{n} \theta_4(a_i)}{\theta_4(0)^{n-4}} +\frac{2\prod_{i=1}^{n} \theta_2(a_i)}{\theta_2(0)^{n-4}}\right) \nonumber\\&\quad -\frac{3\,\theta_2(0)^4}{\eta^{12}}\left(\frac{\prod_{i=1}^{n} \theta_3(a_i)}{\theta_3(0)^{n-4}} + \frac{\prod_{i=1}^{n} \theta_4(a_i)}{\theta_4(0)^{n-4}}\right) \nonumber\\
\varphi_{0,1} &= \frac{1}{\eta^{12}} \left(\frac{\prod_{i=1}^{n} \theta_3(a_i)}{\theta_3(0)^{n-12}} - \frac{\prod_{i=1}^{n} \theta_4(a_i)}{\theta_4(0)^{n-12}} -\frac{\prod_{i=1}^{n} \theta_2(a_i)}{\theta_2(0)^{n-12}}\right)\nonumber.
\end{align}
For $R=E_{n}$, all the $(n+1)$ generators are explicitly constructed in \cite{Sakai:2011xg,Sakai:2017ihc}.
The $F_4$ and $G_2$ generators are obtained from the $A_2$ and $D_4$ generators \cite{MR1163219,Bertola:1999}, e.g.,
\begin{align}
\label{eq:g2-gen}
\varphi_{0,1}^{G_2} = \varphi_{0,1}^{A_2}, \quad \varphi_{-2,1}^{G_2} = \varphi_{-2,1}^{A_2}, \quad \varphi_{-6,1}^{G_2} = \left(\varphi_{-3,1}^{A_2}\right)^2.
\end{align}
All the Weyl invariant Jacobi forms used in this paper will be explicitly displayed in Appendix~\ref{sec:weyl}.
%{\it Write also Weyl invariant forms of SU(n+1) at appendix.}
One typically finds more than one combinations of weak Jacobi forms of weight $(w_d - \frac{n_0}{2})$ and index $(\mathfrak{i} - \mathfrak{i}_d)$. Each of them is a product of the generators of Weyl invariant Jacobi forms for $SU(2)_l$, $SU(2)_d$, $G$, and $F$. Denoting them as $\{\Phi_{1}(\tau, z), \Phi_{2}(\tau, z), \cdots, \Phi_{l}(\tau, z)\}$, the numerator $\mathcal{N}(\tau, z)$ can be generally written as their linear combination, i.e., $\textstyle\mathcal{N}(\tau, z) = \sum_{j=1}^l c_j\, \Phi_{j}(\tau, z)$. A finite number of the numerical coefficients $\{c_1,c_2,\cdots,c_l\}$  will be determined through
comparison with the finite amount of the BPS spectral data in a given 6d theory \cite{DelZotto:2016pvm,Gu:2017ccq}.

\subsection{Test against known examples}
\label{subsec:example}

The conjectured formula \eqref{eq:ansatz} may reduce the problem of obtaining the 6d string elliptic genus
down to the problem of determining a finite number of numerical coefficients.
We will test if \eqref{eq:ansatz} holds for several known elliptic genera in 6d superconformal field theories.
{All of the examples we consider have an alternative gauge theory description.}

The anomaly polynomial of $k$ self-dual strings in $\mathcal{N}=(1,0)$ SCFTs is given by \cite{Kim:2016foj,Shimizu:2016lbw}
\begin{align}
\label{eq:anomaly-string}
\hspace{-0.3cm}k \left(\frac{Q}{4} \text{Tr}(\mathcal{F}_G^2) - \frac{1}{4}\text{Tr}(\mathcal{F}_F^2)- \frac{2-Q}{4}\left(p_1(T_2) -2c_2(l)-2c_2(r)\right) + h^\vee_G c_2(R)\right)
+ \frac{Q k^2}{2} \left(c_2(l) - c_2(r)\right)
\end{align}
where $Q$ is the Dirac pairing of self-dual strings. The field strength $\mathcal{F}_g$ of a Lie algebra $g$ is normalized such that a chiral fermion in a representation $\rho$ contributes $\hat{A}(T_2) \,\text{tr}_\rho (e^{i\mathcal{F}_g})$ to the  anomaly polynomial.
Following \cite{Ohmori:2014kda}, we use the normalized trace `Tr' defined by $\text{tr}_{\rm adj}(\mathcal{F}_g^2) = h_g^\vee\text{Tr}(\mathcal{F}_g^2)$ where $h_g^\vee$ is the dual Coxeter number of $g$.
The conversion factor $s_g$ between $\text{tr}_{\rm fnd}(\mathcal{F}_g^2) = s_g \text{Tr}(\mathcal{F}_g^2)$ is given by
\begin{align}
s_{SU(n)} = \tfrac{1}{2}, s_{SO(n)} = 1,  s_{Sp(n)} = \tfrac{1}{2},  s_{F_4} = 3,  s_{E_6} = 3,  s_{E_7} = 6,  s_{E_8} = 30,  s_{F_4} = 3,  s_{G_2} = 1.
\end{align}
The 2nd Chern class $c_2(g)$ of the $SU(2)$ bundle $g$ can be written as $c_2(g) = \frac{1}{4} \text{Tr}(\mathcal{F}_g^2)$ using the normalized trace.
$h^\vee_\varnothing$ and $\text{Tr}(\mathcal{F}_\varnothing)$ are understood as 1 and 0.
The anomaly polynomial \eqref{eq:anomaly-string} determines the index $\mathfrak{i}(z)$ of the elliptic genus based on \eqref{eq:index-polynomial-free}.
The denominator $\mathcal{D}(\tau,z)$ and the zero point energy $E_0$ of the elliptic genus have been discussed in Section~\ref{subsec:ansatz}.
We summarize $E_0$, $(\mathfrak{i} + \mathfrak{i}_d)$, and $(w_d - \frac{n_0}{2})$ of various elliptic genera in the following table.
\begin{center}
\begin{tabular}{|lll|ccc|l|}\hline
  $G$ & $F$ & $k$& $E_0$   & $n_0$ & $w_d - \frac{n_0}{2}$ & $\mathfrak{i} + \mathfrak{i}_d$ \\ \hline\hline
  $\varnothing$  & $SU(2)$&$1$ & $0$ & $0$ & $-2$ & $m_1^2 + \epsilon_+^2$\\
  $SU(2)$  & $SU(4)$ &$1$& $0$ & $0$ & $-4$ & $2a_1^2 + 4\epsilon_+^2 + \sum_{i \leq j}^3 m_i m_j$ \\\hline
  $\varnothing$  & $SO(16)$ & $1$& $-\frac{1}{2}$ & $-12$ &  $4$ & $\frac{1}{2}\sum_{i=1}^8 m_i^2$\\
  $Sp(1)$  & $SO(20)$ & $1$& $-\frac{1}{2}$& $-12$ & $2$ & $3 a_1^2 + 3\epsilon_+^2 + \frac{1}{2}\sum_{i=1}^{10}m_i^2$\\\hline
  $SU(3)$ & $\varnothing$ & $1$ & $-\frac{1}{2}$& $-12$ & $-2$ & $3 (a_1^2+a_2^2 + a_1a_2) + 12\epsilon_+^2$\\
  $G_2$ & $Sp(1)$ & $1$ & $-\frac{1}{2}$ & $-12$& $-2$ & $m_1^2 + 3 (a_1^2+a_2^2 + a_1a_2) + 11\epsilon_+^2$  \\\hline
  % $SO(8)$ & $\varnothing$ & $1$ & $-1$ & & $4 (\sum_{i=1}^4 a_i^2) + 46\epsilon_+^2$  \\\hline
\end{tabular}
\end{center}
We denote by $a_i$ and $m_j$ the chemical potentials for $G$ and $F$, respectively, which may be subject to the traceless condition $\sum_{i=1}^n a_i = 0$ and/or $\sum_{i=1}^{n'} m_i = 0$ if $G = SU(n)$ and/or $F = SU(n')$.

Let us determine the numerator $\mathcal{N}(\tau,z)$ in an appropriate ring of Weyl invariant Jacobi forms.
For brevity, we denote the $SU(2)_l$ and $SU(2)_d \subset SU(2)_r \times SU(2)_R$ Weyl invariant Jacobi forms by
\begin{align}
\mathfrak{L}_{2} = \varphi_{-2,1}^{SU(2)_l}(\tau,\epsilon_-),\ \mathfrak{L}_{0} = \varphi_{0,1}^{SU(2)_l}(\tau,\epsilon_-),\ \mathfrak{R}_{2} = \varphi_{-2,1}^{SU(2)_d}(\tau,\epsilon_+),\ \mathfrak{R}_{0} = \varphi_{0}^{SU(2)_d}(\tau,\epsilon_+),
\end{align}
and also the Weyl invariant Jacobi forms of $G$ and $F$ by $\mathfrak{g}_{w,m} = \varphi_{-w,m}^{G}(\tau,a_i)$ and $\mathfrak{f}_{w,m} = \varphi_{-w,m}^{F}(\tau,m_j)$.
In some particular cases, e.g., $k=1$ strings in non-Higgsable gauge theories, the numerators $\mathcal{N}(\tau,z)$ are
in the reduced ring of $SU(2)_d$ and $G$ Weyl invariant Jacobi forms, generated by \cite{DelZotto:2016pvm}
\begin{align}
% \tilde{\mathfrak{L}}_{2} = \varphi_{-2,1}^{SU(2)_l}(2\epsilon_-),\ \tilde{\mathfrak{L}}_{0} = \varphi_{0,1}^{SU(2)_l}(2\epsilon_-),\
\tilde{\mathfrak{R}}_{2} = \varphi_{-2,1}^{SU(2)_d}(2\epsilon_+),\ \tilde{\mathfrak{R}}_{0} = \varphi_{0,1}^{SU(2)_d}(2\epsilon_+),\ \mathfrak{g}_{w,m} = \varphi_{-w,m}^{G}(\tau,a).
\end{align}
We now determine the coefficients in  $\mathcal{N}(\tau,z)$ using the initially given BPS data from \cite{Haghighat:2013gba,Haghighat:2013tka,Klemm:1996hh,Kim:2014dza,Kim:2015fxa,Haghighat:2014vxa,Kim:2016foj}.

\paragraph{M-string ($G = \varnothing$, $F = SU(2)$, $k=1$)}
The numerator has 2 coefficients which can be determined through comparison with the initial BPS data at $q^0$ order. It turns out to be
\begin{align}
\mathcal{N} = \tfrac{1}{12}  \mathfrak{R}_0 \mathfrak{f}_{2,1} -\tfrac{1}{12}  \mathfrak{R}_2 \mathfrak{f}_{0,1} = \eta^{-6}\theta_1(m \pm \epsilon_+),
 \end{align}
reproducing the M-string elliptic genus in \cite{Haghighat:2013gba}.
\paragraph{$\bf SU(2)$ string ($G = SU(2)$, $F = SU(4)$, $k=1$)}
The numerator has 34 coefficients. $33$ of them are fixed using the initially given BPS data at $q^0$ order.
The remaining 1 coefficient is determined by the BPS data at $q^1$ order.
The numerator is given by
\begin{align}
\mathcal{N} &= \frac{1}{2^{11}3^5}\Big(
54 E_4^3 \mathfrak{f}_{4,1} \mathfrak{g}_{2,1}^2 \mathfrak{R}_2^4-36 E_4^2 \mathfrak{f}_{4,1} \mathfrak{g}_{2,1}^2 \mathfrak{R}_2^2 \mathfrak{R}_0^2+9 E_4^2 \mathfrak{f}_{2,1} \mathfrak{g}_{2,1}^2 \mathfrak{R}_2^3 \mathfrak{R}_0-48 E_4^2 \mathfrak{f}_{4,1} \mathfrak{g}_{0,1} \mathfrak{g}_{2,1} \mathfrak{R}_2^3 \mathfrak{R}_0-6 E_4^2 \mathfrak{f}_{4,1} \mathfrak{g}_{0,1}^2 \mathfrak{R}_2^4\nonumber\\
&-9 E_4^2 \mathfrak{f}_{2,1} \mathfrak{g}_{0,1} \mathfrak{g}_{2,1} \mathfrak{R}_2^4-32 E_4 E_6 \mathfrak{f}_{4,1} \mathfrak{g}_{2,1}^2 \mathfrak{R}_2^3 \mathfrak{R}_0
-16 E_4 E_6 \mathfrak{f}_{4,1} \mathfrak{g}_{0,1} \mathfrak{g}_{2,1} \mathfrak{R}_2^4-2 E_4 \mathfrak{f}_{4,1} \mathfrak{g}_{2,1}^2 \mathfrak{R}_0^4+3 E_4 \mathfrak{f}_{2,1} \mathfrak{g}_{2,1}^2 \mathfrak{R}_2 \mathfrak{R}_0^3\nonumber\\&
-16 E_4 \mathfrak{f}_{4,1} \mathfrak{g}_{0,1} \mathfrak{g}_{2,1} \mathfrak{R}_2 \mathfrak{R}_0^3-12 E_4 \mathfrak{f}_{4,1} \mathfrak{g}_{0,1}^2 \mathfrak{R}_2^2 \mathfrak{R}_0^2-108 E_4 \mathfrak{f}_{0,1} \mathfrak{g}_{2,1}^2 \mathfrak{R}_2^2 \mathfrak{R}_0^2-3 E_4 \mathfrak{f}_{2,1} \mathfrak{g}_{0,1}^2 \mathfrak{R}_2^3 \mathfrak{R}_0+72 E_4 \mathfrak{f}_{0,1} \mathfrak{g}_{0,1} \mathfrak{g}_{2,1} \mathfrak{R}_2^3 \mathfrak{R}_0\nonumber\\&
+36 E_4 \mathfrak{f}_{0,1} \mathfrak{g}_{0,1}^2 \mathfrak{R}_2^4-64 E_6^2 \mathfrak{f}_{4,1} \mathfrak{g}_{2,1}^2 \mathfrak{R}_2^4
-16 E_6 \mathfrak{f}_{4,1} \mathfrak{g}_{2,1}^2 \mathfrak{R}_2 \mathfrak{R}_0^3+12 E_6 \mathfrak{f}_{2,1} \mathfrak{g}_{2,1}^2 \mathfrak{R}_2^2 \mathfrak{R}_0^2-48 E_6 \mathfrak{f}_{4,1} \mathfrak{g}_{0,1} \mathfrak{g}_{2,1} \mathfrak{R}_2^2 \mathfrak{R}_0^2\nonumber
\\&-16 E_6 \mathfrak{f}_{4,1} \mathfrak{g}_{0,1}^2 \mathfrak{R}_2^3 \mathfrak{R}_0-96 E_6 \mathfrak{f}_{0,1} \mathfrak{g}_{2,1}^2 \mathfrak{R}_2^3 \mathfrak{R}_0-8 E_6 \mathfrak{f}_{2,1} \mathfrak{g}_{0,1} \mathfrak{g}_{2,1} \mathfrak{R}_2^3 \mathfrak{R}_0-4 E_6 \mathfrak{f}_{2,1} \mathfrak{g}_{0,1}^2 \mathfrak{R}_2^4+96 E_6 \mathfrak{f}_{0,1} \mathfrak{g}_{0,1} \mathfrak{g}_{2,1} \mathfrak{R}_2^4\nonumber
\\&+2 \mathfrak{f}_{4,1} \mathfrak{g}_{0,1}^2 \mathfrak{R}_0^4+12 \mathfrak{f}_{0,1} \mathfrak{g}_{2,1}^2 \mathfrak{R}_0^4+\mathfrak{f}_{2,1} \mathfrak{g}_{0,1} \mathfrak{g}_{2,1} \mathfrak{R}_0^4-\mathfrak{f}_{2,1} \mathfrak{g}_{0,1}^2 \mathfrak{R}_2 \mathfrak{R}_0^3+24 \mathfrak{f}_{0,1} \mathfrak{g}_{0,1} \mathfrak{g}_{2,1} \mathfrak{R}_2 \mathfrak{R}_0^3-36 \mathfrak{f}_{0,1} \mathfrak{g}_{0,1}^2 \mathfrak{R}_2^2 \mathfrak{R}_0^2\Big).\nonumber
\end{align}
We checked that this agrees with the previously known result \cite{Haghighat:2013tka} up to $q^3$ order.

%{\it SU(N) with 2N fundamentals?}

\paragraph{E-string ($G = \varnothing$, $F = SO(16)$, $k=1$)} There are five terms in the numerator, given as
\begin{align*}
\mathcal{N}=\frac{1}{2^7 3^3} \big(-\mathfrak{f}_{8,1}  E_4^3+48\mathfrak{f}_{4,1} E_4^2-72\mathfrak{f}_{0,1}  E_4+\mathfrak{f}_{8,1} E_6^2+12 \mathfrak{f}_{2,1}  E_6\big).
\end{align*}
Four coefficients are determined by the BPS data at $q^{-1/2}$ order. The last one is fixed at $q^{1/2}$ order.
We checked its agreement with the known E-string elliptic genus \cite{Klemm:1996hh,Haghighat:2014pva,Kim:2014dza,Kim:2015fxa} up to $q^{5/2}$ order.

\paragraph{$\bf Sp(1)$ string ($G = Sp(1)$, $F = SO(20)$, $k=1$)} The numerator has $91$ coefficients.
We use the initial BPS data at $q^{-1/2}$, $q^{1/2}$, $q^{3/2}$ orders to fix $63$, $27$, $1$ of those coefficients, respectively. After all,
\begin{align*}
\mathcal{N} &= \frac{1}{2^{17} 3^{9}}\big(
    -3^3 \mathfrak{f}_{10,1} \mathfrak{g}_{2,1}^3 \mathfrak{R}_{2}^3 E_4^6+3^2 \mathfrak{f}_{10,1} \mathfrak{g}_{2,1} \mathfrak{R}_{2}^3 \mathfrak{g}_{0,1}^2 E_4^5
     +3^2 \mathfrak{f}_{10,1} \mathfrak{g}_{2,1}^3 \mathfrak{R}_{2} \mathfrak{R}_{0}^2 E_4^5+3^3 \mathfrak{f}_{10,1} \mathfrak{g}_{2,1}^2 \mathfrak{R}_{2}^2 \mathfrak{g}_{0,1} \mathfrak{R}_{0} E_4^5 \\&
     +3 \mathfrak{f}_{10,1} \mathfrak{g}_{2,1}^2 \mathfrak{g}_{0,1} \mathfrak{R}_{0}^3 E_4^4+3^2 \mathfrak{f}_{10,1} \mathfrak{g}_{2,1} \mathfrak{R}_{2} \mathfrak{g}_{0,1}^2 \mathfrak{R}_{0}^2 E_4^4
     -2^6 3^3 \mathfrak{f}_{4,1} \mathfrak{g}_{2,1}^2 \mathfrak{R}_{2}^3 \mathfrak{g}_{0,1} E_4^4+2^2 3^1 \mathfrak{f}_{10,1} E_6 \mathfrak{g}_{2,1}^2 \mathfrak{R}_{2}^3 \mathfrak{g}_{0,1} E_4^4\\&
     +3^1 \mathfrak{f}_{10,1} \mathfrak{R}_{2}^2 \mathfrak{g}_{0,1}^3 \mathfrak{R}_{0} E_4^4
     +2^6 3^3 \mathfrak{f}_{4,1} \mathfrak{g}_{2,1}^3 \mathfrak{R}_{2}^2 \mathfrak{R}_{0} E_4^4+2^2 3^1 \mathfrak{f}_{10,1} E_6 \mathfrak{g}_{2,1}^3 \mathfrak{R}_{2}^2 \mathfrak{R}_{0} E_4^4+59^1 \mathfrak{f}_{10,1} E_6^2 \mathfrak{g}_{2,1}^3 \mathfrak{R}_{2}^3 E_4^3\\&
     -2^6 3^2 \mathfrak{f}_{4,1} \mathfrak{R}_{2}^3 \mathfrak{g}_{0,1}^3 E_4^3+2^1 \mathfrak{f}_{10,1} E_6 \mathfrak{R}_{2}^3 \mathfrak{g}_{0,1}^3 E_4^3+2^6 3^2 \mathfrak{f}_{4,1} \mathfrak{g}_{2,1}^3 \mathfrak{R}_{0}^3 E_4^3+2^1 \mathfrak{f}_{10,1} E_6 \mathfrak{g}_{2,1}^3 \mathfrak{R}_{0}^3 E_4^3-\mathfrak{f}_{10,1} \mathfrak{g}_{0,1}^3 \mathfrak{R}_{0}^3 E_4^3\\&
     -2^4 3^3 \mathfrak{f}_{2,1} \mathfrak{g}_{2,1} \mathfrak{R}_{2}^3 \mathfrak{g}_{0,1}^2 E_4^3+2^4 3^3 \mathfrak{f}_{2,1} \mathfrak{g}_{2,1}^3 \mathfrak{R}_{2} \mathfrak{R}_{0}^2 E_4^3+2^6 3^3 \mathfrak{f}_{4,1} \mathfrak{g}_{2,1}^2 \mathfrak{R}_{2} \mathfrak{g}_{0,1} \mathfrak{R}_{0}^2 E_4^3+2^1 3^2 \mathfrak{f}_{10,1} E_6 \mathfrak{g}_{2,1}^2 \mathfrak{R}_{2} \mathfrak{g}_{0,1} \mathfrak{R}_{0}^2 E_4^3\\&
     -2^5 3^4 \mathfrak{f}_{0,1} \mathfrak{g}_{2,1}^2 \mathfrak{R}_{2}^3 \mathfrak{g}_{0,1} E_4^3+2^5 3^4 \mathfrak{f}_{0,1} \mathfrak{g}_{2,1}^3 \mathfrak{R}_{2}^2 \mathfrak{R}_{0} E_4^3-2^6 3^3 \mathfrak{f}_{4,1} \mathfrak{g}_{2,1} \mathfrak{R}_{2}^2 \mathfrak{g}_{0,1}^2 \mathfrak{R}_{0} E_4^3+2^1 3^2 \mathfrak{f}_{10,1} E_6 \mathfrak{g}_{2,1} \mathfrak{R}_{2}^2 \mathfrak{g}_{0,1}^2 \mathfrak{R}_{0} E_4^3\\&
     +2^6 3^3 \mathfrak{f}_{0,1} \mathfrak{R}_{2}^3 \mathfrak{g}_{0,1}^3 E_4^2-2^6 3^3 \mathfrak{f}_{0,1} \mathfrak{g}_{2,1}^3 \mathfrak{R}_{0}^3 E_4^2-2^6 3^2 \mathfrak{f}_{4,1} \mathfrak{g}_{2,1} \mathfrak{g}_{0,1}^2 \mathfrak{R}_{0}^3 E_4^2+2^4 3^2 \mathfrak{f}_{2,1} \mathfrak{g}_{2,1}^2 \mathfrak{g}_{0,1} \mathfrak{R}_{0}^3 E_4^2\\&
     -3^2 \mathfrak{f}_{10,1} E_6^2 \mathfrak{g}_{2,1} \mathfrak{R}_{2}^3 \mathfrak{g}_{0,1}^2 E_4^2-2^6 3^2 7^1 \mathfrak{f}_{4,1} E_6 \mathfrak{g}_{2,1} \mathfrak{R}_{2}^3 \mathfrak{g}_{0,1}^2 E_4^2+2^6 3^2 \mathfrak{f}_{4,1} \mathfrak{R}_{2} \mathfrak{g}_{0,1}^3 \mathfrak{R}_{0}^2 E_4^2-3^2 \mathfrak{f}_{10,1} E_6^2 \mathfrak{g}_{2,1}^3 \mathfrak{R}_{2} \mathfrak{R}_{0}^2 E_4^2\\&
     +2^6 3^2 7^1 \mathfrak{f}_{4,1} E_6 \mathfrak{g}_{2,1}^3 \mathfrak{R}_{2} \mathfrak{R}_{0}^2 E_4^2-2^6 3^4 \mathfrak{f}_{0,1} \mathfrak{g}_{2,1}^2 \mathfrak{R}_{2} \mathfrak{g}_{0,1} \mathfrak{R}_{0}^2 E_4^2-2^4 3^2 5^1 \mathfrak{f}_{2,1} E_6 \mathfrak{g}_{2,1}^2 \mathfrak{R}_{2}^3 \mathfrak{g}_{0,1} E_4^2-2^4 3^2 \mathfrak{f}_{2,1} \mathfrak{R}_{2}^2 \mathfrak{g}_{0,1}^3 \mathfrak{R}_{0} E_4^2\\&
     +2^4 3^2 5^1 \mathfrak{f}_{2,1} E_6 \mathfrak{g}_{2,1}^3 \mathfrak{R}_{2}^2 \mathfrak{R}_{0} E_4^2+2^6 3^4 \mathfrak{f}_{0,1} \mathfrak{g}_{2,1} \mathfrak{R}_{2}^2 \mathfrak{g}_{0,1}^2 \mathfrak{R}_{0} E_4^2-3^3 \mathfrak{f}_{10,1} E_6^2 \mathfrak{g}_{2,1}^2 \mathfrak{R}_{2}^2 \mathfrak{g}_{0,1} \mathfrak{R}_{0} E_4^2-2^5 3^2 \mathfrak{f}_{2,1} E_6 \mathfrak{R}_{2}^3 \mathfrak{g}_{0,1}^3 E_4\\&
     +2^5 3^2 \mathfrak{f}_{2,1} E_6 \mathfrak{g}_{2,1}^3 \mathfrak{R}_{0}^3 E_4
     +2^5 3^3 \mathfrak{f}_{0,1} \mathfrak{g}_{2,1} \mathfrak{g}_{0,1}^2 \mathfrak{R}_{0}^3 E_4-3^1 \mathfrak{f}_{10,1} E_6^2 \mathfrak{g}_{2,1}^2 \mathfrak{g}_{0,1} \mathfrak{R}_{0}^3 E_4+2^6 3^2 \mathfrak{f}_{4,1} E_6 \mathfrak{g}_{2,1}^2 \mathfrak{g}_{0,1} \mathfrak{R}_{0}^3 E_4\\&
     +2^5 3^3 7^1 \mathfrak{f}_{0,1} E_6 \mathfrak{g}_{2,1} \mathfrak{R}_{2}^3 \mathfrak{g}_{0,1}^2 E_4-2^5 3^3 \mathfrak{f}_{0,1} \mathfrak{R}_{2} \mathfrak{g}_{0,1}^3 \mathfrak{R}_{0}^2 E_4-3^2 \mathfrak{f}_{10,1} E_6^2 \mathfrak{g}_{2,1} \mathfrak{R}_{2} \mathfrak{g}_{0,1}^2 \mathfrak{R}_{0}^2 E_4
     -2^5 3^3 7^1 \mathfrak{f}_{0,1} E_6 \mathfrak{g}_{2,1}^3 \mathfrak{R}_{2} \mathfrak{R}_{0}^2 E_4\\&
     +2^5 3^3 \mathfrak{f}_{2,1} E_6 \mathfrak{g}_{2,1}^2 \mathfrak{R}_{2} \mathfrak{g}_{0,1} \mathfrak{R}_{0}^2 E_4-2^2 3^1 \mathfrak{f}_{10,1} E_6^3 \mathfrak{g}_{2,1}^2 \mathfrak{R}_{2}^3 \mathfrak{g}_{0,1} E_4-2^7 3^2 \mathfrak{f}_{4,1} E_6^2 \mathfrak{g}_{2,1}^2 \mathfrak{R}_{2}^3 \mathfrak{g}_{0,1} E_4-3^1 \mathfrak{f}_{10,1} E_6^2 \mathfrak{R}_{2}^2 \mathfrak{g}_{0,1}^3 \mathfrak{R}_{0} E_4
     \\&-2^6 3^2 \mathfrak{f}_{4,1} E_6 \mathfrak{R}_{2}^2 \mathfrak{g}_{0,1}^3 \mathfrak{R}_{0} E_4-2^2 3^1 \mathfrak{f}_{10,1} E_6^3 \mathfrak{g}_{2,1}^3 \mathfrak{R}_{2}^2 \mathfrak{R}_{0} E_4+2^7 3^2 \mathfrak{f}_{4,1} E_6^2 \mathfrak{g}_{2,1}^3 \mathfrak{R}_{2}^2 \mathfrak{R}_{0} E_4-2^5 3^3 \mathfrak{f}_{2,1} E_6 \mathfrak{g}_{2,1} \mathfrak{R}_{2}^2 \mathfrak{g}_{0,1}^2 \mathfrak{R}_{0} E_4\\&
     -2^5 \mathfrak{f}_{10,1} E_6^4 \mathfrak{g}_{2,1}^3 \mathfrak{R}_{2}^3-2^1 \mathfrak{f}_{10,1} E_6^3 \mathfrak{R}_{2}^3 \mathfrak{g}_{0,1}^3-2^6 3^2 \mathfrak{f}_{4,1} E_6^2 \mathfrak{R}_{2}^3 \mathfrak{g}_{0,1}^3-2^1 \mathfrak{f}_{10,1} E_6^3 \mathfrak{g}_{2,1}^3 \mathfrak{R}_{0}^3+2^6 3^2 \mathfrak{f}_{4,1} E_6^2 \mathfrak{g}_{2,1}^3 \mathfrak{R}_{0}^3\\&
     +\mathfrak{f}_{10,1} E_6^2 \mathfrak{g}_{0,1}^3 \mathfrak{R}_{0}^3-2^4 3^2 \mathfrak{f}_{2,1} E_6 \mathfrak{g}_{2,1} \mathfrak{g}_{0,1}^2 \mathfrak{R}_{0}^3
     -2^5 3^3 \mathfrak{f}_{0,1} E_6 \mathfrak{g}_{2,1}^2 \mathfrak{g}_{0,1} \mathfrak{R}_{0}^3-2^6 3^2 \mathfrak{f}_{2,1} E_6^2 \mathfrak{g}_{2,1} \mathfrak{R}_{2}^3 \mathfrak{g}_{0,1}^2+2^4 3^2 \mathfrak{f}_{2,1} E_6 \mathfrak{R}_{2} \mathfrak{g}_{0,1}^3 \mathfrak{R}_{0}^2\\&
     +2^6 3^2 \mathfrak{f}_{2,1} E_6^2 \mathfrak{g}_{2,1}^3 \mathfrak{R}_{2} \mathfrak{R}_{0}^2-2^1 3^2 \mathfrak{f}_{10,1} E_6^3 \mathfrak{g}_{2,1}^2 \mathfrak{R}_{2} \mathfrak{g}_{0,1} \mathfrak{R}_{0}^2+2^6 3^3 \mathfrak{f}_{4,1} E_6^2 \mathfrak{g}_{2,1}^2 \mathfrak{R}_{2} \mathfrak{g}_{0,1} \mathfrak{R}_{0}^2+2^8 3^3 \mathfrak{f}_{0,1} E_6^2 \mathfrak{g}_{2,1}^2 \mathfrak{R}_{2}^3 \mathfrak{g}_{0,1}\\&
     +2^5 3^3 \mathfrak{f}_{0,1} E_6 \mathfrak{R}_{2}^2 \mathfrak{g}_{0,1}^3 \mathfrak{R}_{0}-2^8 3^3 \mathfrak{f}_{0,1} E_6^2 \mathfrak{g}_{2,1}^3 \mathfrak{R}_{2}^2 \mathfrak{R}_{0}-2^1 3^2 \mathfrak{f}_{10,1} E_6^3 \mathfrak{g}_{2,1} \mathfrak{R}_{2}^2 \mathfrak{g}_{0,1}^2 \mathfrak{R}_{0}-2^6 3^3 \mathfrak{f}_{4,1} E_6^2 \mathfrak{g}_{2,1} \mathfrak{R}_{2}^2 \mathfrak{g}_{0,1}^2 \mathfrak{R}_{0}
   \big).
\end{align*}
We checked that it agrees the known $Sp(1)$ string elliptic genus \cite{Kim:2015fxa} up to $q^{7/2}$ order.

\paragraph{$\bf SU(3)$ string ($G = SU(3)$, $F = \varnothing$, $k=1$)} The numerator has $21$ coefficients.
One can fix $20$ and $1$ of those coefficients using the BPS data at $q^{-1/2}$ and $q^{1/2}$ orders, respectively. We checked that
\begin{align*}
\mathcal{N} &= \frac{1}{2^{12} 3^{4}}\big(
    -24 E_4^3 \tilde{\mathfrak{R}}_2^3 \mathfrak{g}_{2,1} \mathfrak{g}_{3,1}^2+24 E_4^2 \tilde{\mathfrak{R}}_2 \tilde{\mathfrak{R}}_0^2 \mathfrak{g}_{2,1} \mathfrak{g}_{3,1}^2+3 E_4^2 \tilde{\mathfrak{R}}_2^2
   \tilde{\mathfrak{R}}_0 \mathfrak{g}_{2,1}^3-288 E_4^2 \tilde{\mathfrak{R}}_2^2 \tilde{\mathfrak{R}}_0 \mathfrak{g}_{0,1} \mathfrak{g}_{3,1}^2\\&
   +36 E_4^2 \tilde{\mathfrak{R}}_2^3 \mathfrak{g}_{0,1}
   \mathfrak{g}_{2,1}^2+24 E_4 E_6 \tilde{\mathfrak{R}}_2^2 \tilde{\mathfrak{R}}_0 \mathfrak{g}_{2,1} \mathfrak{g}_{3,1}^2-96 E_4 E_6 \tilde{\mathfrak{R}}_2^3 \mathfrak{g}_{0,1}
   \mathfrak{g}_{3,1}^2+E_4 \tilde{\mathfrak{R}}_0^3 \mathfrak{g}_{2,1}^3-96 E_4 \tilde{\mathfrak{R}}_0^3 \mathfrak{g}_{0,1} \mathfrak{g}_{3,1}^2\\&
   -36 E_4 \tilde{\mathfrak{R}}_2 \tilde{\mathfrak{R}}_0^2
   \mathfrak{g}_{0,1} \mathfrak{g}_{2,1}^2-432 E_4 \tilde{\mathfrak{R}}_2^2 \tilde{\mathfrak{R}}_0 \mathfrak{g}_{0,1}^2 \mathfrak{g}_{2,1}+1728 E_4 \tilde{\mathfrak{R}}_2^3 \mathfrak{g}_{0,1}^3+32
   E_6^2 \tilde{\mathfrak{R}}_2^3 \mathfrak{g}_{2,1} \mathfrak{g}_{3,1}^2+8 E_6 \tilde{\mathfrak{R}}_0^3 \mathfrak{g}_{2,1} \mathfrak{g}_{3,1}^2\\&
   +4 E_6 \tilde{\mathfrak{R}}_2 \tilde{\mathfrak{R}}_0^2
   \mathfrak{g}_{2,1}^3-288 E_6 \tilde{\mathfrak{R}}_2 \tilde{\mathfrak{R}}_0^2 \mathfrak{g}_{0,1} \mathfrak{g}_{3,1}^2-576 E_6 \tilde{\mathfrak{R}}_2^3 \mathfrak{g}_{0,1}^2 \mathfrak{g}_{2,1}-144
   \tilde{\mathfrak{R}}_0^3 \mathfrak{g}_{0,1}^2 \mathfrak{g}_{2,1}-1728 \tilde{\mathfrak{R}}_2 \tilde{\mathfrak{R}}_0^2 \mathfrak{g}_{0,1}^3
   \big).
\end{align*}
agrees with the known $SU(3)$ string elliptic genus \cite{Kim:2016foj} up to $q^{7/2}$ order.

\paragraph{$\bf G_2$ string ($G = G_2$, $F = Sp(1)$, $k=1$)}
The 6d theory has the $G_2$ gauge symmetry with 1 hypermultiplet in {$\mathbf{7}$}. Upon the Higgsing, it gives rise to the minimal $SU(3)$ SCFT.
The numerator has $232$ coefficients. For simplicity, we take the $\epsilon_+ \rightarrow 0$ limit which cuts the number of independent Weyl invariant Jacobi forms to $9$.
All of them can be determined through comparison with the initially given BPS data at $q^{-1/2}$ order as follows.
\begin{align}
\mathcal{N} = &\frac{1}{2^8 3^2}
\big(
-96 \mathfrak{g}_{6,2}\mathfrak{g}_{0,1}E_6 \mathfrak{f}_{2,1}-96 \mathfrak{g}_{6,2}\mathfrak{g}_{0,1} E_4 \mathfrak{f}_{0,1}
+8 \mathfrak{g}_{2,1} \mathfrak{g}_{6,2} E_4^2 \mathfrak{f}_{2,1}-4 \mathfrak{g}_{2,1}^3 E_6 \mathfrak{f}_{2,1}\\&
+\mathfrak{g}_{2,1}^3 E_4 \mathfrak{f}_{0,1}+8 \mathfrak{g}_{2,1} \mathfrak{g}_{6,2} E_6 \mathfrak{f}_{0,1}
-5184 \mathfrak{g}_{0,1}^3 \mathfrak{f}_{2,1}-144 \mathfrak{g}_{2,1} \mathfrak{g}_{0,1}^2 \mathfrak{f}_{0,1} + 84 \mathfrak{g}_{2,1}^2 \mathfrak{g}_{0,1} E_4 \mathfrak{f}_{2,1}
\Big)\nonumber
\end{align}
We checked its agreement with \cite{Kim:2017} until $q^{5/2}$ order. If we instead keep $\epsilon_+$ and turn off the $G_2$ gauge holonomy,
the numerator has 30 coefficients. $21$ and $9$ of those coefficients are fixed using the initial BPS data at $q^{-1/2}$ and $q^{1/2}$ orders, respectively.
We checked that the numerator
\begin{align}
\mathcal{N} =&\frac{1}{2^{21} 3^{10}} \Big(
-2^9 E_4 E_6^3 \mathfrak{f}_{2,1} \mathfrak{R}_{2}^{11}+2^3 3^4 E_4^4 E_6 \mathfrak{f}_{2,1} \mathfrak{R}_{2}^{11}
+3^6 E_4^5 \mathfrak{f}_{0,1} \mathfrak{R}_{2}^{11}-3\cdot 2^8 E_4^2 E_6^2 \mathfrak{f}_{0,1} \mathfrak{R}_{2}^{11}+3^5 E_4^5 \mathfrak{f}_{2,1} \mathfrak{R}_{0} \mathfrak{R}_{2}^{10}\nonumber\\&
+3\cdot 2^8 E_4^2 E_6^2 \mathfrak{f}_{2,1} \mathfrak{R}_{0} \mathfrak{R}_{2}^{10}-2^{10} E_6^3 \mathfrak{R}_{0} \mathfrak{f}_{0,1} \mathfrak{R}_{2}^{10}\nonumber
+5\cdot 2^4 3^2 E_4^3 E_6 \mathfrak{R}_{0} \mathfrak{f}_{0,1} \mathfrak{R}_{2}^{10}+3\cdot 2^{10} E_6^3 \mathfrak{f}_{2,1} \mathfrak{R}_{0}^2 \mathfrak{R}_{2}^9\nonumber\\&
+2^4 3^2 E_4^3 E_6 \mathfrak{f}_{2,1} \mathfrak{R}_{0}^2 \mathfrak{R}_{2}^9
-31\cdot 3^3 E_4^4 \mathfrak{R}_{0}^2 \mathfrak{f}_{0,1} \mathfrak{R}_{2}^9-3\cdot 2^6 E_4 E_6^2 \mathfrak{R}_{0}^2 \mathfrak{f}_{0,1} \mathfrak{R}_{2}^9+3^4 E_4^4 \mathfrak{f}_{2,1} \mathfrak{R}_{0}^3 \mathfrak{R}_{2}^8 -33\cdot 2^6 E_6^2 \mathfrak{R}_{0}^4 \mathfrak{f}_{0,1} \mathfrak{R}_{2}^7\nonumber\\&
+87\cdot 2^6 E_4 E_6^2 \mathfrak{f}_{2,1} \mathfrak{R}_{0}^3 \mathfrak{R}_{2}^8-123\cdot 2^4 E_4^2 E_6 \mathfrak{R}_{0}^3 \mathfrak{f}_{0,1} \mathfrak{R}_{2}^8+3\cdot 61^1 2^5 E_4^2 E_6 \mathfrak{f}_{2,1} \mathfrak{R}_{0}^4 \mathfrak{R}_{2}^7-22\cdot 3^2 E_4^3 \mathfrak{R}_{0}^4 \mathfrak{f}_{0,1} \mathfrak{R}_{2}^7\nonumber\\&
+238\cdot 3^2 E_4^3 \mathfrak{f}_{2,1} \mathfrak{R}_{0}^5 \mathfrak{R}_{2}^6+21\ 2^6 E_6^2 \mathfrak{f}_{2,1} \mathfrak{R}_{0}^5 \mathfrak{R}_{2}^6-105\cdot 2^4 E_4 E_6 \mathfrak{R}_{0}^5 \mathfrak{f}_{0,1} \mathfrak{R}_{2}^6
+7\cdot 2^4 3^2 E_4 E_6 \mathfrak{f}_{2,1} \mathfrak{R}_{0}^6 \mathfrak{R}_{2}^5\nonumber\\&
-714 E_4^2 \mathfrak{R}_{0}^6 \mathfrak{f}_{0,1} \mathfrak{R}_{2}^5+66 E_4^2 \mathfrak{f}_{2,1} \mathfrak{R}_{0}^7 \mathfrak{R}_{2}^4 -2^4 3^2 E_6 \mathfrak{R}_{0}^7 \mathfrak{f}_{0,1} \mathfrak{R}_{2}^4+3\cdot 2^3 E_6 \mathfrak{f}_{2,1} \mathfrak{R}_{0}^8 \mathfrak{R}_{2}^3
-3 E_4 \mathfrak{R}_{0}^8 \mathfrak{f}_{0,1} \mathfrak{R}_{2}^3\nonumber\\&
+31 E_4 \mathfrak{f}_{2,1} \mathfrak{R}_{0}^9 \mathfrak{R}_{2}^2-\mathfrak{R}_{0}^{10} \mathfrak{f}_{0,1} \mathfrak{R}_{2}-3 \mathfrak{f}_{2,1} \mathfrak{R}_{0}^{11}
\Big)
\end{align}
agrees with the known $G_2$ string elliptic genus \cite{Kim:2017} up to $q^{5/2}$ order.

\section{Anomaly polynomial of little strings}
\label{sec:anomaly-polynomial}

{In the previous section, we study various 6d SCFTs and work out their BPS spectrum. We naturally expect that the same can be worked out for little string theories.}
Here we initiate our study on LSTs by exploring their 6d/2d anomaly polynomials,
which are necessary to bootstrap the 6d string elliptic genera and the $\mathbf{R}^4 \times T^2$ partition functions.
They are the worldvolume theories of $n$ NS5-branes in the decoupling limit $g_s \rightarrow 0$.
{The corresponding anomaly polynomials will be worked out using the anomaly inflow arguments starting from 10-dimensional string theory.}
% Recall that the 6d string elliptic genera are strongly constrained by their modular behavior, which reflects the chiral anomaly of
% 2d supersymmetric field theories describing 6d strings \cite{DelZotto:2016pvm,Gu:2017ccq}.

Any consistent string theory background should be free from gravitational and gauge anomalies, which are encoded in the 10d anomaly polynomial $I_{12}$.
It is the characteric polynomial made of the Pontryagin class $p_{i}(T_{10})$ of the 10d tangent bundle $T_{10}$ and
the Chern class $c_i(g)$ of the gauge bundle $g$.
The anomaly polynomial $I_{12}$ vanishes for type IIA and IIB string theories, i.e., $I_{12} = 0$.
%  are free from the chiral anomalies, so that
% $F_g$ denotes the curvature 2-form of the gauge bundle $g$.
% The chiral anomalies are absent for type IIA and IIB string theories, so that
For type I and $SO(32)$ heterotic string theories,
\begin{align}
\label{eq:ho-anomaly}
I_{12} = \bigg(-\frac{2p_1 (T_{10}) + \text{Tr}(\mathcal{F}^2)}{4} \bigg)\wedge \bigg(\frac{8\text{Tr}(\mathcal{F}^4) + 2 \text{Tr}(\mathcal{F}^2) p_1(T_{10}) -4 p_2(T_{10}) + 3p_1(T_{10})^2}{192}\bigg)
\end{align}
where $\mathcal{F}$ is the field strength of the $SO(32)$ gauge symmetry. For $E_8 \times E_8$ heterotic string theory,
\begin{align}
\label{eq:he-anomaly}
I_{12} = &\left(-\frac{2p_1 (T_{10}) + \text{Tr}(\mathcal{F}_1^2) + \text{Tr}(\mathcal{F}_2^2)}{4}\right)\  \wedge \nonumber
\bigg( \frac{2 \left(\text{Tr}(\mathcal{F}_1^2)\right)^2+2 \left(\text{Tr}(\mathcal{F}_2^2)\right)^2	- 2 \text{Tr}(\mathcal{F}_1^2) \text{Tr}(\mathcal{F}_2^2) }{192} \\
& +\frac{2 p_1 (T_{10}) \left(\text{Tr}(\mathcal{F}_1^2) + \text{Tr}(\mathcal{F}_2^2)\right)-4 p_2(T_{10}) + 3p_1(T_{10})^2}{192}\bigg)
% =&\sum_{a=1}^2 \left( \frac{-p_1 (T_{10}) - \text{Tr}(F_a^2)}{2}\right)  \wedge \left\{ \frac{1}{24}\left(\frac{-p_1 (T_{10}) - \text{Tr}(F_a^2)}{2}\right)^2  - \frac{4p_2(T_{10}) - p_1(T_{10})^2}{384}\right\}   \nonumber
\end{align}
where $\mathcal{F}_{1}$ and $\mathcal{F}_2$ are the field strengths for the first and second $E_8$ gauge symmetries, respectively.
The normalized trace `Tr' has been explained in Section~\ref{subsec:example}.
 % by $\text{tr}_{\rm adj}({F}_g^2) = h_g^\vee\text{Tr}({F}_g^2)$ with the dual Coxeter number $h_g^\vee$ of $g$.
We note that these anomaly polynomials are factorized into $I_{12} = Y_4 \wedge Y_8$.
The Green-Schwarz mechanism cancels the above 1-loop anomaly by introducing the counter term $\Delta S_{10} = -\int B_2 \wedge Y_8$
and modifying the Bianchi identity of the Kalb-Ramond 2-form $B_2$ to be $dH_3 = Y_4$, where $H_3$ denotes the 3-form field strength of $B_2$.
The equation of motion for $B_2$ accordingly changes to $d(\star H_3) =Y_8$.

In the remaining part of the section, we will denote by $T_6$/$T_2$ the 6d/2d tangent bundles on the worldvolume of NS5-branes/little strings, respectively.
The 10d tangent bundle $T_{10}$ can be decomposed into $T_{10} = T_6 \oplus N$ where $N$ is the $SO(4)_N = SU(2)_F \times SU(2)_R$ normal bundle.
The 6d tangent bundle $T_{6}$ can be further divided into $T_{6} = T_2 \oplus T_4$ where $T_4$ denotes the $SO(4)_T = SU(2)_l \times SU(2)_r$ bundle.
The Pontryagin classes of $T_{10}$ and $T_6$ can be written as
\begin{align}
p_1(T_{10}) &= p_1(T_6) + p_1(N), &
p_2(T_{10}) &= p_2(T_6) + p_2 (N) + p_1(T_6)\, p_1(N),\\
p_1(T_{6}) &= p_1(T_2) + p_1(T_4), &
p_2(T_{6}) &= p_2(T_2) + p_2 (T_4) + p_1(T_2)\, p_1(T_4). \nonumber
\end{align}
For $SO(4)$ bundles, the Pontryagin and Euler classes are written in the Chern classes of $SU(2)$ bundles.
\begin{align}
p_1(T_4) &= -2 c_2 (l) -2 c_2 (r), &
p_2(T_4) &= \chi_4(T_4)^2, &
\chi_4(T_4) &=  c_2 (l) - c_2 (r)\\
p_1(N) &= -2 c_2 (F) -2 c_2 (R), &
p_2(N) &= \chi_4(N)^2, &
\chi_4(N) &=  c_2 (F) - c_2 (R).\nonumber
\end{align}

\subsection{Anomaly on NS5-branes}

The 10d effective action usually includes $-\int B_2 \wedge Y_8$, for which the Bianchi identity of $B_2$ becomes
\begin{align}
\label{eq:bianchi}
dH_3 =
\begin{dcases}
0 &\text{ for type II theories}\\
Y_4 &\text{ for heterotic and type I theories}.
\end{dcases}
\end{align}
Introduction of $n$ NS5-branes adds the delta function source $n \prod_{a=6}^9 \delta(y^a)dy^a$ on the right-hand side of the Bianchi identity.
The 6d inflow anomaly from the bulk action $-\int B_2 \wedge Y_8$ becomes $-n \int Y_6^{(1)}$
where $Y_6^{(1)}$ is obtained from $Y_8$ by the descent formalism \cite{Freed:1998tg,Becker:1999kh}.
Following the prescription of \cite{Freed:1998tg,Becker:1999kh}, the delta function source can be smoothed as
\begin{align}
\label{eq:regul-del}
\prod_{a=6}^9 \delta(y^a)dy^a = \tfrac{1}{2}{d({\rho e_3})}.
\end{align}
$\rho(r)$ is a smooth function of the radial coordinate $r$ for the transverse $\mathbf{R}^4$ plane,
such that $\rho(r) = -1$ at sufficiently small $r$ and $\rho(r) = 0$ at sufficiently large $r$.
The global $S^3$ angular form $e_3$ is normalized to be $\int_{S^3} e_3 = 2$. $e_3$ can be written as \cite{Becker:1999kh}
\begin{align}
e_3 = -\frac{1}{2\pi^2} \epsilon_{abcd}\left[\frac{1}{3}(D\hat{y})^a(D\hat{y})^b(D\hat{y})^c\hat{y}^d - \frac{1}{2}F^{ab}(D\hat{y})^c\hat{y}^d\right]
\end{align}
where $\hat{y}^a = y^a/|y|$. The $SO(4)$ covariant derivative $(D\hat{y})^a$ and curvature $F^{ab}$ are written as
\begin{align}
(D\hat{y})^a \equiv d\hat{y}^a -\Theta^{ab}\hat{y}^b, \quad F^{ab} \equiv d\Theta^{ab} - \Theta^{ac} \wedge\Theta^{cb}
\end{align}
using the global $SO(4)$ connection $\Theta^{ab}$. It was shown in \cite{Harvey:1998bx,Becker:1999kh} that
the angular form $e_3$ is related to the Euler class $\chi_{4} (N)$ of the $SO(4)_N$ normal bundle $N$ by $\tfrac{1}{2}{d{e_3}} = -\chi_{4} (N)$.

The 6d anomaly polynomial $I_8$ is the sum of the 1-loop anomaly polynomial $I_8^{\rm pert}$, the inflow anomaly polynomial $ - n Y_8$, and the possibly existing Green-Schwarz anomaly polynomial $I_8^{\rm GS}$ \cite{Erler:1993zy,Sadov:1996zm}. Since a possible 6d counter term generally
takes the form of $\int B_2 \wedge X_4$ with an exact 4-form $X_4$,
% satisfying $X_4 = dX_3$ and $\delta X_3 = dX_2^{(1)}$,
for consistency, the anomaly polynomial $I_8$ has to be factorized as follows.
\begin{align}
\label{eq:anomaly-factor}
I_8 = I_8^{\rm pert} - n Y_8 + I_8^{\rm GS} =
\begin{dcases}
n \chi_4 (N)  \wedge X_4 &\text{ for type II theories}\\
-\left(Y_4-n \chi_4 (N) \right) \wedge X_4 &\text{ for heterotic and type I theories}.
\end{dcases}
\end{align}
Let us check if such factorization holds true for type II and heterotic NS5-branes.
\begin{itemize}
	\item In type IIA theory, the 1-loop effective action \cite{Vafa:1995fj,Duff:1995wd} induces
	\begin{align}
	-n Y_8 = -\frac{n}{192} \left(p_1(T_{10})^2 - 4p_2(T_{10})\right).
	\end{align}
	$I_8^{\rm pert}$ is the 1-loop anomaly polynomial for $n$ Abelian $\mathcal{N}=(2,0)$ tensor multiplets.
	$I_{8}^{\rm GS}$ denotes the Green-Schwarz anomaly polynomial \cite{Intriligator:2000eq}.
	\begin{align}
	I_8^{\rm pert} = \frac{n}{48}\left[p_2(N) - p_2(T_6)+\frac{1}{4}\left(p_1(T_6)-p_1(N)\right)^2\right],\quad I_{8}^{\rm GS} = n(n^2-1) \frac{p_2(N)}{24}.
	\end{align}
	The total anomaly $I_8$ is factorized as \eqref{eq:anomaly-factor} and removable by the following counter term \cite{Witten:1996hc}.
	\begin{align}
	I_8 = n\chi_4(N) \wedge \left(\frac{n^2}{24}\chi_4(N)\right)
	\end{align}
	
	\item The type IIB NS5-branes does not have the inflow anomaly and Green-Schwarz anomaly \cite{Vafa:1995fj}.
	The total anomaly polynomial $I_8$ only comes from the perturbative contribution of $\mathcal{N}=(1,0)$ vector and adjoint hypermultiplet,
	factorized as follows:
	\begin{align}
	I_8 = I_8^{\rm pert} = n \chi_4(N) \wedge \left(-\frac{np_1(T_6) - np_1(N)}{48}-\frac{\text{Tr}(\hat{\mathcal{F}}^2_{SU(n)})}{4}\right).
	\end{align}
	$\hat{\mathcal{F}}_{SU(n)}$ is the curvature of 6d gauge bundle $G=SU(n)$. In general, for a 6d gauge symmetry $G$,
	\begin{align}
	I_8 =  h_{G}^\vee \chi_4(N) \wedge \left(-\frac{d_G}{h_G^\vee}\frac{p_1(T_6) - p_1(N)}{48}-\frac{\text{Tr}(\hat{\mathcal{F}}^2_{G})}{4}\right).
	\end{align}

	\item The inflow anomaly to $n$  $SO(32)$ 5-branes is given by
	\begin{align}
	-nY_8 = -n\left(\frac{8\text{Tr}(\mathcal{F}^4) + 2 \text{Tr}(\mathcal{F}^2) p_1(T_{10}) -4 p_2(T_{10}) + 3p_1(T_{10})^2}{192}\right),
	\end{align}
	while $I^{\rm GS}_8 = 0$. The 1-loop polynomial $I_8^{\rm pert}$ receives the contribution from a $Sp(n)$ vector multiplet,
	an antisymmetric hypermultiplet, and 16 fundamental hypermultiplets \cite{Witten:1995gx}.
\begin{align}
I_8^{\rm pert} = &\left(-n\chi_4(N)-\frac{p_1(T_{10})}{2}\right) \wedge \bigg(-\frac{n  p_1(T_6) -n p_1(N)}{24}- \frac{\text{Tr} (\hat{\mathcal{F}}_{Sp(n)})^2}{4} \bigg)\nonumber\\&
+\frac{n\left(3p_1(T_{10})^2-4 p_2(T_{10})\right)}{192}+\frac{\left(3\text{Tr}(\hat{\mathcal{F}}_{Sp(n)}^2) + n p_1(T_6)\right)\text{Tr}(\mathcal{F}^2) + 2n \text{Tr}(\mathcal{F}^4)}{48}
\end{align}
% \begin{align}
% I_{v} = &-\frac{n(2n+1)\left(7p_1(T_6)^2-4 p_2(T_6)\right)}{5760}- \frac{\text{tr}_{\rm sym} (\hat{F}_{Sp(n)})^4 + 6c_2(R) \, \text{tr}_{\rm sym} (\hat{F}_{Sp(n)})^2 + n(2n+1) c_2(R)^2}{24} \nonumber\\
% &-\frac{\left(\text{tr}_{\rm sym} (\hat{F}_{Sp(n)})^2 + n(2n+1)\,c_2(R)\right) p_1(T_6)}{48},
% \end{align}
% \begin{align}
% I_{h,a} = &+\frac{n(2n-1)\left(7p_1(T_6)^2-4 p_2(T_6)\right)}{5760}+ \frac{\text{tr}_{\rm anti} (\hat{F}_{Sp(n)})^4 + 6c_2(F) \, \text{tr}_{\rm anti} (\hat{F}_{Sp(n)})^2 + n(2n-1) c_2(F)^2}{24} \nonumber\\
% &+\frac{\left(\text{tr}_{\rm anti} (\hat{F}_{Sp(n)})^2 + n(2n-1)\,c_2(F)\right) p_1(T_6)}{48},
% \end{align}
% \begin{align}
% I_{h,f} = &+\frac{32n\left(7p_1(T_6)^2-4 p_2(T_6)\right)}{5760}+ \frac{16\,\text{tr}_{\rm fnd} (\hat{F}_{Sp(n)})^4 + 3\,\text{tr}_{\rm fnd} (\hat{F}_{Sp(n)})^2 \, \text{tr}_{\rm fnd} (F_{SO(32)})^2 + n\,\text{tr}_{\rm fnd} (F_{SO(32)})^4}{24} \nonumber\\
% &+\frac{\left(16\,\text{tr}_{\rm fnd} (\hat{F}_{Sp(n)})^2 + n\,\text{tr}_{\rm fnd} (F_{SO(32)})^2 \right) p_1(T_6)}{48}.
% \end{align}
Summing them up,
% with the relations $\text{tr}_{\rm sym / anti} (\hat{F}_{Sp(n)})^4  = (2n\pm8)\,\text{tr}_{\rm fnd} (\hat{F}_{Sp(n)})^4  + 3(\text{tr}_{\rm fnd} (\hat{F}_{Sp(n)})^2)^2$ and $\text{tr}_{\rm sym/anti} (\hat{F}_{Sp(n)})^2  = (2n\pm2)\text{tr}_{\rm fnd} (\hat{F}_{Sp(n)})^2$
we find that the total anomaly $I_8$ is in the factorized form \eqref{eq:anomaly-factor} as \cite{Mourad:1997uc}
\begin{align}
I_8 =-\bigg(  Y_4 -n \chi_4 (N)\bigg)
\cdot \bigg(\frac{n  p_1(T_6) -n p_1(N)}{24}+ \frac{\text{Tr} (\hat{\mathcal{F}}_{Sp(n)})^2}{4} \bigg).
\end{align}
\item The inflow anomaly to $n$ $E_8\times E_8$ 5-branes is given by
\begin{align}
-n Y_8 =&
-\tfrac{n}{96} \left(\text{Tr}(\mathcal{F}_1^2)^2+ \text{Tr}(\mathcal{F}_2^2)^2	- \text{Tr}(\mathcal{F}_1^2) \text{Tr}(\mathcal{F}_2^2) \right) \nonumber\\&
-\tfrac{n}{96} \left(p_1 (T_{10}) \left(\text{Tr}(\mathcal{F}_1^2) + \text{Tr}(\mathcal{F}_2^2)\right)\right)-\tfrac{n}{192}\left(-4 p_2(T_{10}) + 3p_1(T_{10})^2\right).
% =&\sum_{a=1}^2 \left( \frac{-p_1 (T_{10}) - \text{Tr}(F_a^2)}{2}\right)  \wedge \left\{ \frac{1}{24}\left(\frac{-p_1 (T_{10}) - \text{Tr}(F_a^2)}{2}\right)^2  - \frac{4p_2(T_{10}) - p_1(T_{10})^2}{384}\right\}   \nonumber
\end{align}
The 1-loop contribution $I_8^\text{pert}$ comes from $\mathcal{N}=(1,0)$ tensor and hypermultiplets.
\begin{align}
I_8^{\rm pert} = \frac{n}{48}\left[p_2(N) - p_2(T_6)+\frac{1}{4}\left(p_1(T_6)-p_1(N)\right)^2\right]
\end{align}
The Green-Schwarz anomaly $I^{GS}_8$ can be found from the tensor branch anomaly matching \cite{Ohmori:2014pca,Intriligator:2014eaa}.
\begin{align}
I^{GS}_8=  \sum_{r=1}^n\frac{1}{2}\left(\frac{p_1(T_{10})}{4} + \frac{\text{Tr}\,\mathcal{F}_1^2}{4} + (r-\tfrac{1}{2})\chi_4(N)\right)^2
\end{align}
Combining them all, the total anomaly $I_8$ can be factorized like \eqref{eq:anomaly-factor} as required \cite{Monnier:2013rpa}.
\begin{align}
I_8 = -\left( Y_4 -n \chi_4 (N)  \right)
\wedge \bigg(\frac{  2n \text{Tr}\mathcal{F}_{1}^2 - n \text{Tr}\mathcal{F}_{2}^2 +n p_1(T_{10})}{24}+\frac{n^2 \chi_4(N)}{6} \bigg).
\end{align}
	\end{itemize}

\subsection{Anomaly on little strings}

The anomaly polynomial $I_4$ of strings in 6d superconformal field theories was studied in \cite{Kim:2016foj,Shimizu:2016lbw}
based on the anomaly inflow mechanism. The self-dual string is the source of the 2-form potential $C_2^i$,
whose field strength $G_3$ satisfies the 6d self-duality $G_3^i = \star G_3^i$ and  the Bianchi identity $dG_3^i = J_4^i$.
The index $i=1,\,\cdots,\, n$ labels all tensor multiplets and self-dual strings in a given 6d SCFT.
The right-hand side of the Bianchi identify constitutes the Green-Schwarz term $I^{\rm GS}_8 = \frac{1}{2}\Omega_{ij} J_4^i J_4^j$
of the 6d anomaly polynomial $I_8$, where $\Omega_{ij}$ denotes the Dirac pairing between $i$-th and $j$-th self-dual strings.
Introducing $(\bar{k}^1, \bar{k}^2, \cdots, \bar{k}^n)$ strings adds the source term $\bar{k}^i \prod_{a=2}^5 \delta{(y^a)}dy^a$ to the Bianchi identity, such that
\begin{align}
\label{eq:6d-scft-bianchi-modified}
d G_3^i = J^i_4 + \bar{k}^i \prod_{a=2}^5 \delta{(y^a)}dy^a.
\end{align}
We again smooth the delta function source as in \eqref{eq:regul-del} using the $SO(4)$ angular form $e_3$ satisfying $de_3 = -2\chi_4(T_4)$ \cite{Freed:1998tg,Becker:1999kh}. The 6d effective action $\Omega_{ij} \int \frac{1}{2} G_3^i \wedge \star G_3^j + \Omega_{ij}  \int C_2^i\wedge J_4^j$
% with the Green-Schwarz coupling induces
induces the inflow anomaly on the string worldsheet, which can be encoded into the following 4-form polynomial \cite{Kim:2016foj,Shimizu:2016lbw}
\begin{align}
\label{eq:inflow-string}
Z_4 = \frac{\Omega_{ij}\bar{k}^i \bar{k}^j}{2}\chi_4(T_4) + \Omega_{ij} \bar{k}^i J_4^j.
\end{align}

A little string theory can be viewed as an  affine extension of 6d superconformal field theories \cite{Bhardwaj:2015oru}
by the background 2-form potential and the massive string $m \sim \alpha'^{-1/2}$, which we call the \emph{full} winding string.
They are inherited from the ten-dimensional 2-form tensor $B_2$ and the fundamental string \cite{Seiberg:1997zk}.
The worldsheet coupling of $\bar{k}$ full strings to the bulk 2-form $B_2$ is given by  $\bar{k}\int  B_2$.
Since the Bianchi identity of $B_2$ has been modified by $n$ NS5-branes,
\begin{align}
\label{eq:bianchi-modified}
dH_3 =
\begin{dcases}
n\textstyle\prod_{a=6}^9 \delta(y^a)dy^a &\text{ for type II theories}\\
Y_4 + n\textstyle\prod_{a=6}^9 \delta(y^a)dy^a &\text{ for heterotic and type I theories},
\end{dcases}
\end{align}
the worldvolume coupling $\bar{k}\int  B_2$ contributes to $I_4$ by $\bar{k}  (Y_4 - n\chi_4(N))$. Combining with \eqref{eq:inflow-string}, we find
\begin{align}
\label{eq:anomaly-general}
I_4 = \bar{k} (Y_4 - n\chi_4(N)) + \frac{\Omega_{ij}\bar{k}^i \bar{k}^j}{2}\chi_4(T_4) + \Omega_{ij} \bar{k}^i J_4^j
\end{align}
as the entire 2d anomaly polynomial of $(\bar{k}^1,\cdots,\bar{k}^n)$ strings and $\bar{k}$ full strings  in a general 6d LST.
For later discussions in Section~\ref{sec:examples}, let us explicitly write $I_4$ for LSTs on type II and heterotic NS5-branes.

\begin{figure}[t!]
    \centering
    \begin{subfigure}[t!]{0.5\textwidth}
        \centering
        \includegraphics[width=8.2cm]{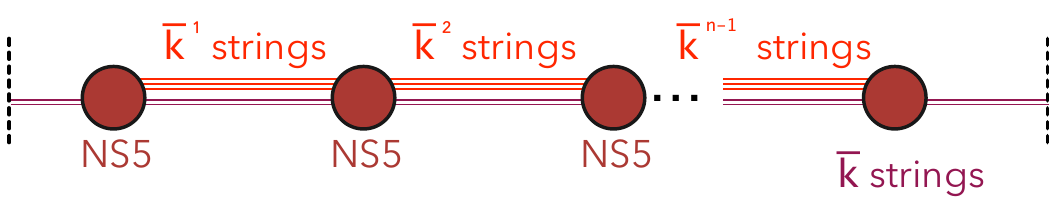}
        \caption{Type $A_{n-1}$}% The dashed lines are the two boundaries identified to each other.
        \label{subfig:20LSTtypeA}
    \end{subfigure}%
    \hspace*{\fill}
    \begin{subfigure}[t!]{0.5\textwidth}
        \centering
        \includegraphics[width=8.5cm]{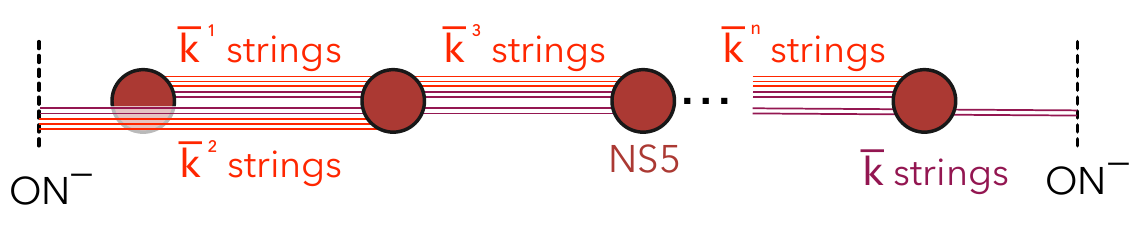}
        \caption{Type $D_{n}$}
        \label{subfig:20LSTtypeD}
    \end{subfigure}
    \caption{Fractional and full strings in $\mathcal{N}=(2,0)$ LSTs}
    \label{fig:20LST}
\end{figure}

% comes from the bulk action $\frac{1}{2}\int H_3 \wedge \star H_3$ with the modified equation of motion $d(\star H_3) = Y_8 + k\prod_{a=2}^{9}\delta(y^a) dy^a$
\paragraph{(2,0) LST} A parallel stack of type IIA $n$ NS5-branes engineers $\mathcal{N}=(2,0)$ LST of $\hat{A}_{n-1}$ type,
for which $Y_4 = 0$, $\Omega_{ij}$ is the $A_{n-1}$ Cartan matrix, and  $J_4^i = -\rho^i\, \chi_4(N)$ with the $A_{n-1}$ Weyl vector $\rho^i$.
We separate the NS5-branes along the transverse circle of radius $\sim g_s \alpha'^{1/2}$, which sets the length scale of the LST.
Let us denote the worldvolume coordinates of the NS5-branes and the circle coordinate by $x^{012345}$ and $x^6$, respectively.
All strings are realized as D2-branes along the $x^{01}$ and $x^6$ directions, suspended between a pair of NS5-branes.
One can view a long string as a composition of short strings which interconnect adjacent pairs of NS5-branes.
Especially the \emph{full} strings are those which completely wind the $x^6$ circle,
consisting of $n$ different types of \emph{fractional} short strings.
For $(\bar{k}^1,\cdots,\bar{k}^n)$ strings and $\bar{k}$ full strings, illustated in Figure~\ref{subfig:20LSTtypeA},
it is often more convenient to express the anomaly polynomial $I_4$ in terms of the numbers of fractional short strings,
$k^{n} \equiv \bar{k}$ and $k^i \equiv \bar{k}^i + \bar{k}$ for $i<n$, such that
\begin{align}
\label{eq:20anomaly}
I_4 = \sum_{i,j=1}^{n}\frac{\hat{\Omega}_{ij}k^i k^j}{2}\left(c_2(l) - c_2(r)\right)+ \sum_{i=1}^{n} k^i \left(c_2(R) - c_2(F)\right),
\end{align}
where $\hat{\Omega}_{ij}$ is the Cartan matrix of affine $\hat{A}_{n-1}^{(1)}$ Lie algebra.

To engineer $\mathcal{N}=(2,0)$ LST of $\hat{D}_{n}$ type, for which $\rho^i$ and $\Omega_{ij}$ are replaced with the $D_n$ Weyl vector and Cartan matrix,
we introduce two ON${}^-$ planes \cite{Sen:1998ii,Kapustin:1998fa} parallel to $n$ NS5-branes.
They are orbifold planes which change the transverse $x^6$ circle to be $S^1/\mathbf{Z}_2$ and sit at both ends of the $S^1/\mathbf{Z}_2$ segment.
The total NS5-brane charge is $(2n-2)$ since an ON${}^-$ plane carries a negative unit.
One can regard a long string as a composition of short strings suspended between adjacent pairs of NS5-branes.
The full $\bar{k}$ little strings (or equivalently, $2\bar{k}$ half strings stuck on the orbifold planes) are therefore
the collection of $r$ different types of \emph{fractional} short strings, where $r$ is the rank of the affine $\hat{D}_{n}^{(1)}$ algebra.
For $(\bar{k}^1,\cdots,\bar{k}^n)$ fractional strings and $\bar{k}$ full strings, illustrated in Figure~\ref{subfig:20LSTtypeD},
the numbers of fractional short strings are given by
\begin{align}
n=2:&&& k^i \equiv \bar{k}^i + \bar{k} \ \text{ for } i=1,2,3, && k^{4} \equiv \bar{k}\\
n>2:&&& k^i \equiv \bar{k}^i + \bar{k} \ \text{ for } i=1,2,2n, && k^i \equiv \bar{k}^i + 2\bar{k} \ \text{ for } i=3,\cdots,(2n-1), \quad k^{2n+1} \equiv \bar{k} \nonumber
\end{align}
Then the anomaly polynomial $I_4$ becomes \eqref{eq:20anomaly} in which $\hat{\Omega}_{ij}$ means the affine $\hat{D}_{n}^{(1)}$ Cartan matrix.
We expect that \eqref{eq:20anomaly} also holds for $(2,0)$ $\hat{E}_n$ LST by replacing $\hat{\Omega}_{ij}$ with the affine $\hat{E}_{n}^{(1)}$ Cartan matrix.

\paragraph{(1,1) LST} The worldvolume theory of type IIB $n$ NS5-branes is the maximally supersymmetric $U(n)$ Yang-Mills theory,
in which $Y_4 =0$. It has only one type of strings with zero Dirac self-pairing, which couples to the background 2-form tensor $B_2$.
This is the instanton string of 6d maximal SYM.
The anomaly polynomial of $\bar{k}$ strings is given by
\begin{align}
\label{eq:11anomaly}
I_4 = \bar{k}h^\vee \left(c_2 (R) - c_2(F)\right).
\end{align}
where we replace the NS5-brane charge $n$ by $h^\vee = n$, the dual Coexter number of $A_{n-1}$.

By suitably introducing an orbifold 5-plane, one can engineer  $(1,1)$ LSTs of $\hat{B}_n$, $\hat{C}_n$, $\hat{D}_n$-type,
whose 5-brane charges are respectively $(2n-1)$, $(2n+2)$, and $(2n-2)$.
The dual Coexter numbers of $B_n$, $C_n$, $D_n$ Lie algebras are given by
\begin{align}
h^\vee = 2n-1 \ \text{ for }\ B_n, \qquad h^\vee = n+1 \ \text{ for }\ C_n, \qquad h^\vee = 2n-2 \ \text{ for }\ D_n.
\end{align}
We find that \eqref{eq:11anomaly} gives the anomaly polynomial of $\bar{k}$ strings in $\hat{B}_n$, $\hat{C}_n$, $\hat{D}_n$ LSTs. Here $\bar{k}$ must be understood as the number of half strings stuck on the ON${}^+$ plane that engineers $\hat{C}_n$ LST.
We expect that \eqref{eq:11anomaly} holds true for an other $(1,1)$ LST with an exceptional gauge symmetry $G$,
by replacing $h^\vee$ with the dual Coexter number of Lie algebra $G$.

\paragraph{SO(32) LST} A stack of $n$ NS5-branes in $SO(32)$ heterotic string theory engineers
$\mathcal{N}=(1,0)$ LST with $Sp(n)$ gauge symmetry and $SO(32)$ flavor symmetry.
It allows only one type of strings with zero Dirac self-pairing which is the instanton string of $Sp(n)$ gauge theory.
Denoting the 2-form curvature of the $SO(32)$ bundle by $\mathcal{F}$, the anomaly polynomial $I_4$ of $\bar{k}$ strings is given by
% Similar to $\mathcal{N}=(1,1)$ little string theory, $\mathcal{N}=(1,0)$ $SO(32)$ little string theory
% does not have a dynamical tensor $C_2$. Only one type of strings can exist, coupled to the bulk 2-form $B_2$.
% For $SO(32)$ theory with $n$ NS5-branes, the anomaly polynomial $I_4$ of  little strings is given by
\begin{align}
I_4 = \bar{k} (Y_4 - n\chi_4(N)) =  \bar{k}\bigg( c_2(l)+c_2(r)  - (n-1) c_2(F) +(n+1)c_2(R)  -\frac{p_1(T_{2})}{2} -\frac{\text{Tr}\mathcal{F}^2}{4} \bigg).
\end{align}

\begin{figure}[t]
	\centering
	\includegraphics[height=2cm]{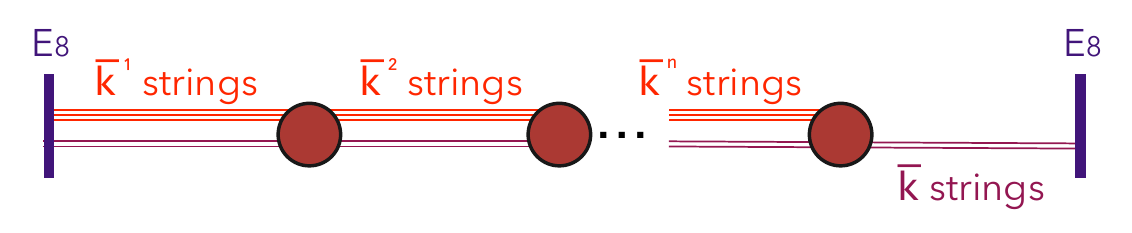}
	\caption{Fractional and full strings in $\mathcal{N}=(1,0)$ $E_8 \times E_8$ heterotic LSTs}
	\label{fig:10LSTtypeE8}
\end{figure}

\paragraph{$\bf E_8 \times E_8$ LST}
The worldvolume theory of $n$  $E_8 \times E_8$ heterotic NS5-branes is the rank-$n$ $(1,0)$ LST with $E_8 \times E_8$ flavor symmetry
which contains $n$ dynamical tensor multiplets. After S-duality transformation, we obtain the configuration of IIA NS5-branes probing the $S^1/ \mathbf{Z}_2$ orbifold
parametrized by the $x^6$ coordinate. All strings are realized by D2-branes filling the $x^{01}$ and $x^6$ directions,
suspended between a pair of NS5-branes. Regarding a long BPS string as a combination of short strings which connects adjacent NS5-branes,
the $\bar{k}$ half strings stuck on the orbifold fixed plane will be equivalent to the composition of $(n+1)$ different types of short strings. All Dirac pairings between the short strings are recorded in the $(n+1) \times (n+1)$ matrix entries as follows \cite{Bhardwaj:2015oru}.
\begin{align}
\label{eq:dirac-pairing-e8}
&\Omega_{ij} = \begin{dcases} +1& i=j=1 \text{ or } 2\\
-1& i = j+1 \text{ or } i = j-1
\end{dcases} &&\text{for $n=1$}\\
&\Omega_{ij} = \begin{dcases} +1& i=j=1 \text{ or } (n+1)\\
+2& i=j =2,\cdots, n\\
-1& i = j+1 \text{ or } i = j-1
\end{dcases} &&\text{for $n>1$}\nonumber
\end{align}
Computing the inflow from the 6d Green-Schwarz term \cite{Ohmori:2014pca,Intriligator:2014eaa}, for $(\bar{k}^1,\cdots,\bar{k}^n)$ short strings alone,
\begin{align}
Z_4 = &\left(\sum_{i=1}^{n-1}\frac{(\bar{k}_{i}-\bar{k}_{i+1})^2}{2} + \frac{\bar{k}_n^2}{2}\right) \chi_4(T_4) + \sum_{i=1}^{n-1} (\bar{k}_i - \bar{k}_{i+1}) \left(-\frac{p_1(T_{10}) + \text{Tr}\mathcal{F}_1^2}{4} - (i-\tfrac{1}{2})\chi_4(N)\right)\nonumber\\
&+\bar{k}_n\left(-\frac{p_1(T_{10}) + \text{Tr}\mathcal{F}_1^2}{4} - (n-\tfrac{1}{2})\chi_4(N)\right).
\end{align}
Introduction of the $\bar{k}$ long strings contributes to the anomaly polynomial $I_4$ by
\begin{align}
\bar{k} \left(-n \chi_4 (N)  -\frac{2p_1(T_{10}) + \text{Tr}\mathcal{F}_{1}^2 +  \text{Tr}\mathcal{F}_{2}^2}{4} \right)
\end{align}
where $\mathcal{F}_1$ and $\mathcal{F}_2$ denote the curvature 2-forms of two $E_8$ gauge bundles.
If we rewrite the anomaly polynomial $I_4$ using the numbers of short strings,
$k^{n+1} \equiv \bar{k}$ and $k^{i} \equiv \bar{k}^i + \bar{k}$ for $i=1,\cdots,n$,
\begin{align}
I_4 &=\frac{\hat{\Omega}^{ij}k_i k_j}{2}\left(c_2(l) - c_2(r)\right) + k_1 \left( -\frac{p_1(T_{2}) + \text{Tr}F_1^2}{4} + \frac{c_2(l) + c_2(r)}{2} + c_2(R) \right) \\
&+\sum_{i=2}^n k_i \left(c_2(R) - c_2(F)\right) + k_{n+1} \left( -\frac{p_1(T_{2}) + \text{Tr}F_2^2}{4} + \frac{c_2(l) + c_2(r)}{2} + c_2(R) \right)\nonumber
\end{align}
which follows the general form of \eqref{eq:anomaly-general}. See Figure~\ref{fig:10LSTtypeE8} for the illustration.

\section{BPS spectra from T-duality}
\label{sec:examples}

In this section, we will study the $\mathbf{R}^4 \times T^2$ partition function of circle compactified LSTs,
based on their T-duality relation and the modular bootstrap of the little string elliptic genera.
It is defined as a grand canonical partition function that displays the BPS spectrum for an infinite number of strings, counting the bound states between winding and/or momentum modes. More precisely, \cite{Nekrasov:2002qd,Nekrasov:2003rj}
\begin{align}
  \label{eq:6d-nekrasov}\textstyle
Z_{6d} \equiv \text{Tr}_{\mathcal{H}_{6d}}\,\left[ (-1)^F
 e^{2\pi i (\tau H_L - \bar{\tau} H_R)}\, e^{2\pi i \epsilon_+(J_{r} + J_{R})} e^{2\pi i \epsilon_- J_{l}} e^{2\pi i z\cdot J_z} \prod_{i=1}^{n}
\mathfrak{n}_i^{k_i}
   \right]
\end{align}
where the trace is taken over the entire 6d BPS Hilbert space $\mathcal{H}_{6d}$. The integral charge $k_i$ is conjugate to the winding fugacity $\mathfrak{n}_i$,
counting the number of the $i$-th strings coupled to the $i$-th tensor multiplet.
All other chemical potentials and conjugate charges have been introduced in Section~\ref{sec:modular-anomaly}.
 We will also frequently use the fugacity variables
\begin{align}
q =e^{2\pi i \tau}, \quad t = e^{2\pi i \epsilon_+}, \quad u = e^{2\pi i \epsilon_-}, \quad w_j^{(i)} = e^{2\pi i a_j^{(i)}}, \quad \mathfrak{n}_i = e^{- \text{vol}(T^2) \cdot \langle\Phi_i \rangle}
\end{align}
where $\langle\Phi_i \rangle$ and $a_j^{(i)}$ denote the scalar VEV of the $i$-th tensor multiplet and the $i$-th gauge holonomy, respectively.
In particular, the non-zero gauge holonomy allows the \emph{fractional} circle momentum mode, leading us to interpret
$q$ and $w_j^{(i)}$  as the momentum fugacities. As the tensor branch observable, the $\mathbf{R}^4 \times T^2$ partition function will be expanded mainly in the winding fugacities $\mathfrak{n}_1,\cdots,\mathfrak{n}_n$ such that
\begin{align}
\label{eq:6d-nekrasov-fac}
	\textstyle Z_{6d} = \mathcal{I}_{0}\cdot \left(1 + \sum_{\{k_i\}} 	 \mathfrak{n}_1^{k_1} \cdots  \mathfrak{n}_n^{k_n}\cdot \mathcal{I}_{\{k_1, \cdots, k_n\}}\right).	
\end{align}
We remark that the individual coefficient $\mathcal{I}_{\{k_1, \cdots, k_n\}}$ corresponds to the elliptic genus of $\{k_1, \cdots, k_n\}$ strings,
satisfying the modular property \eqref{eq:modular-property}. The overall dressing factor $\mathcal{I}_{0}$ is the BPS partition function for the pure momentum sector,
capturing all the multi-trace letter operators \cite{Bhattacharya:2008zy} made of the elementary fields and the gauge covariant derivatives.
All of them are given as closed-form expressions in the momentum fugacities made of $q$ and $w_j^{(i)}$.

T-duality is a distinctive feature of the LST that identifies two apparently different LSTs on $S^1$,
at different circle radii $R' = \alpha'/R$, by exchanging the winding and momentum modes \cite{Seiberg:1997zk}.
Since the BPS spectrum is insensitive to the circle radius,
it will be incarnated as the equivalence between the dual $\mathbf{R}^4 \times T^2$ partition functions
after suitably mapping the winding/momentum fugacities on one side to the momentum/winding fugacities on the other side.
This has been confirmed in several examples, i.e., the maximally supersymmetric LSTs of A-type \cite{Kim:2015gha} and their orbifold variations \cite{Hohenegger:2016eqy,Kim:2017xan}, engineered from type IIA and IIB NS5-branes probing the $\mathbf{R}^4$ and $\mathbf{R}^4/\Gamma_\text{AD}$ backgrounds. For these theories, the underlying 2d gauge theory description is known. Thus one can completely determine the elliptic genus of little strings in type IIA and IIB NS5-branes separately, thereby showing the T-duality of type IIA and IIB 5-branes. 
However we will turn the logic around and will assume the equivalence of the BPS spectra for T-dual pairs of $S^1$ compactified LSTs.
Since the BPS partition function $\mathcal{I}_{0}$/$\mathcal{I}_{\{k_1, \cdots, k_n\}}$ for a given winding sector captures an arbitrary number of momentum modes,
it provides the BPS data with a given momentum unit for all individual winding sectors in the T-dual version of the LST.
We will start from $\mathcal{I}_{0}$ for the pure momentum sector which can be easily obtained by counting the BPS letters.
Recall that the modular bootstrap based on \eqref{eq:modular-property} and \eqref{eq:ansatz} has reduced
the problem of obtaining the 6d string elliptic genera down to the problem of finding the sufficient amount of BPS coefficients \cite{Haghighat:2014pva,Cai:2014vka,Huang:2015sta,Huang:2015ada,Haghighat:2015ega,DelZotto:2016pvm,Gu:2017ccq}.
Knowing the BPS data for zero momentum modes would be sufficient to bootstrap some elliptic genera with low winding numbers,
so that $\mathcal{I}_{0}$ could determine those elliptic genera in the T-dual LST.
The newly found elliptic genera will be closed-form expressions in the dual momentum fugacities, providing
the additional BPS data with certain momentum modes for all winding sectors in the original LST.
Using these coefficients, one could find the elliptic genera for some winding sectors
which will yield again the BPS data with higher momentum modes in the T-dual description of the LST.
Repeating this procedure, we would obtain the infinite tower of the 6d string elliptic genera
composing the $\mathbf{R}^4 \times T^2$ partition function of the LST. This approach is quite effective for the theories which do
not have the gauge theory realization.

For the iterated bootstrap of the 6d string elliptic genera, it is necessary to know the precise map
between the winding/momentum fugacities on one side to the momentum/winding fugacities on the other side.
We will particularly focus on the maximally supersymmetric LSTs of AD-types as well as
the $\mathcal{N}=(1,0)$ heterotic LSTs with $SO(32)$ and $E_8 \times E_8$ global symmetries.
 \begin{itemize}
 	\item $(2,0)$ LST of $\hat{A}_{n-1}$ type contains $n$ different types of fractional strings, illustrated in Figure~\ref{subfig:20LSTtypeA},
 	and an integral unit of the circle momentum. As we denote by $\mathfrak{n}_1,  \cdots, \mathfrak{n}_n$ the fractional winding fugacities,
 	the combination $\mathfrak{n}_1\mathfrak{n}_2 \cdots \mathfrak{n}_n$ corresponds to the full winding mode. 
 	On the contrary, $(1,1)$ LST of
 	$\hat{A}_{n-1}$ type has $n$ different units of fractional momentum, due to the $SU(n)$ gauge holonomy,
 	and only one type of full winding modes carrying the Yang-Mills instanton charge. The fractional momentum fugacities are labeled by the simple roots of
 	the affine $\hat{A}_{n-1}$ algebra,
 	\begin{align}
 	% q = e^{2\pi i \tau} = \mathfrak{n},\
 	q_1' = \frac{w_1'}{w_2'}, ~~ q_2' = \frac{w_2'}{w_3'}, ~~ \cdots, ~~q_n' = \frac{q' w_n'}{w_1'}  \qquad \text{ with }\qquad \prod_{i=1}^n w_i' = 1,
 	% \ \mathfrak{n}_{n-1} = e^{2\pi i(a_{n-1} - a_n)}, \ \mathfrak{n}_{n} = e^{2\pi i(\tau' - a_1 + a_n)
 	\end{align}
 	where all variables in $(1,1)$ LST are primed for distinction. T-duality implies the fugacity map
	\begin{align}
	\label{eq:t-duality-sun}
	q =  \mathfrak{n}',\quad  \mathfrak{n}_1 = q_1',\quad \cdots, \quad \mathfrak{n}_{n-1} = q_{n-1}', \quad \mathfrak{n}_{n} = q'_{n}.
	\end{align}

 	\item The fugacity map between $(2,0)$ and $(1,1)$ LSTs of $\hat{D}_{n}$ type can be derived in an analogous way.\\
 	The $(2,0)$ LST has $r = \text{rank}(\hat{D}_n)$ different kinds of fractional strings,
 	as depicted in Figure~\ref{subfig:20LSTtypeD}, whose Dirac pairings are given by the Cartan matrix of affine $\hat{D}_{n}$ algebra.
 	Denoting the fractional winding fugacities by $\mathfrak{n}_1,  \cdots, \mathfrak{n}_r$, the full winding fugacity is
 	the combination $\mathfrak{n}_1\mathfrak{n}_2 \mathfrak{n}_{r-1}\mathfrak{n}_r \prod_{i=3}^{r-2}\mathfrak{n}_i^2$.
 	The $(1,1)$ LST has $r = \text{rank}(\hat{D}_n)$ different units of fractional momentum due to the $SO(2n)$ gauge holonomy.
 	Their conjugate fugacities are labeled by the simple roots of the $\hat{D}_{n}$ algebra.
 	\begin{align}
        \label{eq:so2n-mom-fug}
	 &D_2:&& q_1' = \frac{w_1'}{w_2'}, ~~ q_2' = \frac{q'w_2'}{w_1'}, ~~   q_3' = w_1' w_2', ~~ q_4' = \frac{q'}{w_1' w_2'}\\
	 &D_3:&& q_1' = \frac{w_1'}{w_2'},~~ q_2' = \frac{w_2'}{w_3'}, ~~ q_3'  = \frac{w_3'}{w_4'}, ~~ q_4' = \frac{q' w_4'}{w_1'}\qquad \text{ with }\qquad \prod_{i=1}^4 w_i' = 1 \nonumber\\
	 &D_{n\geq4}:&& q_1' = \frac{w_1'}{w_2'},~~ q_2' = \frac{q'}{w_1'w_2'}, ~~ q_3'  = \frac{w_2'}{w_3'}, ~~ \cdots, ~~ q_{n-1}' = \frac{w_{n-2}'}{w_{n-1}'},
	 ~~ q_{n}' = \frac{w_{n-1}'}{w_{n}'}, ~~ q_{n+1}' = {w_{n-1}'}{w_{n}'}\nonumber
	\end{align}
	T-duality imposes the following map between winding and momentum fugacities
	\begin{align}
        \label{eq:t-duality-so2n}
	q =  \mathfrak{n}',\quad  \mathfrak{n}_1 = q_1',\quad \cdots, \quad \mathfrak{n}_{n} = q_{n}', \quad \mathfrak{n}_{n+1} = q'_{n+1}.
	\end{align}	

	\item The rank-$n$ $E_8 \times E_8$ heterotic LST has $(n+1)$ different types of fractional strings, illustrated in Figure~\ref{fig:10LSTtypeE8},
	and an integral unit of the circle momentum. The Dirac pairings between those strings are noted in the matrix \eqref{eq:dirac-pairing-e8}.
	Denoting the fractional winding fugacities by $\mathfrak{n}_1, \mathfrak{n}_2, \cdots, \mathfrak{n}_{n+1}$,
	the combination $\prod_{i=1}^{n+1} \mathfrak{n}_i^2$ is conjugate to the full string that wraps the transverse circle. In contrast,
	the rank-$n$ $SO(32)$ heterotic LST has $(n+1)$ different units of fractional momentum, due to the $Sp(n)$ gauge holonomy,
	and only one type of full strings carrying the instanton charge.
	The fractional momentum fugacities are labeled by the simple roots of the affine $\hat{C}_{n}$ algebra, i.e.,
     \begin{align}
        \label{eq:het-mom-fug}
 	% q = e^{2\pi i \tau} = \mathfrak{n},\
 	q_1' = \frac{q'}{w_1'^2}, \quad q_2' = \frac{w_1'}{w_2'}, \quad \cdots, \quad  q_n' = \frac{w_{n-1}'}{w_{n}'}, \quad  q_{n+1}' = w_{n}'^2.
 	% \ \mathfrak{n}_{n-1} = e^{2\pi i(a_{n-1} - a_n)}, \ \mathfrak{n}_{n} = e^{2\pi i(\tau' - a_1 + a_n)
 	\end{align}
 	% \cmt{here}
 	The circle compactified $E_8 \times E_8$ and $SO(32)$ heterotic LSTs are identified by T-duality which requires
 	the Wilson lines along $S^1_x \subset T^2$ breaking the flavor symmetry to $SO(16) \times SO(16)$:
 	\begin{align}
  	\label{eq:het-wilson-line}
 	RA^{SO(32)}_x= (0^8,\tfrac{1}{2}^8), \quad RA^{E_8 \times E_8}_x = (0^7, 1, 0^7, 1).
 	\end{align}
 	These background Wilson lines produce shifts in the left-moving momentum $H_L$ and the flavor charges $f_a$, depending on the winding number(s) $k_i$ \cite{Narain:1986am}.
 \begin{align}
   \label{eq:charge-relation}
	E_8 \times E_8: &\qquad
	\tilde{H}_L = 2H_L +  (k_1 + k_{n+1}) - 2 (f_8 + f_{16}), &&
	\tilde{f}_8 = f_8- k_1, \quad \tilde{f}_{16} = f_{16}- k_2. \\
	SO(32): & \qquad   \tilde{H}_L = 2H_L +  2k -  \textstyle\sum_{a=9}^{16} f_a, &&
     \tilde{f}_a =  f_a- \tfrac{1}{2}k\ \ \ (\text{for all }9\leq a\leq 16). \nonumber
 \end{align}
The new momentum and flavor charges are distinguished from the original ones by the tilde. 
To establish T-duality, we express the $\mathbf{R}^4 \times T^2$ partition functions in the new fugacities conjugate to the shifted charges. The relation between the original and new variables can be derived from 
\begin{align}
   \label{eq:fugacity-relation}
   {\textstyle
     {q}^{H_L} \prod_{i}{\frak{n}_i}^{k_i}  \prod_{l=1}^{16} y_a^{f_a}  \equiv
     {\tilde{q}}^{\tilde{H}_L} \prod_{i}{\tilde{\frak{n}}_i}^{k_i}  \prod_{l=1}^{16} \tilde{y}_a^{\tilde{f}_a}},
\end{align}
where the right-hand side should appear in the definition of the $\mathbf{R}^4\times T^2$ partition functions with the $SO(16) \times SO(16)$ Wilson line.
Combining \eqref{eq:charge-relation} and \eqref{eq:fugacity-relation}, we find that 
 \begin{align}
   \label{eq:charge-relation-2}
	E_8 \times E_8: & \qquad \textstyle\tilde{\mathfrak{n}}_1 = \mathfrak{n}_1\, q \, y_8 ,&& \tilde{\mathfrak{n}}_{n+1} = \mathfrak{n}_{n+1}\, q \, y_{16} , && \tilde{y}_a = y_a \,q \ (a =8,16), && \tilde{q} = \sqrt{q}\\
	SO(32): &\qquad\textstyle\tilde{\mathfrak{n}}' = \mathfrak{n}'\, q' \prod_{l=9}^{16} y'_a{}^{1/2},&&&&\tilde{y}_a' = y_a' \,\sqrt{q}' \ (a \geq 9), && \tilde{q}' = \sqrt{q}'\nonumber.
 \end{align}
Especially the momentum and winding fugacities of both LSTs are identified as
\begin{align}
\label{eq:het-fug-tdual}
\tilde{q} =  \tilde{\mathfrak{n}}',\quad  \tilde{\mathfrak{n}}_1 = \frac{\tilde{q}'}{w_1'},\quad \tilde{\mathfrak{n}}_{2} = \frac{w_1'}{w_2'}, \quad \tilde{\mathfrak{n}}_{3} = \frac{w_2'}{w_3'},\quad \cdots, \quad \tilde{\mathfrak{n}}_{n} = \frac{w_{n-1}'}{w_{n}'}, \quad \tilde{\mathfrak{n}}_{n+1} = w_n'.
\end{align}	

\end{itemize}

For the rest of this section, we will study the $\mathbf{R}^4 \times T^2$ partition functions of the above LSTs through the iterated bootstrap
of the 6d string elliptic genera. Here we briefly summarize the results. First, we successfully construct the $\mathbf{R}^4 \times T^2$ partition functions
of several $\hat{A}$-type $(2,0)$ and $(1,1)$ LSTs based on \eqref{eq:modular-property} and \eqref{eq:ansatz}. They agree with the results of \cite{Kim:2015gha} which
 obtain the little string elliptic genera using the worldsheet gauge theories of type IIA and IIB little strings.
Second, we show the existence of additional bosonic zero modes in the 2d SCFTs of \emph{full} strings
in maximally supersymmetric LSTs of $\hat{D}_{n\geq 4}$-type as well as heterotic LSTs.
The extra zero modes correspond to the string motion moving away from the NS5-branes,
developing the tubelike region in the target space.
They are lifted by the chemical potentials $-\epsilon_+ \pm m$ where $m$ has been introduced for $SU(2)_F \subset SO(4)_N$ of these LSTs. 
Here $SO(4)_N$ is the $\mathbf{R}^4$ rotation of the transverse directions to NS5-branes. The conjectured form of the denominator $\mathcal{D}(\tau,z)$ in \eqref{eq:ansatz}
must be appropriately modified by additional factors, e.g., $\theta_1(-\epsilon_+ \pm m)$, for successful bootstrapping of the elliptic genera.
We propose the modified denominator by considering the $q' \rightarrow 0$ limit of those elliptic genera.
Third, we construct the $\mathbf{R}^4 \times T^2$ partition functions of $\hat{D}_{n\geq 4}$-type
 $(2,0)$ and $(1,1)$ LSTs and heterotic LSTs based on \eqref{eq:modular-property} and \eqref{eq:ansatz} with the modified denominators.
These 6d partition functions include the novel elliptic genera of \emph{fractional} strings, some of which also appear in D-type $(2,0)$ SCFTs,
lacking the 2d gauge theory descriptions.

\subsection{$(2,0)$ and $(1,1)$ LSTs of $\hat{A}_{n-1}$-type}
\label{subsec:20an}

Bootstrapping the $\mathbf{R}^4 \times T^2$ partition function starts from the index $\mathcal{I}_0$ of the pure momentum sector,
decoupled from stringy excitations. It can be easily obtained by counting the multi-trace BPS letter operators made of the elementary fields and
the gauge covariant derivatives \cite{Bhattacharya:2008zy}. As the multi-letter partition function is the Plethystic exponential of the single-letter partition function $f_0 (\tau,z)$, i.e., \cite{Feng:2007ur}
\begin{align}
	\label{eq:pert-PE}
	 \textstyle\mathcal{I}_{0}(\tau,z) = \text{PE} \left[ f_{0} (\tau,z ) \right] \equiv \exp \left(\sum_{p=1}^\infty \frac{1}{p} \cdot f_{0}(p\tau,pz )	\right),
\end{align}
we compute $f_0 \equiv \text{tr}\left[(-1)^F q^{H_L} \, t^{J_{r} + J_{R}} u^{J_{l}} v^{J_F} \, \prod_{i=1}^{n}w_i^{G_i}
\right]$ over the single BPS letters.
$J_F$ and $G_1, \cdots, G_{n}$ denote the Cartan generators of the $SU(2)_F$ flavor and gauge symmetries, respectively,
whose conjugate fugacities are $v \equiv e^{2\pi i m}$ and $w_i =e^{2\pi i a_i}$. 
The $SU(N)$ gauge fugacities are subject to the traceless condition $\prod_{i=1}^n w_i = 1$.
The trace over the $(1,0)$ supermultiplet and its derivatives takes the form of
\begin{align}
	\label{eq:pert-trace}
	\text{tensor}:& \quad -\frac{t (u + u^{-1})}{(1- tu) (1-tu^{-1})}\textstyle\left(\sum_{n=-\infty}^{\infty} q^n\right)^+\\
	\text{vector}:& \quad -\frac{t (t + t^{-1})}{(1- tu) (1-tu^{-1})}\textstyle \textstyle\left(\chi_{\mathbf{adj}}\,(w_{i}) \sum_{n=-\infty}^{\infty} q^n\right)^+ \nonumber \\
	\text{hyper}\,: & \quad+\frac{t (v + v^{-1})}{(1- tu) (1-tu^{-1})}\textstyle \textstyle\left(\chi_{\mathbf{R}}\,(w_{i}) \sum_{n=-\infty}^{\infty} q^n\right)^+ \nonumber
\end{align}
where the $+$ superscript in the parenthesis indicates that all non-positive momentum states have been discarded.
$\chi_{\mathbf{R}}$ is the irreducible character for a gauge representation $\mathbf{R}$ of a given supermultiplet.
For the $SU(n)$ adjoint representation, the irreducible character $\chi_{\mathbf{adj}}$ is given by $\chi_{\mathbf{adj}} (w_i) = \sum_{i,j=1}^n\frac{w_i}{w_j} - 1$.
For the stack of $n$ IIA NS5-branes, engineering $\hat{A}_{n-1}$-type $(2,0)$ LST plus a free $(2,0)$ tensor multiplet,
\begin{align}
\label{eq:20partpert-su2}
\mathcal{I}_{0} = \text{PE}\bigg[  \frac{n \cdot t \left(v + v^{-1} - u - u^{-1}\right)}{(1-tu)(1-tu^{-1})} \frac{q}{1-q}\bigg].
\end{align}
For the stack of $n$ IIB NS5-branes, engineering $\hat{A}_{n-1}$-type $(1,1)$ LST and a free $(1,1)$ vector multiplet,
\begin{align}
\label{eq:11partpert-su2}
\mathcal{I}_{0}' &= \text{PE}\bigg[  \frac{ t\left(v + v^{-1} - t - t^{-1}\right)}{(1-tu)(1-tu^{-1})}\bigg({\sum_{i<j}\frac{w_i'}{w_j'} + \sum_{i>j}\frac{q'w_i'}{w_j'}} + nq'\bigg) \frac{1}{1-q'}\bigg]\\
&= \text{PE}\bigg[\frac{t \left(v + v^{-1} - t - t^{-1}\right)}{(1-tu)(1-tu^{-1})} \bigg( \sum_{i \leq j}^{n-1}\prod_{k=i}^{j}q_k' + \sum_{j \leq i}\Big(\prod_{k=i+1}^{n}q_k' \cdot \prod_{l=1}^{j-1}q_l' \Big)+ n q' \bigg)\frac{1}{1-q'} \bigg] \nonumber
\end{align}
where $q' = \prod_{i=1}^n q'_i$. We prime the fugacity variables of the $(1,1)$ LST for distinction.

The indices $\mathcal{I}_0$ and $\mathcal{I}_0'$ display the infinite towers of the pure momentum states.
According to the relation \eqref{eq:t-duality-sun} between the dual fugacity variables, they supply the BPS data with zero momentum
for all distinct winding sectors in the T-dual descriptions. For example, the BPS data from $\mathcal{I}_0'$ is sufficient
to determine the elliptic genus of $\{k_1,k_2,\cdots, k_n\}$ string chain if all $k_i \leq 1$.
For maximally supersymmetric LSTs, the Casimir energy of the elliptic genera must be always zero, i.e., $n_0 = 0$.
We summarize the index $\mathfrak{i}(z)$, the denominator $\mathcal{D}(\tau,z)$, and the numerator $\mathcal{N}(\tau,z)$
of the elliptic genera for various string chains in Table~\ref{tbl:20chain}.
\begin{table}[t]
    \centering
\begin{tabular}{>{\centering\arraybackslash}m{3cm}ccc}\toprule
        String chain & Index $\mathfrak{i}(z)$ & Denominator $\mathcal{D}(\tau,z)$ & Numerator $\mathcal{N}(\tau,z)$\\\midrule
        \vspace{0.15cm}\includegraphics[scale=0.4]{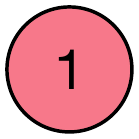} & $m^2 - \epsilon_-^2$ & $ \eta^{-6} \theta_1 (\epsilon_+ \pm \epsilon_-)$ & $\eta^{-6} \theta_1(m \pm \epsilon_+)$ \\\midrule
        \vspace{0.15cm}\includegraphics[scale=0.4]{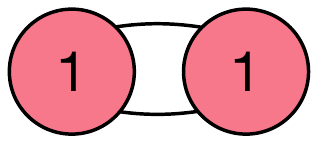}& $2m^2 - 2\epsilon_+^2$ & $ \eta^{-12} \theta_1 (\epsilon_+ \pm \epsilon_-)^2$ & $\eta^{-12} \theta_1(m \pm \epsilon_-)^2$ \\\midrule
        \vspace{0.15cm}\includegraphics[scale=0.4]{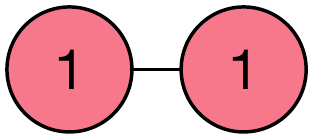} & $2m^2 - \epsilon_+^2 -\epsilon_-^2$ & $ \eta^{-12} \theta_1 (\epsilon_+ \pm \epsilon_-)^2$ & $\eta^{-12} \theta_1(m \pm \epsilon_+)\theta_1(m \pm \epsilon_-)$ \\\midrule
        \vspace{0.05cm}\includegraphics[scale=0.4]{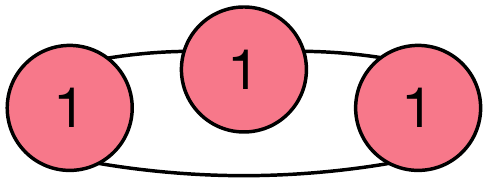}& $3m^2 - 3\epsilon_+^2$ & $ \eta^{-18} \theta_1 (\epsilon_+ \pm \epsilon_-)^3$ & $\eta^{-18} \theta_1(m \pm \epsilon_+)^3$ \\\midrule
        \vspace{0.15cm}\includegraphics[scale=0.4]{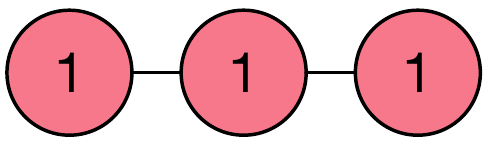}& $3m^2 - 2\epsilon_+^2 - \epsilon_-^2$ & $ \eta^{-18} \theta_1 (\epsilon_+ \pm \epsilon_-)^3$ & $\eta^{-18}\theta_1(m \pm \epsilon_+) \theta_1(m \pm \epsilon_-)^2$\\\midrule
        \vspace{0.10cm}\includegraphics[scale=0.4]{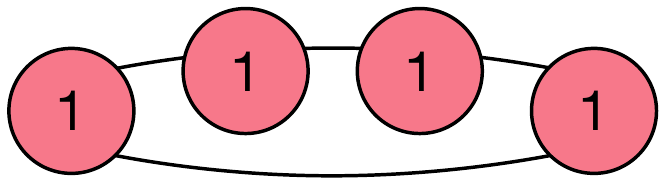} & $4m^2 - 4\epsilon_+^2$ & $ \eta^{-24} \theta_1 (\epsilon_+ \pm \epsilon_-)^4$ & $ \eta^{-24} \theta_1 (\epsilon_+ \pm \epsilon_-)^4$\\\bottomrule
    \end{tabular}
    \caption{Elliptic genera of $(2,0)$ string chains determined from $\mathcal{I}_0'$}
        \label{tbl:20chain}\vspace{1cm}
        \setcounter{table}{2}
                \begin{tabular}{>{\centering\arraybackslash}m{0.8cm}>{\centering\arraybackslash}m{1.5cm}m{8.2cm}>{\centering\arraybackslash}m{1cm}m{3cm}}
	\toprule
	\multirow{2}{*}{\includegraphics[width=0.8cm]{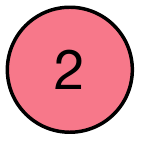}} &  $\mathcal{D}(\tau,z)$ & $\eta^{-12}\theta_1 (\epsilon_+ \pm \epsilon_-)\theta_1 (2\epsilon_+ \pm 2\epsilon_-) $  &  $\mathfrak{i}(z)$ &  $2m^2 + 2\epsilon_+^2 - 4\epsilon_-^2$  \\ \cmidrule(l){2-5} & $\mathcal{N}(\tau,z)$ & \multicolumn{3}{m{13.7cm}@{}}{
	$\begingroup \renewcommand*{\arraystretch}{1.2} \begin{array}{l@{}}
2^{-11}3^{-6}\eta^{-12}\theta_1(m \pm \epsilon_+) \theta_1(m \pm \epsilon_-) \times \big( 27 E_4^3 \mathfrak{R}_2^6-45 E_4^2 \mathfrak{R}_2^4 \mathfrak{R}_0^2-24 E_4 E_6
   \mathfrak{R}_2^5 \mathfrak{R}_0\\
-15 E_4 \mathfrak{R}_2^2 \mathfrak{R}_0^4-32 E_6^2\mathfrak{R}_2^6-40 E_6 \mathfrak{R}_2^3 \mathfrak{R}_0^3+\mathfrak{R}_0^6\big)
	\end{array} \endgroup$
	}\\ \midrule
	\multirow{2}{*}{\includegraphics[width=0.8cm]{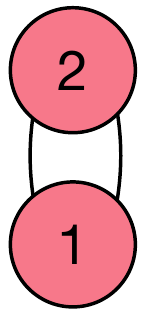}} &  $\mathcal{D}(\tau,z)$ & $\eta^{-18}\theta_1 (\epsilon_+ \pm \epsilon_-)^2 \theta_1 (2\epsilon_+ \pm 2\epsilon_-) $  &  $\mathfrak{i}(z)$ &  $3m^2 - 2\epsilon_+^2 - \epsilon_-^2$  \\
\bottomrule
    \end{tabular}
    \caption{Elliptic genera of $(2,0)$ string chains determined from $\mathcal{I}_0'$ and Table~\ref{tbl:11atype}}
        \label{tbl:20chain2}
\end{table}

\begin{table}[p!]
 \setcounter{table}{1}
    \begin{tabular}{>{\centering\arraybackslash}m{0.5cm}>{\centering\arraybackslash}m{1.5cm}m{8.5cm}>{\centering\arraybackslash}m{1cm}m{3cm}}
	\toprule
	\multirow{2}{*}{$A_1$} &  $\mathcal{D}(\tau,z)$ & $\eta^{-12}\theta_1 (\epsilon_+ \pm \epsilon_-) \theta_1(2\epsilon_+ \pm (a_1 - a_2)) $  &  $\mathfrak{i}(z)$ &  $2 m^2 - 2\epsilon_+^2$  \\ \cmidrule(l){2-5} & $\mathcal{N}(\tau,z)$ & \multicolumn{3}{m{14cm}@{}}{
	$\begingroup \renewcommand*{\arraystretch}{1.2} \begin{array}{l@{}}
 	2^{-7} 3^{-4} \eta^{-6} \theta_1(m \pm\epsilon_+)\big(-9 E_4^2 \mathfrak{f}_{2,1} \mathfrak{g}_{2,1} \mathfrak{R}_2^3-3 E_4 \mathfrak{f}_{2,1} \mathfrak{g}_{2,1} \mathfrak{R}_2 \mathfrak{R}_0^2-9 E_4 \mathfrak{f}_{2,1} \mathfrak{g}_{0,1} \mathfrak{R}_2^2 \mathfrak{R}_0 \nonumber\\
 	-3 E_4 \mathfrak{f}_{0,1} \mathfrak{g}_{0,1} \mathfrak{R}_2^3 -12 E_6 \mathfrak{f}_{2,1} \mathfrak{g}_{2,1} \mathfrak{R}_2^2 \mathfrak{R}_0-8 E_6 \mathfrak{f}_{2,1} \mathfrak{g}_{0,1} \mathfrak{R}_2^3-4 E_6 \mathfrak{f}_{0,1} \mathfrak{g}_{2,1} \mathfrak{R}_2^3+\mathfrak{f}_{2,1} \mathfrak{g}_{0,1} \mathfrak{R}_0^3\\
 	-\mathfrak{f}_{0,1} \mathfrak{g}_{2,1} \mathfrak{R}_0^3+3 \mathfrak{f}_{0,1} \mathfrak{g}_{0,1} \mathfrak{R}_2 \mathfrak{R}_0^2 -3 E_4 \mathfrak{f}_{0,1} \mathfrak{g}_{2,1} \mathfrak{R}_2^2 \mathfrak{R}_0 \big) \nonumber
	\end{array} \endgroup$ }\\ \midrule
	\multirow{2}{*}{$A_2$} &  $\mathcal{D}(\tau,z)$ & $\eta^{-24}\theta_1 (\pm \epsilon_-) \prod_{i<j}^3 \theta_1(\pm (a_i - a_j)) $  &  $\mathfrak{i}(z)$ &  $3 m^2 - 3\epsilon_+^2 \rightarrow 3m^2$  \\
\bottomrule
	\end{tabular}
    \caption{Elliptic genera of $(1,1)$ little strings determined from $\mathcal{I}_{\{k_1,\cdots,k_n\}}$ with all $k_i \leq 1$}
	\label{tbl:11atype}
	 \setcounter{table}{3}
\end{table}

As the elliptic genera of $(2,0)$ string chains capture the infinite tower of momentum modes,
they supply the BPS data with given momentum modes for all winding sectors in $(1,1)$ LSTs.
One can completely determine the elliptic genus of one instanton string based on the provided BPS data.
We summarize the index $\mathfrak{i}(z)$, the denominator $\mathcal{D}(\tau,z)$, the numerator $\mathcal{N}(\tau,z)$
of the elliptic genera for single $SU(2)$, $SU(3)$, $SU(4)$ instanton strings in Table~\ref{tbl:11atype}.
To keep the expressions simpler, we have turned off $\epsilon_+ = 0$ for $SU(3)$ and $SU(4)$ instanton strings.
See Appendix~\ref{sec:weyl} for the explicit expressions for all Weyl invariant Jacobi forms used in this paper.

One can further iterate the procedure to obtain the elliptic genera for higher winding sectors. For instance, we determine the elliptic genera of $\{2,0\}$ and $\{2,1\}$ string chains in $(2,0)$ LST of $\hat{A}_1$-type, using the initial data supplied by the BPS index \eqref{eq:11partpert-su2} of the pure momentum sector and the elliptic genus of the $SU(2)$ single instanton string expressed in Table~\ref{tbl:11atype}. We summarize them in Table~\ref{tbl:20chain2}. Such iteration enables us to find out the infinite tower of the 6d string elliptic genera which constitute the $\mathbf{R}^4\times T^2$ partition function of the maximally supersymmetric LST of $\hat{A}_{n-1}$-type. All of these results precisely match the results of \cite{Kim:2015gha} which were computed from the worldsheet UV gauge theories of type IIA and IIB little strings \cite{Haghighat:2013tka,Witten:1997yu}. Furthermore, the iterated bootstrap requires no more inputs than the modular/analytic properties of the elliptic genera as well as the T-duality relation, so that it can be applied to a broader class of theories for which the worldsheet gauge theory description is unknown. We will focus on two such examples: $(2,0)$ LSTs of $\hat{D}_{n \geq 4}$-type and $E_8 \times E_8$ heterotic LSTs.

\subsection{$(2,0)$ and $(1,1)$ LSTs of $\hat{D}_{n}$-type}

To initiate the modular bootstrap of the $\mathbf{R}^4 \times T^2$ partition functions, we first study the BPS indices $\mathcal{I}_0$ and $\mathcal{I}_0'$ of the pure momentum states in $(2,0)$ and $(1,1)$ LSTs. They are the multi-trace indices which can be obtained from the single-letter partition functions $f_0(\tau,z)$ using \eqref{eq:pert-PE}. The contribution to the single-letter partition function from each $(1,0)$ supermultiplet has been summarized in \eqref{eq:pert-trace}.
$(2,0)$ LST of $\hat{D}_n$ type has $r = \text{rank}(\hat{D}_n)$ free $(2,0)$ tensor multiplets, which gives
\begin{align}
\label{eq:20partpert-dn}
\mathcal{I}_{0} = \text{PE}\bigg[  \frac{r \cdot t \left(v + v^{-1} - u - u^{-1}\right)}{(1-tu)(1-tu^{-1})} \frac{q}{1-q}\bigg].
\end{align}
$(1,1)$ LST of $\hat{D}_n$ type has a $(1,1)$ vector multiplet in the $SO(2n)$ adjoint representation, which gives
\begin{align}
\label{eq:11partpert-dn}
{\mathcal{I}}_{0}' &= \text{PE}\bigg[\frac{t \left(v + v^{-1} - t - t^{-1}\right)}{(1-tu)(1-tu^{-1})} \prod_{i<j}^n\left(w_i' w_j' + w_i'/w_j' + q' /w_i' w_j' + q' w_j' /w_i' + rq'\right)\frac{1}{1-q'}\bigg].
\end{align}
Here we recall that the fractional momentum fugacities are identified in \eqref{eq:so2n-mom-fug} as combinations of the $SO(2n)$ gauge fugacities $w_i'$ and the full momentum fugacity $q'$.

The above indices $\mathcal{I}_0$ and $\mathcal{I}_0'$ capture the BPS spectra of the pure momentum sectors
decoupled from stringy excitations. Using the fugacity relation \eqref{eq:t-duality-so2n} imposed by T-duality,
they provide the BPS data with zero momentum for all winding sectors in the dual descriptions.
One can particularly determine the elliptic genus of $\{k_1,k_2,\cdots, k_{n+1}\}$ string chain for all $k_i \leq 1$ using the BPS index
$\mathcal{I}_0'$.
For instance, those elliptic genera for connected string chains with $k_i \leq 1$ in $(2,0)$ LST of $\hat{D}_4$-type
are summarized in Table~\ref{tbl:20chain}~and~\ref{tbl:20chain3}.
\begin{table}[tbp]
    \renewcommand{\arraystretch}{1.1}
    \begin{tabular}{>{\arraybackslash}m{1.5cm}>{\centering\arraybackslash}m{1.0cm}m{4.5cm}>{\arraybackslash}m{1.5cm}>{\centering\arraybackslash}m{1.0cm}m{5.0cm}}
	\toprule
    \multirow{3}{*}{\includegraphics[width=1.6cm]{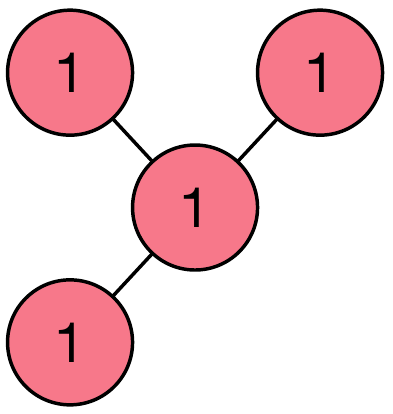}} &  $\mathcal{D}(\tau,z)$ & $\eta^{-24}\theta_1 (\epsilon_+ \pm \epsilon_-)^4$ & \multirow{3}{*}{\includegraphics[width=1.6cm]{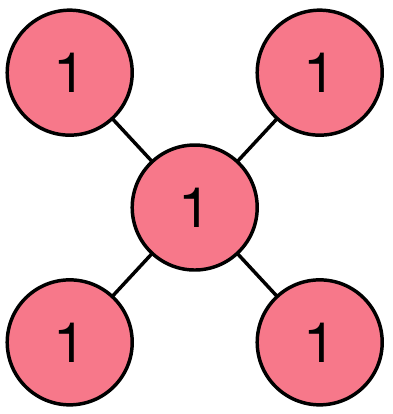}} &  $\mathcal{D}(\tau,z)$  & $\eta^{-30}\theta_1 (\epsilon_+ \pm \epsilon_-)^5$
    \\ \cmidrule(lr){2-3}\cmidrule(l){5-6}  & $\mathfrak{i}(z)$ &  $4m^2 - 3\epsilon_+^2 - \epsilon_-^2$ &   & $\mathfrak{i}(z)$& $5m^2 - 4\epsilon_+^2 - \epsilon_-^2$
    \\ \cmidrule(lr){2-3}\cmidrule(l){5-6}  & $\mathcal{N}(\tau,z)$ & $\eta^{-24}\theta_1(m\pm \epsilon_+)\theta_1(m\pm \epsilon_-)^3$ & &  $\mathcal{N}(\tau,z)$ & $-\eta^{-30}\theta_1(m\pm \epsilon_+)\theta_1(m\pm \epsilon_-)^4$ \\\bottomrule
    \end{tabular}
    \caption{Elliptic genera of $(2,0)$ string chains determined from $\mathcal{I}_0'$}
        \label{tbl:20chain3}
\end{table}
As a next step towards the entire $\mathbf{R}^4 \times T^2$ partition function,
we attempt to bootstrap the elliptic genus of the single $SO(2n)$ instanton string, utilizing the BPS data
given by $\mathcal{I}_0$ and the elliptic genera of $(2,0)$ string chains.
For the specific cases of $\hat{D}_2 = \hat{A}_1^2$, $\hat{D}_3 = \hat{A}_3$ instantons,
the single string elliptic genera have already been constructed in Section~\ref{subsec:20an}, i.e.,
\begin{align}
    \label{eq:so4so6-elliptic}
    \mathcal{I}_1'^{D_2}(\tau,z) &= \left.\frac{\mathcal{N}^{A_1}(\tau,z)}{\mathcal{D}^{A_1}(\tau,z)}\right|_{2a_1 \rightarrow a_1 - a_2} + \left.\frac{\mathcal{N}^{A_1}(\tau,z)}{\mathcal{D}^{A_1}(\tau,z)}\right|_{2a_1 \rightarrow a_1 + a_2}\\
    \mathcal{I}_1'^{D_3}(\tau,z) &= \left.\frac{\mathcal{N}^{A_3}(\tau,z)}{\mathcal{D}^{A_3}(\tau,z)}\right|_{2a_1 \rightarrow a_1 +a_2 - a_3,\ 2a_2 \rightarrow a_1 - a_2 + a_3,\ 2a_3 \rightarrow -a_1 + a_2 + a_3} \nonumber.
\end{align}
For the generic case of $\hat{D}_{n \geq 4}$ instantons, we report that the conjectured form \eqref{eq:ansatz} of the 6d string elliptic genus is \emph{not} compatible with T-duality of the LST. Let us illustrate this point in detail.

Applying the moduli space approximation, the low energy dynamics of instanton strings are described by
the supersymmetric non-linear sigma model onto the instanton moduli space.
Classically the instanton moduli space has a singular point called the small instanton singularity,
which will be replaced by a semi-infinite tube at a quantum level.
As one moves down the tube, the strings gets ejected from NS5-branes as dynamical objects.
This can be viewed as passing from the Higgs branch to the Coulomb branch in the ADHM gauge theory,
whose Higgs branch realizes the instanton moduli space and whose Coulomb branch corresponds to the strings runaway from NS5-branes.
Although the two branches are infinitely far away \cite{Witten:1997yu}, the presence of the semi-infinite tube in the quantum Higgs branch develops the continuum in the spectrum of the Higgs branch CFT \cite{Seiberg:1999xz,Aharony:1999dw}.
For instance, some `throat' states propagating in the tube were identified as supergravity particles \cite{Aharony:1999dw}.
We generally expect such throat states to be captured in the elliptic genera \eqref{eq:elliptic-genera}, defined with
the $SO(4)_N = SU(2)_F \times SU(2)_R$ chemical potentials gapping the continuum states.
For the $\hat{A}_n$ instantons, the throat continuum can be removed by turning on the $SU(2)_r$ triplet of
FI term $\xi^{1,2,3}$ and the theta angle $\theta$, deforming the Higgs branch to be detached from the Coulomb branch.
Such deformation is what the Ansatz \eqref{eq:ansatz} implicitly do.
One has to suitably extend \eqref{eq:ansatz} for the general cases where the Higgs branch cannot be deformed.

For the $\hat{D}_2$ and $\hat{D}_3$ instantons, the throat states' contribution can be isolated by comparing
\eqref{eq:so4so6-elliptic} with the elliptic genera of the ADHM gauge theory for maximally supersymmetric $SO(2n)$ instantons.
We consider $\mathcal{N}=(4,4)$ $Sp(k)$ gauge theory with the following matter contents:
% for $SO(4)$ and $SO(6)$ instantons.
% The ADHM construction \cite{Atiyah:1978ri} is a method for constructing the moduli space of self-dual instantons.
% With stringy realization of self-dual instantons as dynamical D-branes, it provides a useful prescription for a UV gauge theory which flows to the IR non-linear sigma model onto the instanton moduli space.
\begin{align}
\label{eq:spk-adhm-gauge}
\begin{array}{ll}
\text{an $Sp(k)$ symmetric vector multiplet} & A_\mu,\ \varphi_{a A},\ \lambda_{+}^{A\dot{\alpha}},\ \lambda^{\dot{\alpha}a}_{-}\\
\text{an $Sp(k)$ antisymmetric hypermultiplet} & a_{\alpha\dot{\beta}},\ \lambda^{A\alpha}_-, \ \lambda_+^{\alpha a}\\
\text{an $Sp(k)\times SO(2n)$ bifundamental hypermultiplet} & q_{\dot\alpha},\ \psi_{-}^A,\ \psi_{+}^a
\end{array}
\end{align}
where $k$ is the instanton number. Its elliptic genus can be obtained via localizing the gauge theory path integral,
following the formalism of \cite{Benini:2013nda,Benini:2013xpa}. The single string elliptic genus is given by
\begin{align}
    \label{eq:sp1-elliptic}
    \bar{\mathcal{I}}_{1}^{D_n} &= \frac{\theta_1(m \pm \epsilon_-)}{2\theta_1(\epsilon_+ \pm \epsilon_-)} \Bigg[ \sum_{i=1}^n\bigg\{\Big(\frac{\theta_1(4\epsilon_+ - 2a_i)\theta_1(2\epsilon_+ - 2a_i) \prod_{j \neq i}^n \theta_1(m \pm (\epsilon_+ - a_i) \pm a_j)}{\theta_1(m\pm (3\epsilon_+-2a_i) ) \prod_{j \neq i}^n \theta_1(a_i \pm a_j)\theta_1(2\epsilon_+ - a_i \pm a_j) }+ (a_i \rightarrow -a_i) \Big) \bigg\}\nonumber \\&\qquad\qquad\qquad\quad
	- \bigg\{\sum_{p=1}^4 \frac{\theta_1(m - 3\epsilon_+)\theta_1(m - \epsilon_+) \prod_{i=1}^n \theta_p(\frac{3m-\epsilon_+}{2} \pm a_i)}{2\theta_1(2m) \theta_1(2m-2\epsilon_+)  \prod_{i=1}^n \theta_p(\frac{m-3\epsilon_+}{2} \pm a_i)} + (m \rightarrow -m)\bigg\}\Bigg].
\end{align}
We observe that the difference between  \eqref{eq:so4so6-elliptic} and \eqref{eq:sp1-elliptic} only arise in the \emph{full} momentum sector, being independent of the $SO(4)$ and $SO(6)$ holonomies. Taking the limit $q' \rightarrow 0$ which truncates the tower of momentum modes,
the difference between two indices are
\begin{align}
    \label{eq:throat-dn}
    \bar{\mathcal{I}}_1^{D_2} - \mathcal{I}_1'^{D_2} &\rightarrow \frac{t^2(v+v^{-1}-u-u^{-1})}{(1-tu^\pm)(1+tv^\pm)} - \frac{2t(v+v^{-1}-u-u^{-1})}{(1-tu^\pm)}\\
    \bar{\mathcal{I}}_1^{D_3} - \mathcal{I}_1'^{D_3} &\rightarrow \frac{t^2(v+v^{-1}-u-u^{-1})}{(1-tu^\pm)(1+tv^\pm)} - \frac{t(v+v^{-1}-u-u^{-1})}{(1-tu^\pm)}\nonumber.
\end{align}
The first term captures the single-letter operators with a unit momentum for 11d gravity multiplet in
$\mathbf{R}^{1,4} \times (\mathbf{R}^4 \times S^1) / \mathbf{Z}_2 \times S^1$, which can be dualized to
type IIB ON5${}^-$ plane engineering $\hat{D}_n$ LSTs~\cite{Hwang:2016gfw}.
% This shows that the throat states do not fully belong to the 6d BPS spectrum.
The denominator of the throat states' index is
$ \sinh(\tfrac{\epsilon_+ \pm \epsilon_-}{2})\,\cosh(\tfrac{-\epsilon_+ \pm m}{2})$, which will be promoted to
\begin{align}
    \mathcal{D}^\text{bulk}(\tau,z) = \frac{\theta_1(\epsilon_+ \pm \epsilon_-)}{\eta^6}\frac{\theta_1(-2\epsilon_+ \pm 2m)}{\theta_1(-\epsilon_+ \pm m)}.
\end{align}
The BPS indices for the throat states must share the same modularity \eqref{eq:modular-property} with \eqref{eq:so4so6-elliptic} and \eqref{eq:sp1-elliptic}. Here we attempt to recast them into the form of \eqref{eq:ansatz} with the new denominator $\mathcal{D}^\text{bulk}(\tau,z)$.
\begin{align}
 \bar{\mathcal{I}}_1^{D_2} - \mathcal{I}_1'^{D_2} &= 	 \frac{1}{2^{11}3^5}\frac{\theta_1(m \pm \epsilon_-)\theta_1(-\epsilon_+ \pm m)}{\theta_1(\epsilon_+ \pm \epsilon_-)\theta_1(-2\epsilon_+ \pm 2m)} \big(27 E_4^3 \mathfrak{R}_2^2 \mathfrak{f}_{2,1}^4-18 E_4^2 \mathfrak{R}_2^2 \mathfrak{f}_{2,1}^2 \mathfrak{f}_{0,1}^2-24 E_4^2 \mathfrak{R}_0 \mathfrak{R}_2 \mathfrak{f}_{2,1}^3
    \mathfrak{f}_{0,1}\nonumber\\&
    -16 E_4 E_6 \mathfrak{R}_2^2 \mathfrak{f}_{2,1}^3 \mathfrak{f}_{0,1}-8 E_4 E_6 \mathfrak{R}_0
    \mathfrak{R}_2 \mathfrak{f}_{2,1}^4-E_4 \mathfrak{R}_2^2 \mathfrak{f}_{0,1}^4-8 E_4 \mathfrak{R}_0 \mathfrak{R}_2 \mathfrak{f}_{2,1} \mathfrak{f}_{0,1}^3-6 E_4 \mathfrak{R}_0^2
    \mathfrak{f}_{2,1}^2 \mathfrak{f}_{0,1}^2\nonumber\\&-32 E_6^2 \mathfrak{R}_2^2 \mathfrak{f}_{2,1}^4-8 E_6 \mathfrak{R}_2^2 \mathfrak{f}_{2,1} \mathfrak{f}_{0,1}^3-24 E_6 \mathfrak{R}_0
    \mathfrak{R}_2 \mathfrak{f}_{2,1}^2 \mathfrak{f}_{0,1}^2-8 E_6 \mathfrak{R}_0^2 \mathfrak{f}_{2,1}^3 \mathfrak{f}_{0,1}+\mathfrak{R}_0^2 \mathfrak{f}_{0,1}^4-3 E_4^2 \mathfrak{R}_0^2 \mathfrak{f}_{2,1}^4\big). \nonumber\\
 \bar{\mathcal{I}}_1^{D_3} - \mathcal{I}_1'^{D_3} &= \frac{1}{2^{11}3^6}\frac{\theta_1(m \pm \epsilon_-)\theta_1(-\epsilon_+ \pm m)}{\theta_1(\epsilon_+ \pm \epsilon_-)\theta_1(-2\epsilon_+ \pm 2m)}  \big(
 27 E_4^3 \mathfrak{f}_{2,1}^6-45 E_4^2 \mathfrak{f}_{2,1}^4
    \mathfrak{f}_{0,1}^2-24 E_4 E_6 \mathfrak{f}_{2,1}^5 \mathfrak{f}_{0,1} \nonumber\\
	 & -15 E_4 \mathfrak{f}_{2,1}^2 \mathfrak{f}_{0,1}^4 -32 E_6^2 \mathfrak{f}_{2,1}^6-40 E_6 \mathfrak{f}_{2,1}^3 \mathfrak{f}_{0,1}^3+\mathfrak{f}_{0,1}^6
	 \big)
     \label{eq:throat-dn-modular}
\end{align}
Starting from $\hat{D}_4$ instantons, it is not possible to separate the throat states' contribution from \eqref{eq:sp1-elliptic}.
If that were possible, one could bootstrap the BPS index for the throat states based on \eqref{eq:ansatz} with the new denominator  $\mathcal{D}^\text{bulk}(\tau,z)$, satisfying the modularity \eqref{eq:modular-property} with index $\mathfrak{i} = (2n-2)(m^2 - \epsilon_+^2)$.
As the denominator $\mathcal{D}^\text{bulk}(\tau,z)$ is a weak Jacobi form of weight $-2$ and index $\mathfrak{i}_d = 3m^2 + 4\epsilon_+^2 + \epsilon_-^2$, the numerator $\mathcal{N}(\tau,z)$ would be a weak Jacobi form of index $\mathfrak{i} + \mathfrak{i}_d = \epsilon_-^2 + (1+2n) m^2 + (6-2n) \epsilon_+^4$. \\
This cannot exist for $n \geq 4$, although the $Sp(1)$ elliptic genus still includes the first term in \eqref{eq:throat-dn} counting the throat states.
We conclude that the throat states’ contribution cannot be isolated out.

Accordingly, the elliptic genera of the $SO(2n)$ instanton strings involve the extra factor $\frac{\theta_1(-2\epsilon_+ \pm 2m)}{\theta_1(-\epsilon_+ \pm m)}$ in the denominator.
It would be desirable if one could precisely distinguish the bulk states from the 6d LST spectrum.
For now, we continue to bootstrap the elliptic genus of the single $SO(2n)$ instanton string for $n \geq 4$
with the modified denominator
\begin{align}
    \mathcal{D}^\text{bulk}(\tau,z) = \mathcal{D}(\tau,z) \cdot \frac{\theta_1(-2\epsilon_+ \pm 2m)}{\theta_1(-\epsilon_+ \pm m)}.
\end{align}
As the bootstrapped elliptic genus $\mathcal{I}_1'^{D_n}$ share the same modular and analytic properties with \eqref{eq:sp1-elliptic} and has to display the same BPS data for fractional momentum modes, it must inevitably agree with the $Sp(1)$ elliptic genus \eqref{eq:sp1-elliptic}. We determined the coefficients in $n=4$ case up to $q^1$ order and found the agreement (after turning off $\epsilon_+ = 0$).
To move on to the next, we extract the BPS data for fractional momentum modes from \eqref{eq:sp1-elliptic} and study the elliptic genera of fractional string chains in $(2,0)$ LST, which are summarized in Table~\ref{tbl:20chain2}~and~\ref{tbl:20chain5}.
One can study the higher winding sector by the iterated bootstrap. For $k$ strings, the elliptic genus can be bootstrapped with the denominator
\begin{align}
    \mathcal{D}^\text{bulk}(\tau,z) = \mathcal{D}(\tau,z) \cdot\prod_{n=1}^k \frac{\theta_1(-2n\epsilon_+ \pm 2nm)}{\theta_1(-n\epsilon_+ \pm n m)}
\end{align}
which must reproduce the $Sp(k)$ elliptic genus of the ADHM gauge theory \eqref{eq:spk-adhm-gauge}. Utilizing the BPS data with fractional momentum modes in the $Sp(2)$ elliptic genus, we obtain the elliptic genera of higher winding modes displayed in Table~\ref{tbl:20chain6}.
These novel indices are expected to be a useful probe to find the UV gauge theory for DE-type $(2,0)$ string chains generalizing \cite{Haghighat:2013tka}.

\begin{table}
\begin{tabular}{>{\centering\arraybackslash}m{1.3cm}>{\centering\arraybackslash}m{1.5cm}m{7.7cm}>{\centering\arraybackslash}m{1cm}m{3cm}}
	\toprule
		\multirow{2}{*}{\includegraphics[width=1.4cm]{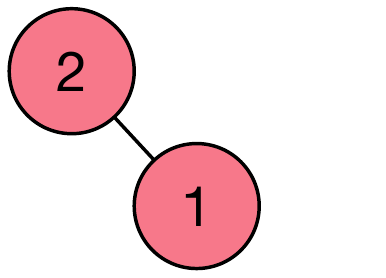}} &  $\mathcal{D}(\tau,z)$ & $\eta^{-18}\theta_1 (\epsilon_+ \pm \epsilon_-)^2\theta_1 (2\epsilon_+ \pm 2\epsilon_-) $  &  $\mathfrak{i}(z)$ &  $3 m^2-3 \epsilon _-^2$
    \\
\bottomrule
    \end{tabular}
    \caption{Elliptic genera of $(2,0)$ string chains determined from $\mathcal{I}_0'$ and $\bar{\mathcal{I}}_1$}
        \label{tbl:20chain5}
\end{table}

\begin{table}
\begin{tabular}{>{\centering\arraybackslash}m{1.3cm}>{\centering\arraybackslash}m{1.5cm}m{7.7cm}>{\centering\arraybackslash}m{1cm}m{3cm}}
	\toprule
	\multirow{2}{*}{\includegraphics[width=1.3cm]{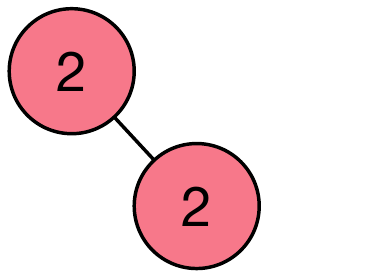}} &  $\mathcal{D}(\tau,z)$ & $\eta^{-24}\theta_1 (\epsilon_+ \pm \epsilon_-)^2\theta_1 (2\epsilon_+ \pm 2\epsilon_-)^2$  &  $\mathfrak{i}(z)$ &  $4 m^2-4 \epsilon _-^2$  \\
\bottomrule
    \end{tabular}
    \caption{Elliptic genera of $(2,0)$ string chains determined from $\mathcal{I}_0'$, $\bar{\mathcal{I}}_1$, $\bar{\mathcal{I}}_2$}
    \label{tbl:20chain6}
\end{table}

\subsection{Heterotic little strings}

To construct the $\mathbf{R}^4 \times T^2$ partition functions of heterotic LSTs, let us first consider the BPS indices of their pure momentum sectors. They are the multi-letter BPS indices, which can be computed by taking the Plethystic exponential on the single-letter indices 
\begin{align}
    f_0(\tau, z)  \equiv \text{tr}\left[(-1)^F q^{H_L} \, t^{J_{r} + J_{R}} u^{J_{l}} v^{J_F} \, \prod_{i=1}^{n}w_i^{G_i}\,\prod_{a=1}^{16} y_a^{f_a}
    \right].
\end{align}
$G_i$ and $f_{a}$ are the Cartan generators of the $Sp(n)$ gauge symmetry and the $SO(32)$ or $E_8 \times E_8$ flavor symmetry. Each $(1,0)$ supermultiplet contributes to the single-letter partition function by
\begin{align}
	\text{tensor}:& \quad -\frac{t (u + u^{-1})}{(1- tu) (1-tu^{-1})}\textstyle\left(\sum_{n=-\infty}^{\infty} q^n\right)^+\\
	\text{vector}:& \quad -\frac{t (t + t^{-1})}{(1- tu) (1-tu^{-1})}\textstyle \textstyle\left(\chi_{\mathbf{adj}}\,(w_{i}) \sum_{n=-\infty}^{\infty} q^n\right)^+ \nonumber \\
	\text{$\tfrac{1}{2}$-hyper}\,: & \quad+\frac{t \cdot \chi_{\bf F}(v, y_a)}{(1- tu) (1-tu^{-1})}\textstyle \textstyle\left(\chi_{\mathbf{R}}\,(w_{i}) \sum_{n=-\infty}^{\infty} q^n\right)^+ \nonumber
\end{align}
where $\chi_{\bf F}$ denotes the irreducible character for a flavor representation $\mathbf{F}$ of a given $\frac{1}{2}$-hypermultiplet.

The stack of $n$ heterotic NS5-branes in $E_8 \times E_8$ string theory engineers the rank-$n$ heterotic LST with $E_8 \times E_8$ flavor symmetry. It has $n$ tensor multiplets and $n$ free hypermultiplets. These elementary fields and their derivatives lead to the following multi-letter index:
\begin{align}
	\label{eq:he-pure-momentum}
    \mathcal{I}_0 = \text{PE} \bigg[ - \frac{n \, t \cdot(u+u^{-1})}{(1-tu^\pm)}\frac{q}{ 1-q}  +  \frac{n\, t \cdot(v+v^{-1})
%    \cdot ( \chi_{\bf adj}^{E_8} (y_{1, \cdots, 8})  +  \chi_{\bf adj}^{E_8} (y_{9,\cdots,16}) )
	}{(1-tu^\pm)}\frac{q}{ 1-q}\bigg]
\end{align}
%where $\chi_{\bf adj}^{E_8}$ denotes the $E_8$ adjoint character.
On the other hand, the $SO(32)$ heterotic LST of rank-$n$ has a vector multiplet in the $Sp(n)$ adjoint representation, a half-hypermultiplet in the $Sp(n)$ antisymmetric representation, and a hypermultiplet in the $Sp(n) \times SO(32)$ bifundamental representation. The corresponding multi-trace partition function is given by
\begin{align}
	\label{eq:ho-pure-momentum}
    \mathcal{I}_0' = \text{PE}\Bigg[ \frac{t \cdot \chi_{\bf fnd}^{SO(32)} }{(1-tu^\pm)}
    \bigg(\frac{\chi_{\bf fnd}^{Sp(n)}}{ 1-q'}\bigg)^+ - \frac{t \cdot(t+t^{-1})}{(1-tu^\pm)}\bigg(\frac{\chi_{\bf sym}^{Sp(n)}}{1-q'}\bigg)^+ + \frac{t \cdot(v+v^{-1})}{(1-tu^\pm)}\bigg(\frac{\chi_{\bf anti}^{Sp(n)}}{1-q'}\bigg)^+ \Bigg]
\end{align}
where the irreducible characters for $Sp(n)$ and $SO(32)$ representations are 
\begin{align}   
    &\chi_{\bf sym}^{Sp(n)}(w'_i) = \sum_{i \leq j}^n \Big(w'_i w'_j + \frac{1}{w'_i w'_j} + \frac{w_i'}{w_j'} + \frac{w'_j}{w'_i}\Big) - n,  &&\chi_{\bf fnd}^{Sp(n)}(w_i') = \sum_{i=1}^n (w_i' + w_i'^{-1})\\&
    \chi_{\bf anti}^{Sp(n)}(w'_i) = \sum_{i < j}^n \Big(w'_i w'_j + \frac{1}{w'_i w'_j} + \frac{w'_i}{w'_j} + \frac{w_j'}{w_i'}\Big) + n,  &&\chi_{\bf fnd}^{SO(32)}(y_a') = \sum_{a=1}^{16} (y_a' + y_a'^{-1}).
\end{align}
Here we recall that the fractional momentum fugacities have been identified in \eqref{eq:het-mom-fug} using the full momentum fugacity $q'$ and the $Sp(n)$ gauge fugacities $w_i'$. All gauge and flavor fugacity variables in $SO(32)$ LST are primed for distinction.

As the T-duality between the two LSTs involves the Wilson lines \eqref{eq:het-wilson-line} preserving the $SO(16)\times SO(16)$ flavor symmetry, it is more convenient to express the indices \eqref{eq:he-pure-momentum} and \eqref{eq:ho-pure-momentum} in terms of the $SO(16) \times SO(16)$ flavor fugacities. Recall that the background Wilson line $R A_a$ shuffles the  momentum $H_L$, the flavor charges $f_a$, the winding number(s) $k_i$ \cite{Narain:1986am}. The new fugacity variables conjugate to the shifted charges are identified in \eqref{eq:charge-relation-2}. By replacing the original variables with the new ones, then dropping out the tildes for simplicity, the indices $\mathcal{I}_0$ and $\mathcal{I}'_0$ for the pure momentum sectors become
\begin{align}
	\label{eq:pure-he-wilson}
    \mathcal{I}_0 = \text{PE} \bigg[& - \frac{n \, t \cdot(u+u^{-1} -v - v^{-1})}{(1-tu^\pm)}\frac{q^2}{ 1-q^2} \bigg]\\
	\label{eq:pure-ho-wilson}
\mathcal{I}_0' = \text{PE}\Bigg[& \frac{t \cdot \chi_{\bf fnd}^{SO(16)}(y_{1, \cdots, 8}) }{(1-tu^\pm)}
    \frac{\sum_{i=1}^n (w_i' + q'^2/w_i')}{ 1-q'^2} + \frac{t \cdot \chi_{\bf fnd}^{SO(16)}(y_{9, \cdots, 16}) }{(1-tu^\pm)}
    \frac{\sum_{i=1}^n (q' w_i' + q'/w_i')}{ 1-q'^2}  \nonumber\\&
    - \frac{t \cdot(t+t^{-1})}{(1-tu^\pm)}\frac{\sum_{i \leq j}^n (w'_i w'_j + q'^2/{w'_i w'_j}) + \sum_{i < j}^n ( w_i'/w_j' + q'^2 w'_j/w'_i) + n q'^2}{1-q'^2} \nonumber\\&
    + \frac{t \cdot(v+v^{-1})}{(1-tu^\pm)}\frac{\sum_{i < j}^n (w'_i w'_j + q'^2/{w'_i w'_j} + w_i'/w_j' + q'^2 w'_j/w'_i) + n q'^2}{1-q'^2} \Bigg].
\end{align}	

The indices $\mathcal{I}_0$ and $\mathcal{I}'_0$ capture the infinite towers of the pure momentum states. Based on the T-duality relation \eqref{eq:het-fug-tdual}, the BPS data supplied by $\mathcal{I}_0$ and $\mathcal{I}'_0$ are used to determine the numerical coefficients in the elliptic genera \eqref{eq:ansatz} of various winding sectors. We should first replace the chemical potentials in the ansatz \eqref{eq:ansatz} following \eqref{eq:charge-relation-2}, expand it in the momentum fugacities, then compare it with the BPS data from $\mathcal{I}_0$ and $\mathcal{I}'_0$ to determine the numerical coefficients. The zero point energy of the elliptic genus matches with
\begin{align}
n_0 = 
\begin{dcases*}
-12 (k_1 + k_{n+1}) & for $\{k_1, k_2, \cdots, k_{n+1}\}$ $E_8 \times E_8$ string chains \\
-24k & for $k$ $SO(32)$ heterotic strings.
\end{dcases*}
\end{align}
In particular, the BPS data from \eqref{eq:pure-he-wilson} are sufficient to determine the elliptic genus of $\{k_1, k_2, \cdots, k_{n+1}\}$ string chain if $k_{i} \leq 1$ for all $i$ and $\prod_{j=1}^{n+1} k_j = 0$. The index $\mathfrak{i}(z)$, the denominator $\mathcal{D}(\tau,z)$, the numerator $\mathcal{N}(\tau,z)$ of the elliptic genera for some of such string chains are summarized in Table~\ref{tbl:table-he} \cite{Gadde:2015tra,Kim:2015fxa,Gu:2017ccq}.
\begin{table}[tbp]
    \centering
     \begin{tabular}{>{\arraybackslash}p{3cm}>{\centering\arraybackslash}m{1.5cm}m{4cm}>{\centering\arraybackslash}m{1cm}m{5cm}}
	\toprule
	\multirow{2}{*}{\parbox[c]{3cm}{\vspace{-0.4cm}\hspace{-0.6cm}\includegraphics[height=2cm]{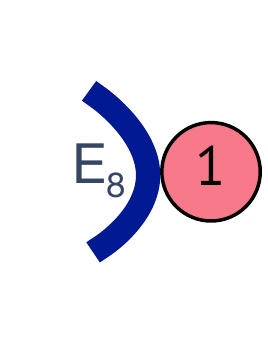}}} &  $\mathcal{D}(\tau,z)$ & $\eta^{-6}\theta_1 (\epsilon_+ \pm \epsilon_-) $  &  $\mathfrak{i}(z)$ &  $ -\epsilon_+^2 - \epsilon_-^2 + \frac{1}{2}\sum_{i=1}^{8}m_i^2$  \\ \cmidrule(l){2-5} & $\mathcal{N}(\tau,z)$ & \multicolumn{3}{m{11.5cm}@{}}{$-\mathfrak{f}_{-4,1}^{E_8}$  }\\ \midrule
	\multirow{2}{*}{\parbox[c]{3cm}{\vspace{-0.4cm}\hspace{-0.6cm}\includegraphics[height=2cm]{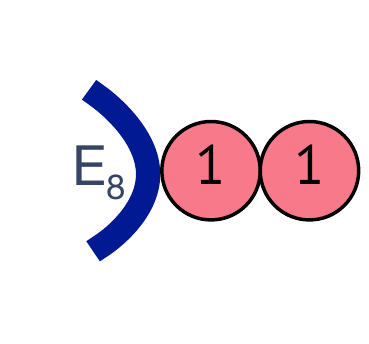}}} &  $\mathcal{D}(\tau,z)$ & $\eta^{-12}\theta_1 (\epsilon_+ \pm \epsilon_-)^2 $  &  $\mathfrak{i}(z)$ &  $m^2 - 2\epsilon_+^2 - \epsilon_-^2 + \frac{1}{2}\sum_{i=1}^{8}m_i^2$  \\ \cmidrule(l){2-5} & $\mathcal{N}(\tau,z)$ &\multicolumn{3}{m{11.5cm}@{}}{$-  \mathfrak{f}_{-4,1}^{E_8} \; \eta^{-6}\theta_1( m \pm \epsilon_-)$} \\ \midrule
	\multirow{2}{*}{\parbox[c]{3cm}{\vspace{-0.4cm}\hspace{-0.6cm}\includegraphics[height=2cm]{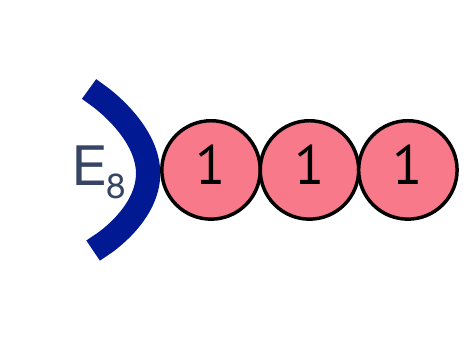}}} &  $\mathcal{D}(\tau,z)$ & $\eta^{-18}\theta_1 (\epsilon_+ \pm \epsilon_-)^3 $  &  $\mathfrak{i}(z)$ &  $2m^2 - 3\epsilon_+^2 - \epsilon_-^2 + \frac{1}{2}\sum_{i=1}^{8}m_i^2$  \\ \cmidrule(l){2-5} & $\mathcal{N}(\tau,z)$ &\multicolumn{3}{m{11.5cm}@{}}{$-  \mathfrak{f}_{-4,1}^{E_8} \; \eta^{-12}\theta_1( m \pm \epsilon_-)^2$} \\ \midrule
	\multirow{2}{*}{\parbox[c]{3cm}{\vspace{-0.4cm}\hspace{-0.6cm}\includegraphics[height=2cm]{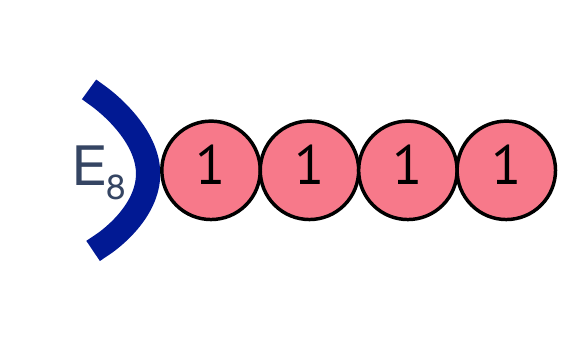}}} &  $\mathcal{D}(\tau,z)$ & $\eta^{-24}\theta_1 (\epsilon_+ \pm \epsilon_-)^4 $  &  $\mathfrak{i}(z)$ &  $3m^2 - 4\epsilon_+^2 - \epsilon_-^2 + \frac{1}{2}\sum_{i=1}^{8}m_i^2$  \\ \cmidrule(l){2-5} & $\mathcal{N}(\tau,z)$ &\multicolumn{3}{m{11.5cm}@{}}{$-  \mathfrak{f}_{-4,1}^{E_8} \; \eta^{-18}\theta_1( m \pm \epsilon_-)^3$} \\ \bottomrule
    \end{tabular}
    \caption{Elliptic genera of string chains in $E_8 \times E_8$ heterotic LST}
    \label{tbl:table-he}
\end{table}

Here we comment about the elliptic genera of the full string chains where $k_1 = \cdots = k_{n+1} = k > 0$. Once we attempt to write their elliptic genera based on \eqref{eq:ansatz}, the whole elliptic genus would be a weak Jacobi form of weight $0$ and index $\mathfrak{i} = k\left( -\epsilon_-^2 - (n+2)\epsilon_+^2 + (n-1)m^2 \right)$. Moreover, its denominator $\mathcal{D}(\tau,z)$ would have weight $-2k(n+1)$ and 	index $\mathfrak{i}_d = \frac{(n+1)\cdot k(k+1)(2k+1)}{12}(\epsilon_+^2 + \epsilon_-^2)$. For the $k=1$ case, this implies that the numerator $\mathcal{N}(z,\tau)$ must be a weak Jacobi form of weight $(-2n+4)$ and index $\mathfrak{i}_n = (n-1)m^2 +(\frac{n-1}{2}) \epsilon_-^2 -(\frac{n+3}{2}) \epsilon_+^2$ which does not exist. We interpret it as an inevitable appearance of the throat states, for which one has to conjecture the new denominator $\mathcal{D}^{\rm bulk}(\tau,z)$ associated to the extra bosonic zero modes. The throat states correspond to the strings runaway from NS5-branes. But their contribution to the elliptic genus cannot be separated from the states localized on NS5-branes.

We make the similar observation for the elliptic genera of $SO(32)$ heterotic little strings. One may continue the iterated bootstrap to study the elliptic genera of $SO(32)$ heterotic little strings, using the ansatz \eqref{eq:ansatz} and the BPS data from $\mathcal{I}_{\{k_1, k_2, \cdots, k_{n+1}\}}$ of  fractional strings with $k_{i} \leq 1$ and $\prod_{j=1}^{n+1} k_j = 0$.  However, the conjectured form \eqref{eq:ansatz} of the elliptic genus is not compatible with T-duality, because the overdetermined set of equations for numerical coefficients, provided by T-duality relation \eqref{eq:het-fug-tdual}, is inconsistent. We again interpret it as an unavoidable presence of the throat states corresponding to the strings escaping from NS5-branes. Let us take the 5d limit $q' \rightarrow 0$ to find the new denominator $\mathcal{D}^{\rm bulk}(\tau,z)$ including the extra bosonic zero modes. What remains is 5d $\mathcal{N}=1$ $Sp(n)$ gauge theory with 1 antisymmetric and 8 fundamental hypermultiplets. \cite{Hwang:2014uwa} computed its Omega-deformed $\mathbf{R}^4 \times S^1$ partition function. In particular, the $Sp(n)$ neutral states contributes to the partition function by \cite{Hwang:2014uwa}
\begin{align}
	\text{PE}\bigg[-\frac{t^2 (t+t^{-1})(u+u^{-1}+v+v^{-1})}{2(1-tu^\pm)(1-tv^\pm)}\frac{\mathfrak{n}'^2}{1-\mathfrak{n}'^2} - \frac{t^2 (\chi_{\bf 128}^{SO(16)}(y_{i}')\, \mathfrak{n}' + \chi_{\bf 120}^{SO(16)}(y_{i}')\, \mathfrak{n}'^2 )}{(1-tu^\pm)(1-tv^\pm)}\frac{1}{1-\mathfrak{n}'^2}\bigg].
\end{align}
The first term is the single-letter index for the 11d gravity multiplet on $\mathbf{R}^{1,8} \times S^1 \times \mathbf{R}^+$. The second term comes from the single-letter operators of the 10d $E_8$ gauge theory with the $E_8 \rightarrow SO(16)$ Wilson line. The 6d partition function should also contain these bulk states having denominator $\prod_{n=1}^k \sinh(\frac{n\epsilon_+ \pm n\epsilon_-}{2}) \, \sinh(\frac{-n\epsilon_+ \pm n m}{2})$. Reflecting this observation, we propose the new denominator $\mathcal{D}^{\rm bulk}(\tau,z)$ of the $k$ string elliptic genus to be 
\begin{align}
\label{eq:den-new}
\mathcal{D}^{\rm bulk}(\tau,z) = \mathcal{D}(\tau,z) \cdot \prod_{n=1}^k\frac{\theta_1(-n\epsilon_+ \pm nm)}{\eta^6}.
\end{align}
The bootstrapped elliptic genus $\mathcal{I}_k'$ shares the same modular and analytic properties with the elliptic genus of $\mathcal{N}=(0,4)$ $O(k)$ gauge theory having the following field contents:
\begin{align}	
\begin{array}{ll}
\text{an $O(k)$ adjoint vector multiplet} & A_\mu,\ \lambda_{+}^{A\dot{\alpha}}\\
\text{an $O(k)$ symmetric hypermultiplet} & a_{\alpha\dot{\beta}},\ \lambda^{A\alpha}_-\\
\text{an $O(k)$ symmetric twisted hypermultiplet} & \varphi_{a A},\ \lambda^{\dot{\alpha}a}_{-}\\
\text{an $O(k)$ antisymmetric Fermi multiplet} & \lambda_{+}^{\alpha a}\\
\text{an $O(k) \times Sp(N)$ bifundamental hypermultiplet} & q_{\dot\alpha},\ \psi_{-}^A\\
\text{an $O(k) \times Sp(N)$ bifundamental Fermi multiplet} & \psi_{+}^a\\
\text{an $ O(k) \times SO(16)$ bifundamental Fermi multiplet} & \Psi_l
\end{array}
\end{align}
This is the 2d ADHM gauge theory of $k$ F1 and $n$ NS5-branes in $SO(32)$ heterotic string theory \cite{Johnson:1998yw}. In $k=1$ case, by localizing the gauge theory path integral, one can write the elliptic genus as follows.
\begin{align}
	\mathcal{I}'_1 = \frac{\eta^2}{2 \theta_1(\epsilon_+ \pm \epsilon_-)\theta_1(-\epsilon_+ \pm m)}\sum_{i=1}^4 \prod_{j=1}^n \frac{\theta_i (m \pm a_j)}{\theta_i (\epsilon_+ \pm a_j)}\prod_{l=1}^{16}\frac{\theta_i(m_l)}{\eta}.
\end{align}
For example, if $n=1$, one can see that its denominator agrees with \eqref{eq:den-new} using $\prod_{i=1}^4\theta_i (z) = \eta^3 \theta_1 (2z)$. 
It also captures the correct BPS data of all fractional winding modes in the dual theory, such as Table~\ref{tbl:table-he}. It would be desirable if one could precisely distinguish the bulk states from the 6d LST spectrum in the elliptic genera. One can still continue to study the elliptic genera of fractional string chains in $E_8 \times E_8$ little string theory,  based on the BPS data provided by $\mathcal{I}_1'$. This procedure can be iterated up to as high winding numbers as we want.

\section{Concluding remarks}
\label{sec:conclusion}

In this work, we studied the elliptic genera of 6d strings using their modular properties. They are weak Jacobi forms of weight $0$ and index $\mathfrak{i}(z)$ which can be derived from the anomaly polynomial of 6d strings \cite{DelZotto:2016pvm,Gu:2017ccq}. The conjectured form of the 6d string elliptic genera respects the analytic structure of the $\mathbf{R}^4 \times T^2$ instanton partition function \cite{DelZotto:2016pvm}. Given a finite amount of initial BPS data, we constructed the elliptic genera of 6d strings in various 6d SCFTs \cite{DelZotto:2016pvm,Gu:2017ccq}.

We also applied the general ansatz for the 6d string elliptic genera to study the  little string theories. T-duality of little string theories is an equivalence between two circle compactified LSTs, interchanging the winding and momentum modes, when their circle radii $R$ and $\tilde{R}$  are related as $\tilde{R} =  \alpha'/R$ \cite{Seiberg:1997zk}. The $\mathbf{R}^4 \times T^2$ partition functions for T-dual LSTs should agree with each other, after imposing a fugacity relation which identifies the winding/momentum fugacities on one side with the momentum/winding fugacities on the other side \cite{Kim:2015gha,Hohenegger:2016eqy,Kim:2017xan}. Once we know the elliptic genus at a given winding number, it can supply the BPS data for the dual elliptic genera at any winding number but a given circle momentum. We summarized the fugacity maps for $\mathcal{N}=(2,0)$ and $(1,1)$ LSTs of $\hat{A}_n$ and $\hat{D}_n$ types as well as $\mathcal{N}=(1,0)$ $E_8 \times E_8$ and $SO(32)$ heterotic LSTs. We also worked out the anomaly polynomials of strings in those LSTs, to derive the modular properties of the little string elliptic genera. Collecting these pieces of information, the elliptic genera of various winding modes in LSTs can be constructed. We initially prepare the BPS indices for the pure momentum sectors, then utilize their BPS data to fix all numerical coefficients in the dual elliptic genera. Then the obtained elliptic genera yields more BPS data to fix the dual elliptic genera with higher winding numbers. One can gain more BPS data for each iteration. In principle, the entire LST partition functions on $\mathbf{R}^4 \times T^2$ can be constructed from the iterated bootstrap. We successfully bootstrapped the elliptic genera of various fractional string chains in $\mathcal{N}=(2,0)$ LSTs of $\hat{A}_n$ and $\hat{D}_n$ types and $\mathcal{N}=(1,0)$ $E_8 \times E_8$ heterotic LST. 

For some little string theories, the \emph{full} string elliptic genera may include an additional contribution that comes from the bulk bound states  unrelated to the 6d physics. These states are localized in the throat continuum of the target space, which is a quantum resolution of the point-like  singularity in the classical moduli space of 6d strings \cite{Seiberg:1999xz,Aharony:1999dw}. Unless we suppress the emergence of the throat region by Fayet-Iliopoulos deformation, just as we did in $\mathcal{N}=(2,0)$ and $(1,1)$ LSTs of $\hat{A}_n$ and $\hat{D}_{2,3}$ types, the \emph{full} strings may escape from NS5-branes by moving down the throat region. For $(1,1)$ LSTs of $\hat{D}_n$ types and $SO(32)$ heterotic LSTs, we proposed the new ansatz for the full string elliptic genera to include the extra bosonic zero modes  parameterizing the string movement transverse to NS5-branes. With the new ansatz, the bootstrapped elliptic genera agree with those of the ADHM gauge theories. It would be desirable to separate out the throat states from the bound states localized on NS5-branes. To achieve this, one might examine the $T^2$ partition function of the ADHM gauge theories with NS-NS boundary condition \cite{Gadde:2013ftv}. Each term in the partition function may have an  interpretation as a gauge invariant operator, while it uniquely maps to a term in the ADHM elliptic genus. This analysis would be helpful to distinguish the throat states in the full string elliptic genus, identifying the entire BPS spectrum of the little string theory on $\mathbf{R}^4 \times T^2$. We hope to solve this problem in a near future.

\vspace{0.5cm}
\noindent{\bf\large Acknowledgements}

\noindent
We thank Seok Kim, Sung-Soo Kim, Antonio Sciarappa for helpful comments and discussions.  KL is supported in part by the National Research Foundation of Korea Grant NRF-2017R1D1A1B06034369. JP is supported in part by the NRF Grant 2015R1A2A2A01007058.

\appendix
\section{Weyl invariant Jacobi forms}
\label{sec:weyl}

In this Appendix, we collect the explicit expressions for the generators of the Weyl invariant Jacobi forms used in the paper. We refer to the literatures such as \cite{MR1163219,Bertola:1999,Sakai:2011xg,Sakai:2017ihc} for the detailed explanations. For $SU(N+1)$ and $SO(2N+1)$ Weyl groups, $(N+1)$ generators of Weyl invariant Jacobi forms are obtained from the following generating functions \cite{Bertola:1999}:
 \begin{align}
 SU(N+1):& \quad \prod_{i=1}^N \frac{ \theta_1(a_i+v)}{\eta^3}\bigg|_{\sum a_i = 0} = \left(\frac{ \theta_1(v)}{\eta^3}\right)^N \sum_{i=0}^N   \wp^{(i-2)}(v) \varphi_{-i,1} (a_i)\\
 SO(2N+1):& \quad \prod_{i=1}^N \frac{ \theta_1(a_i+v)\theta_1(-a_i+v)}{\eta^6} = \left(\frac{ \theta_1(v)}{\eta^3}\right)^{2N} \sum_{i=0}^N   \wp^{(2i-2)}(v) \varphi_{-2i,1}(a_i)
 \end{align}
 where $a_i$ denote the $SU(N+1)$/$SO(2N+1)$ chemical potentials. $\wp^{(-2)}(v)$ must be understood as $1$. The Weierstrass $\wp$ function is a weak Jacobi form of weight $2$ and index $0$ which can be expressed
 using Jacobi theta functions as follows:
 \begin{align}
 {\displaystyle \wp (z)=\frac{\theta_3(0)^{2}\theta_2(0)^2}{4}{\theta_4(z)^2 \over \theta_1(z)^2}-{1\over 12}\left[\theta_3(0)^4 + \theta_2(0)^4\right]}.
 \end{align}
 The following identities are useful for writing the explicit expressions for the generators \cite{Bertola:1999}.
 \begin{align}
 \frac{\prod_{i=1}^N \eta^3 \, \theta_1(a_i - v)}{\prod_{i=1}^N  \theta_1(a_i ) \theta_1(v)} &= -\frac{1}{2^{N-2} (N-1)!} \frac{\text{det}\left(\begin{matrix} 1 & \wp (v) & \wp' (v) & \cdots & \wp^{(N-2)}(v)\\
 1 & \wp (a_1) & \wp' (a_1) & \cdots & \wp^{(N-2)}(a_1)\\
 \vdots & \vdots & \vdots &  & \vdots \\
 1 & \wp (a_{N-1}) & \wp' (a_{N-1}) & \cdots & \wp^{(N-2)}(a_{N-1})\end{matrix}\right)}{\text{det}\left(\begin{matrix} 
 1 & \wp (a_1) & \wp' (a_1) & \cdots & \wp^{(N-3)}(a_1)\\
 \vdots & \vdots & \vdots &  & \vdots \\
 1 & \wp (a_{N-1}) & \wp' (a_{N-1}) & \cdots & \wp^{(N-3)}(a_{N-1})\end{matrix}\right)}\\
 \frac{\prod_{i=1}^N \eta^6 \, \theta_1(\pm a_i + v) }{\prod_{i=1}^N  \theta_1(a_i )^2 \theta_1(v)^2} &= -\frac{1}{2^{2N-2} (2N-1)!} \frac{\text{det}\left(\begin{matrix} 1 & \wp (v) & \wp'' (v) & \cdots & \wp^{(2N-2)}(v)\\
 1 & \wp (a_1) & \wp'' (a_1) & \cdots & \wp^{(2N-2)}(a_1)\\
 \vdots & \vdots & \vdots &  & \vdots \\
 1 & \wp (a_{N}) & \wp'' (a_{N}) & \cdots & \wp^{(2N-2)}(a_{N})\end{matrix}\right)}{\text{det}\left(\begin{matrix} 
 1 & \wp (a_1) & \wp'' (a_1) & \cdots & \wp^{(2N-4)}(a_1)\\
 \vdots & \vdots & \vdots &  & \vdots \\
 1 & \wp (a_{N}) & \wp'' (a_{N}) & \cdots & \wp^{(2N-4)}(a_{N})\end{matrix}\right)}
 \end{align}

We particularly consider the generators of $SU(2)$, $SU(3)$, $SU(4)$ Weyl invariant Jacobi forms used in Section~\ref{subsec:example}. Imposing $a_1 + a_2 = 0$ for $SU(2)$, we obtain 
 \begin{align}
 \varphi_{-2,1} = \frac{\theta_1(a_1)^2}{\eta^6},\quad \varphi_{0,1} = \frac{12 \,\theta_1( a_1)^2}{\eta^6}\wp(a_1).
 \end{align}
For $SU(3)$, imposing the traceless condition $a_1 + a_2 + a_3 = 0$, the generators can be written as
 \begin{align}
 \label{eq:su3-gen}
 \varphi_{0,1} &=\frac{1}{2}\frac{
    \left(\wp \left(a_1\right) \wp '\left(a_2\right)-\wp \left(a_2\right) \wp
    '\left(a_1\right)\right)}{  \wp \left(a_2\right)-\wp
    \left(a_1\right)}\prod_{i=1}^3 \frac{\theta_1(a_i)}{\eta^3}\\
 \varphi_{-2,1} &=-\frac{1}{2}\frac{
    \wp '\left(a_1\right)-\wp '\left(a_2\right)}{\wp
    \left(a_1\right)-\wp \left(a_2\right)}\prod_{i=1}^3 \frac{\theta_1(a_i)}{\eta^3},\quad
 \varphi_{-3,1} = \frac{1}{2}\prod_{i=1}^3 \frac{\theta_1(a_i)}{\eta^3}\nonumber
 \end{align}
For $SU(4)$, the Weyl invariant Jacobi forms are generated by 
\begin{align}
\varphi_{0,1} &= \frac{1}{6}\frac{\sum_{(x,y,z)}\wp(a_x)\left(\wp'(a_y)\wp''(a_z)- \wp'(a_z)\wp''(a_y)\right)}{ \sum_{(x,y,z)}\wp(a_x)\left(\wp'(a_y)-\wp'(a_z)\right)} \prod_{i=1}^4 \frac{\theta_1(a_i)}{\eta^3}\\
\varphi_{-2,1} &=  \frac{1}{6}\frac{\sum_{(x,y,z)}\wp'(a_x)\left(\wp''(a_y)-\wp''(a_z)\right)}{ \sum_{(x,y,z)}\wp(a_x)\left(\wp'(a_y)-\wp'(a_z)\right)} \prod_{i=1}^4 \frac{\theta_1(a_i)}{\eta^3}\nonumber\\
\varphi_{-3,1} &=  \frac{1}{6}\frac{\sum_{(x,y,z)}\wp(a_x)\left(\wp''(a_y)-\wp''(a_z)\right)}{ \sum_{(x,y,z)}\wp(a_x)\left(\wp'(a_y)-\wp'(a_z)\right)} \prod_{i=1}^4 \frac{\theta_1(a_i)}{\eta^3}, \quad
\varphi_{-4,1} = \frac{1}{6}\prod_{i=1}^4 \frac{\theta_1(a_i)}{\eta^3} \nonumber\end{align}
where $(x,y,z)$ runs over $\{(1,2,3),(2,3,1),(3,1,2)\}$. Based on the $SU(3)$ generators and \eqref{eq:g2-gen}, one can write the generators of the $G_2$ Weyl invariant Jacobi forms \cite{MR1163219,Bertola:1999}. We also found the expressions \eqref{eq:dn-gen} for the generators of the $D_n$ Weyl invariant Jacobi forms, by generalizing the $SO(8)$ generators given in \cite{Bertola:1999}. Finally, all the generators of the $E_n$ Weyl invariant Jacobi forms are given in \cite{Sakai:2011xg,Sakai:2017ihc}.

\providecommand{\href}[2]{#2}\begingroup\raggedright\endgroup


\begin{thebibliography}{10}

\bibitem{Witten:1995zh}
E.~Witten, ``{Some comments on string dynamics},'' in {\em {Future perspectives
  in string theory. Proceedings, Conference, Strings'95, Los Angeles, USA,
  March 13-18, 1995}}, pp.~501--523.
\newblock 1995.
\newblock
\href{http://arxiv.org/abs/hep-th/9507121}{{\ttfamily arXiv:hep-th/9507121
  [hep-th]}}.
\newblock
%%CITATION = HEP-TH/9507121;%%.

\bibitem{Strominger:1995ac}
A.~Strominger, ``{Open p-branes},''
  \href{http://dx.doi.org/10.1016/0370-2693(96)00712-5}{{\em Phys. Lett.}
  {\bfseries B383} (1996) 44--47},
\href{http://arxiv.org/abs/hep-th/9512059}{{\ttfamily arXiv:hep-th/9512059
  [hep-th]}}.
%%CITATION = HEP-TH/9512059;%%.

\bibitem{Nekrasov:2002qd}
N.~A. Nekrasov, ``{Seiberg-Witten prepotential from instanton counting},''
  \href{http://dx.doi.org/10.4310/ATMP.2003.v7.n5.a4}{{\em Adv. Theor. Math.
  Phys.} {\bfseries 7} no.~5, (2003) 831--864},
\href{http://arxiv.org/abs/hep-th/0206161}{{\ttfamily arXiv:hep-th/0206161
  [hep-th]}}.
%%CITATION = HEP-TH/0206161;%%.

\bibitem{Nekrasov:2003rj}
N.~Nekrasov and A.~Okounkov, ``{Seiberg-Witten theory and random partitions},''
  \href{http://dx.doi.org/10.1007/0-8176-4467-9_15}{{\em Prog. Math.}
  {\bfseries 244} (2006) 525--596},
\href{http://arxiv.org/abs/hep-th/0306238}{{\ttfamily arXiv:hep-th/0306238
  [hep-th]}}.
%%CITATION = HEP-TH/0306238;%%.

\bibitem{Seiberg:1994rs}
N.~Seiberg and E.~Witten, ``{Electric - magnetic duality, monopole
  condensation, and confinement in N=2 supersymmetric Yang-Mills theory},''
  \href{http://dx.doi.org/10.1016/0550-3213(94)90124-4,
  10.1016/0550-3213(94)00449-8}{{\em Nucl. Phys.} {\bfseries B426} (1994)
  19--52}, \href{http://arxiv.org/abs/hep-th/9407087}{{\ttfamily
  arXiv:hep-th/9407087 [hep-th]}}.
[Erratum: Nucl. Phys.B430,485(1994)].
%%CITATION = HEP-TH/9407087;%%.

\bibitem{Benini:2013nda}
F.~Benini, R.~Eager, K.~Hori, and Y.~Tachikawa, ``{Elliptic genera of
  two-dimensional N=2 gauge theories with rank-one gauge groups},''
  \href{http://dx.doi.org/10.1007/s11005-013-0673-y}{{\em Lett. Math. Phys.}
  {\bfseries 104} (2014) 465--493},
\href{http://arxiv.org/abs/1305.0533}{{\ttfamily arXiv:1305.0533 [hep-th]}}.
%%CITATION = ARXIV:1305.0533;%%.

\bibitem{Benini:2013xpa}
F.~Benini, R.~Eager, K.~Hori, and Y.~Tachikawa, ``{Elliptic Genera of 2d
  ${\mathcal{N}}$ = 2 Gauge Theories},''
  \href{http://dx.doi.org/10.1007/s00220-014-2210-y}{{\em Commun. Math. Phys.}
  {\bfseries 333} no.~3, (2015) 1241--1286},
\href{http://arxiv.org/abs/1308.4896}{{\ttfamily arXiv:1308.4896 [hep-th]}}.
%%CITATION = ARXIV:1308.4896;%%.

\bibitem{Haghighat:2014pva}
B.~Haghighat, G.~Lockhart, and C.~Vafa, ``{Fusing E-strings to heterotic
  strings: E+E→H},'' \href{http://dx.doi.org/10.1103/PhysRevD.90.126012}{{\em
  Phys. Rev.} {\bfseries D90} no.~12, (2014) 126012},
\href{http://arxiv.org/abs/1406.0850}{{\ttfamily arXiv:1406.0850 [hep-th]}}.
%%CITATION = ARXIV:1406.0850;%%.

\bibitem{Cai:2014vka}
W.~Cai, M.-x. Huang, and K.~Sun, ``{On the Elliptic Genus of Three E-strings
  and Heterotic Strings},''
  \href{http://dx.doi.org/10.1007/JHEP01(2015)079}{{\em JHEP} {\bfseries 01}
  (2015) 079},
\href{http://arxiv.org/abs/1411.2801}{{\ttfamily arXiv:1411.2801 [hep-th]}}.
%%CITATION = ARXIV:1411.2801;%%.

\bibitem{Huang:2015sta}
M.-x. Huang, S.~Katz, and A.~Klemm, ``{Topological String on elliptic CY
  3-folds and the ring of Jacobi forms},''
  \href{http://dx.doi.org/10.1007/JHEP10(2015)125}{{\em JHEP} {\bfseries 10}
  (2015) 125},
\href{http://arxiv.org/abs/1501.04891}{{\ttfamily arXiv:1501.04891 [hep-th]}}.
%%CITATION = ARXIV:1501.04891;%%.

\bibitem{Huang:2015ada}
M.-x. Huang, S.~Katz, and A.~Klemm, ``{Elliptically fibered Calabi--Yau
  manifolds and the ring of Jacobi forms},''
\href{http://dx.doi.org/10.1016/j.nuclphysb.2015.06.020}{{\em Nucl. Phys.}
  {\bfseries B898} (2015) 681--692}.
%%CITATION = NUPHA,B898,681;%%.

\bibitem{Haghighat:2015ega}
B.~Haghighat, S.~Murthy, C.~Vafa, and S.~Vandoren, ``{F-Theory, Spinning Black
  Holes and Multi-string Branches},''
  \href{http://dx.doi.org/10.1007/JHEP01(2016)009}{{\em JHEP} {\bfseries 01}
  (2016) 009},
\href{http://arxiv.org/abs/1509.00455}{{\ttfamily arXiv:1509.00455 [hep-th]}}.
%%CITATION = ARXIV:1509.00455;%%.

\bibitem{DelZotto:2016pvm}
M.~Del~Zotto and G.~Lockhart, ``{On Exceptional Instanton Strings},''
\href{http://arxiv.org/abs/1609.00310}{{\ttfamily arXiv:1609.00310 [hep-th]}}.
%%CITATION = ARXIV:1609.00310;%%.

\bibitem{Gu:2017ccq}
J.~Gu, M.-x. Huang, A.-K. Kashani-Poor, and A.~Klemm, ``{Refined BPS invariants
  of 6d {SCFTs} from anomalies and modularity},''
  \href{http://dx.doi.org/10.1007/JHEP05(2017)130}{{\em JHEP} {\bfseries 05}
  (2017) 130},
\href{http://arxiv.org/abs/1701.00764}{{\ttfamily arXiv:1701.00764 [hep-th]}}.
%%CITATION = ARXIV:1701.00764;%%.

\bibitem{Kim:2015gha}
J.~Kim, S.~Kim, and K.~Lee, ``{Little strings and T-duality},''
  \href{http://dx.doi.org/10.1007/JHEP02(2016)170}{{\em JHEP} {\bfseries 02}
  (2016) 170},
\href{http://arxiv.org/abs/1503.07277}{{\ttfamily arXiv:1503.07277 [hep-th]}}.
%%CITATION = ARXIV:1503.07277;%%.

\bibitem{Hohenegger:2016eqy}
S.~Hohenegger, A.~Iqbal, and S.-J. Rey, ``{Self-Duality and Self-Similarity of
  Little String Orbifolds},''
  \href{http://dx.doi.org/10.1103/PhysRevD.94.046006}{{\em Phys. Rev.}
  {\bfseries D94} no.~4, (2016) 046006},
\href{http://arxiv.org/abs/1605.02591}{{\ttfamily arXiv:1605.02591 [hep-th]}}.
%%CITATION = ARXIV:1605.02591;%%.

\bibitem{Kim:2017xan}
J.~Kim and K.~Lee, ``{Little strings on D$_{n}$ orbifolds},''
  \href{http://dx.doi.org/10.1007/JHEP10(2017)045}{{\em JHEP} {\bfseries 10}
  (2017) 045},
\href{http://arxiv.org/abs/1702.03116}{{\ttfamily arXiv:1702.03116 [hep-th]}}.
%%CITATION = ARXIV:1702.03116;%%.

\bibitem{Seiberg:1997zk}
N.~Seiberg, ``{New theories in six-dimensions and matrix description of M
  theory on T**5 and T**5 / Z(2)},''
  \href{http://dx.doi.org/10.1016/S0370-2693(97)00805-8}{{\em Phys. Lett.}
  {\bfseries B408} (1997) 98--104},
\href{http://arxiv.org/abs/hep-th/9705221}{{\ttfamily arXiv:hep-th/9705221
  [hep-th]}}.
%%CITATION = HEP-TH/9705221;%%.

\bibitem{Gopakumar:1998jq}
R.~Gopakumar and C.~Vafa, ``{M theory and topological strings. 2.},''
\href{http://arxiv.org/abs/hep-th/9812127}{{\ttfamily arXiv:hep-th/9812127
  [hep-th]}}.
%%CITATION = HEP-TH/9812127;%%.

\bibitem{Witten:1995gx}
E.~Witten, ``{Small instantons in string theory},''
  \href{http://dx.doi.org/10.1016/0550-3213(95)00625-7}{{\em Nucl. Phys.}
  {\bfseries B460} (1996) 541--559},
\href{http://arxiv.org/abs/hep-th/9511030}{{\ttfamily arXiv:hep-th/9511030
  [hep-th]}}.
%%CITATION = HEP-TH/9511030;%%.

\bibitem{Witten:1997yu}
E.~Witten, ``{On the conformal field theory of the Higgs branch},''
  \href{http://dx.doi.org/10.1088/1126-6708/1997/07/003}{{\em JHEP} {\bfseries
  07} (1997) 003},
\href{http://arxiv.org/abs/hep-th/9707093}{{\ttfamily arXiv:hep-th/9707093
  [hep-th]}}.
%%CITATION = HEP-TH/9707093;%%.

\bibitem{Diaconescu:1997gu}
D.-E. Diaconescu and N.~Seiberg, ``{The Coulomb branch of (4,4) supersymmetric
  field theories in two-dimensions},''
  \href{http://dx.doi.org/10.1088/1126-6708/1997/07/001}{{\em JHEP} {\bfseries
  07} (1997) 001},
\href{http://arxiv.org/abs/hep-th/9707158}{{\ttfamily arXiv:hep-th/9707158
  [hep-th]}}.
%%CITATION = HEP-TH/9707158;%%.

\bibitem{Seiberg:1999xz}
N.~Seiberg and E.~Witten, ``{The D1 / D5 system and singular CFT},''
  \href{http://dx.doi.org/10.1088/1126-6708/1999/04/017}{{\em JHEP} {\bfseries
  04} (1999) 017},
\href{http://arxiv.org/abs/hep-th/9903224}{{\ttfamily arXiv:hep-th/9903224
  [hep-th]}}.
%%CITATION = HEP-TH/9903224;%%.

\bibitem{Aharony:1999dw}
O.~Aharony and M.~Berkooz, ``{IR dynamics of D = 2, N=(4,4) gauge theories and
  DLCQ of 'little string theories'},''
  \href{http://dx.doi.org/10.1088/1126-6708/1999/10/030}{{\em JHEP} {\bfseries
  10} (1999) 030},
\href{http://arxiv.org/abs/hep-th/9909101}{{\ttfamily arXiv:hep-th/9909101
  [hep-th]}}.
%%CITATION = HEP-TH/9909101;%%.

\bibitem{Hwang:2016gfw}
Y.~Hwang, J.~Kim, and S.~Kim, ``{M5-branes, orientifolds, and S-duality},''
  \href{http://dx.doi.org/10.1007/JHEP12(2016)148}{{\em JHEP} {\bfseries 12}
  (2016) 148},
\href{http://arxiv.org/abs/1607.08557}{{\ttfamily arXiv:1607.08557 [hep-th]}}.
%%CITATION = ARXIV:1607.08557;%%.

\bibitem{Hwang:2014uwa}
C.~Hwang, J.~Kim, S.~Kim, and J.~Park, ``{General instanton counting and 5d
  SCFT},'' \href{http://dx.doi.org/10.1007/JHEP07(2015)063,
  10.1007/JHEP04(2016)094}{{\em JHEP} {\bfseries 07} (2015) 063},
  \href{http://arxiv.org/abs/1406.6793}{{\ttfamily arXiv:1406.6793 [hep-th]}}.
[Addendum: JHEP04,094(2016)].
%%CITATION = ARXIV:1406.6793;%%.

\bibitem{Gadde:2015tra}
A.~Gadde, B.~Haghighat, J.~Kim, S.~Kim, G.~Lockhart, and C.~Vafa, ``{6d String
  Chains},''
\href{http://arxiv.org/abs/1504.04614}{{\ttfamily arXiv:1504.04614 [hep-th]}}.
%%CITATION = ARXIV:1504.04614;%%.

\bibitem{Kim:2015fxa}
J.~Kim, S.~Kim, and K.~Lee, ``{Higgsing towards E-strings},''
\href{http://arxiv.org/abs/1510.03128}{{\ttfamily arXiv:1510.03128 [hep-th]}}.
%%CITATION = ARXIV:1510.03128;%%.

\bibitem{DelZotto:2017mee}
M.~Del~Zotto, J.~Gu, M.-x. Huang, A.-K. Kashani-Poor, A.~Klemm, and
  G.~Lockhart, ``{Topological Strings on Singular Elliptic Calabi-Yau 3-folds
  and Minimal 6d SCFTs},''
\href{http://arxiv.org/abs/1712.07017}{{\ttfamily arXiv:1712.07017 [hep-th]}}.
%%CITATION = ARXIV:1712.07017;%%.

\bibitem{DiPietro:2014bca}
L.~Di~Pietro and Z.~Komargodski, ``{Cardy formulae for SUSY theories in $d =$ 4
  and $d =$ 6},'' \href{http://dx.doi.org/10.1007/JHEP12(2014)031}{{\em JHEP}
  {\bfseries 12} (2014) 031},
\href{http://arxiv.org/abs/1407.6061}{{\ttfamily arXiv:1407.6061 [hep-th]}}.
%%CITATION = ARXIV:1407.6061;%%.

\bibitem{Golkar:2015oxw}
S.~Golkar and S.~Sethi, ``{Global Anomalies and Effective Field Theory},''
  \href{http://dx.doi.org/10.1007/JHEP05(2016)105}{{\em JHEP} {\bfseries 05}
  (2016) 105},
\href{http://arxiv.org/abs/1512.02607}{{\ttfamily arXiv:1512.02607 [hep-th]}}.
%%CITATION = ARXIV:1512.02607;%%.

\bibitem{Kim:2017zyo}
S.~Kim and J.~Nahmgoong, ``{Asymptotic M5-brane entropy from S-duality},''
\href{http://arxiv.org/abs/1702.04058}{{\ttfamily arXiv:1702.04058 [hep-th]}}.
%%CITATION = ARXIV:1702.04058;%%.

\bibitem{Banerjee:2012iz}
N.~Banerjee, J.~Bhattacharya, S.~Bhattacharyya, S.~Jain, S.~Minwalla, and
  T.~Sharma, ``{Constraints on Fluid Dynamics from Equilibrium Partition
  Functions},'' \href{http://dx.doi.org/10.1007/JHEP09(2012)046}{{\em JHEP}
  {\bfseries 09} (2012) 046},
\href{http://arxiv.org/abs/1203.3544}{{\ttfamily arXiv:1203.3544 [hep-th]}}.
%%CITATION = ARXIV:1203.3544;%%.

\bibitem{Hollowood:2003cv}
T.~J. Hollowood, A.~Iqbal, and C.~Vafa, ``{Matrix models, geometric engineering
  and elliptic genera},''
  \href{http://dx.doi.org/10.1088/1126-6708/2008/03/069}{{\em JHEP} {\bfseries
  03} (2008) 069},
\href{http://arxiv.org/abs/hep-th/0310272}{{\ttfamily arXiv:hep-th/0310272
  [hep-th]}}.
%%CITATION = HEP-TH/0310272;%%.

\bibitem{Belavin:1975fg}
A.~A. Belavin, A.~M. Polyakov, A.~S. Schwartz, and {\relax Yu}.~S. Tyupkin,
  ``{Pseudoparticle Solutions of the Yang-Mills Equations},''
\href{http://dx.doi.org/10.1016/0370-2693(75)90163-X}{{\em Phys. Lett.}
  {\bfseries B59} (1975) 85--87}.
%%CITATION = PHLTA,B59,85;%%.

\bibitem{Bernard:1977nr}
C.~W. Bernard, N.~H. Christ, A.~H. Guth, and E.~J. Weinberg, ``{Instanton
  Parameters for Arbitrary Gauge Groups},''
\href{http://dx.doi.org/10.1103/PhysRevD.16.2967}{{\em Phys. Rev.} {\bfseries
  D16} (1977) 2967}.
%%CITATION = PHRVA,D16,2967;%%.

\bibitem{Cremonesi:2014xha}
S.~Cremonesi, G.~Ferlito, A.~Hanany, and N.~Mekareeya, ``{Coulomb Branch and
  The Moduli Space of Instantons},''
  \href{http://dx.doi.org/10.1007/JHEP12(2014)103}{{\em JHEP} {\bfseries 12}
  (2014) 103},
\href{http://arxiv.org/abs/1408.6835}{{\ttfamily arXiv:1408.6835 [hep-th]}}.
%%CITATION = ARXIV:1408.6835;%%.

\bibitem{Hayashi:2017jze}
H.~Hayashi and K.~Ohmori, ``{5d/6d DE instantons from trivalent gluing of web
  diagrams},''
\href{http://arxiv.org/abs/1702.07263}{{\ttfamily arXiv:1702.07263 [hep-th]}}.
%%CITATION = ARXIV:1702.07263;%%.

\bibitem{Kim:2017}
H.-C. Kim, J.~Kim, S.~Kim, K.-H. Lee, and J.~Park, ``6d strings and exceptional
  instantons,'' {\em work in progress} .

\bibitem{MR781735}
M.~Eichler and D.~Zagier,
  \href{http://dx.doi.org/10.1007/978-1-4684-9162-3}{{\em The theory of
  {J}acobi forms}}, vol.~55 of {\em Progress in Mathematics}.
\newblock Birkh\"auser Boston, Inc., Boston, MA, 1985.

\bibitem{MR1163219}
K.~Wirthm{\"u}ller, ``Root systems and {J}acobi forms,'' {\em Compositio Math.}
  {\bfseries 82} no.~3, (1992) 293--354.
  \url{http://www.numdam.org/item?id=CM_1992__82_3_293_0}.

\bibitem{Sakai:2011xg}
K.~Sakai, ``{Topological string amplitudes for the local $\frac{1}{2}$K3
  surface},'' \href{http://dx.doi.org/10.1093/ptep/ptx027}{{\em PTEP}
  {\bfseries 2017} no.~3, (2017) 033B09},
\href{http://arxiv.org/abs/1111.3967}{{\ttfamily arXiv:1111.3967 [hep-th]}}.
%%CITATION = ARXIV:1111.3967;%%.

\bibitem{Bertola:1999}
M.~Bertola, {\em Jacobi Groups, Jacobi Forms and Their Applications}.
\newblock PhD thesis, SISSA, 1999.

\bibitem{Sakai:2017ihc}
K.~Sakai, ``{$E_n$ Jacobi forms and Seiberg-Witten curves},''
\href{http://arxiv.org/abs/1706.04619}{{\ttfamily arXiv:1706.04619 [hep-th]}}.
%%CITATION = ARXIV:1706.04619;%%.

\bibitem{Kim:2016foj}
H.-C. Kim, S.~Kim, and J.~Park, ``{6d strings from new chiral gauge
  theories},''
\href{http://arxiv.org/abs/1608.03919}{{\ttfamily arXiv:1608.03919 [hep-th]}}.
%%CITATION = ARXIV:1608.03919;%%.

\bibitem{Shimizu:2016lbw}
H.~Shimizu and Y.~Tachikawa, ``{Anomaly of strings of 6d
  $\mathcal{N}=\left(1,0\right)$ theories},''
  \href{http://dx.doi.org/10.1007/JHEP11(2016)165}{{\em JHEP} {\bfseries 11}
  (2016) 165},
\href{http://arxiv.org/abs/1608.05894}{{\ttfamily arXiv:1608.05894 [hep-th]}}.
%%CITATION = ARXIV:1608.05894;%%.

\bibitem{Ohmori:2014kda}
K.~Ohmori, H.~Shimizu, Y.~Tachikawa, and K.~Yonekura, ``{Anomaly polynomial of
  general 6d SCFTs},'' \href{http://dx.doi.org/10.1093/ptep/ptu140}{{\em PTEP}
  {\bfseries 2014} no.~10, (2014) 103B07},
\href{http://arxiv.org/abs/1408.5572}{{\ttfamily arXiv:1408.5572 [hep-th]}}.
%%CITATION = ARXIV:1408.5572;%%.

\bibitem{Haghighat:2013gba}
B.~Haghighat, A.~Iqbal, C.~Koz{\c c}az, G.~Lockhart, and C.~Vafa,
  ``{M-Strings},'' \href{http://dx.doi.org/10.1007/s00220-014-2139-1}{{\em
  Commun. Math. Phys.} {\bfseries 334} no.~2, (2015) 779--842},
\href{http://arxiv.org/abs/1305.6322}{{\ttfamily arXiv:1305.6322 [hep-th]}}.
%%CITATION = ARXIV:1305.6322;%%.

\bibitem{Haghighat:2013tka}
B.~Haghighat, C.~Kozcaz, G.~Lockhart, and C.~Vafa, ``{Orbifolds of
  M-strings},'' \href{http://dx.doi.org/10.1103/PhysRevD.89.046003}{{\em Phys.
  Rev.} {\bfseries D89} no.~4, (2014) 046003},
\href{http://arxiv.org/abs/1310.1185}{{\ttfamily arXiv:1310.1185 [hep-th]}}.
%%CITATION = ARXIV:1310.1185;%%.

\bibitem{Klemm:1996hh}
A.~Klemm, P.~Mayr, and C.~Vafa, ``{BPS states of exceptional noncritical
  strings},'' \href{http://arxiv.org/abs/hep-th/9607139}{{\ttfamily
  arXiv:hep-th/9607139 [hep-th]}}.
[Nucl. Phys. Proc. Suppl.58,177(1997)].
%%CITATION = HEP-TH/9607139;%%.

\bibitem{Kim:2014dza}
J.~Kim, S.~Kim, K.~Lee, J.~Park, and C.~Vafa, ``{Elliptic Genus of
  E-strings},''
\href{http://arxiv.org/abs/1411.2324}{{\ttfamily arXiv:1411.2324 [hep-th]}}.
%%CITATION = ARXIV:1411.2324;%%.

\bibitem{Haghighat:2014vxa}
B.~Haghighat, A.~Klemm, G.~Lockhart, and C.~Vafa, ``{Strings of Minimal 6d
  SCFTs},'' \href{http://dx.doi.org/10.1002/prop.201500014}{{\em Fortsch.
  Phys.} {\bfseries 63} (2015) 294--322},
\href{http://arxiv.org/abs/1412.3152}{{\ttfamily arXiv:1412.3152 [hep-th]}}.
%%CITATION = ARXIV:1412.3152;%%.

\bibitem{Freed:1998tg}
D.~Freed, J.~A. Harvey, R.~Minasian, and G.~W. Moore, ``{Gravitational anomaly
  cancellation for M theory five-branes},'' {\em Adv. Theor. Math. Phys.}
  {\bfseries 2} (1998) 601--618,
\href{http://arxiv.org/abs/hep-th/9803205}{{\ttfamily arXiv:hep-th/9803205
  [hep-th]}}.
%%CITATION = HEP-TH/9803205;%%.

\bibitem{Becker:1999kh}
K.~Becker and M.~Becker, ``{Five-brane gravitational anomalies},''
  \href{http://dx.doi.org/10.1016/S0550-3213(00)00153-X}{{\em Nucl. Phys.}
  {\bfseries B577} (2000) 156--170},
\href{http://arxiv.org/abs/hep-th/9911138}{{\ttfamily arXiv:hep-th/9911138
  [hep-th]}}.
%%CITATION = HEP-TH/9911138;%%.

\bibitem{Harvey:1998bx}
J.~A. Harvey, R.~Minasian, and G.~W. Moore, ``{NonAbelian tensor multiplet
  anomalies},'' \href{http://dx.doi.org/10.1088/1126-6708/1998/09/004}{{\em
  JHEP} {\bfseries 09} (1998) 004},
\href{http://arxiv.org/abs/hep-th/9808060}{{\ttfamily arXiv:hep-th/9808060
  [hep-th]}}.
%%CITATION = HEP-TH/9808060;%%.

\bibitem{Erler:1993zy}
J.~Erler, ``{Anomaly cancellation in six-dimensions},''
  \href{http://dx.doi.org/10.1063/1.530885}{{\em J. Math. Phys.} {\bfseries 35}
  (1994) 1819--1833},
\href{http://arxiv.org/abs/hep-th/9304104}{{\ttfamily arXiv:hep-th/9304104
  [hep-th]}}.
%%CITATION = HEP-TH/9304104;%%.

\bibitem{Sadov:1996zm}
V.~Sadov, ``{Generalized Green-Schwarz mechanism in F theory},''
  \href{http://dx.doi.org/10.1016/0370-2693(96)01134-3}{{\em Phys. Lett.}
  {\bfseries B388} (1996) 45--50},
\href{http://arxiv.org/abs/hep-th/9606008}{{\ttfamily arXiv:hep-th/9606008
  [hep-th]}}.
%%CITATION = HEP-TH/9606008;%%.

\bibitem{Vafa:1995fj}
C.~Vafa and E.~Witten, ``{A One loop test of string duality},''
  \href{http://dx.doi.org/10.1016/0550-3213(95)00280-6}{{\em Nucl. Phys.}
  {\bfseries B447} (1995) 261--270},
\href{http://arxiv.org/abs/hep-th/9505053}{{\ttfamily arXiv:hep-th/9505053
  [hep-th]}}.
%%CITATION = HEP-TH/9505053;%%.

\bibitem{Duff:1995wd}
M.~J. Duff, J.~T. Liu, and R.~Minasian, ``{Eleven-dimensional origin of
  string-string duality: A One loop test},''
  \href{http://dx.doi.org/10.1016/0550-3213(95)00368-3}{{\em Nucl. Phys.}
  {\bfseries B452} (1995) 261--282},
\href{http://arxiv.org/abs/hep-th/9506126}{{\ttfamily arXiv:hep-th/9506126
  [hep-th]}}.
%%CITATION = HEP-TH/9506126;%%.

\bibitem{Intriligator:2000eq}
K.~A. Intriligator, ``{Anomaly matching and a Hopf-Wess-Zumino term in 6d,
  N=(2,0) field theories},''
  \href{http://dx.doi.org/10.1016/S0550-3213(00)00148-6}{{\em Nucl. Phys.}
  {\bfseries B581} (2000) 257--273},
\href{http://arxiv.org/abs/hep-th/0001205}{{\ttfamily arXiv:hep-th/0001205
  [hep-th]}}.
%%CITATION = HEP-TH/0001205;%%.

\bibitem{Witten:1996hc}
E.~Witten, ``{Five-brane effective action in M theory},''
  \href{http://dx.doi.org/10.1016/S0393-0440(97)80160-X}{{\em J. Geom. Phys.}
  {\bfseries 22} (1997) 103--133},
\href{http://arxiv.org/abs/hep-th/9610234}{{\ttfamily arXiv:hep-th/9610234
  [hep-th]}}.
%%CITATION = HEP-TH/9610234;%%.

\bibitem{Mourad:1997uc}
J.~Mourad, ``{Anomalies of the SO(32) five-brane and their cancellation},''
  \href{http://dx.doi.org/10.1016/S0550-3213(97)00774-8}{{\em Nucl. Phys.}
  {\bfseries B512} (1998) 199--208},
\href{http://arxiv.org/abs/hep-th/9709012}{{\ttfamily arXiv:hep-th/9709012
  [hep-th]}}.
%%CITATION = HEP-TH/9709012;%%.

\bibitem{Ohmori:2014pca}
K.~Ohmori, H.~Shimizu, and Y.~Tachikawa, ``{Anomaly polynomial of E-string
  theories},'' \href{http://dx.doi.org/10.1007/JHEP08(2014)002}{{\em JHEP}
  {\bfseries 08} (2014) 002},
\href{http://arxiv.org/abs/1404.3887}{{\ttfamily arXiv:1404.3887 [hep-th]}}.
%%CITATION = ARXIV:1404.3887;%%.

\bibitem{Intriligator:2014eaa}
K.~Intriligator, ``{6d, $ \mathcal{N}=\left(1,\;0\right) $ Coulomb branch
  anomaly matching},'' \href{http://dx.doi.org/10.1007/JHEP10(2014)162}{{\em
  JHEP} {\bfseries 10} (2014) 162},
\href{http://arxiv.org/abs/1408.6745}{{\ttfamily arXiv:1408.6745 [hep-th]}}.
%%CITATION = ARXIV:1408.6745;%%.

\bibitem{Monnier:2013rpa}
S.~Monnier, ``{Global gravitational anomaly cancellation for five-branes},''
  \href{http://dx.doi.org/10.4310/ATMP.2015.v19.n3.a5}{{\em Adv. Theor. Math.
  Phys.} {\bfseries 19} (2015) 701--724},
\href{http://arxiv.org/abs/1310.2250}{{\ttfamily arXiv:1310.2250 [hep-th]}}.
%%CITATION = ARXIV:1310.2250;%%.

\bibitem{Bhardwaj:2015oru}
L.~Bhardwaj, M.~Del~Zotto, J.~J. Heckman, D.~R. Morrison, T.~Rudelius, and
  C.~Vafa, ``{F-theory and the Classification of Little Strings},''
  \href{http://dx.doi.org/10.1103/PhysRevD.93.086002}{{\em Phys. Rev.}
  {\bfseries D93} no.~8, (2016) 086002},
\href{http://arxiv.org/abs/1511.05565}{{\ttfamily arXiv:1511.05565 [hep-th]}}.
%%CITATION = ARXIV:1511.05565;%%.

\bibitem{Sen:1998ii}
A.~Sen, ``{Stable nonBPS bound states of BPS D-branes},''
  \href{http://dx.doi.org/10.1088/1126-6708/1998/08/010}{{\em JHEP} {\bfseries
  08} (1998) 010},
\href{http://arxiv.org/abs/hep-th/9805019}{{\ttfamily arXiv:hep-th/9805019
  [hep-th]}}.
%%CITATION = HEP-TH/9805019;%%.

\bibitem{Kapustin:1998fa}
A.~Kapustin, ``{D(n) quivers from branes},''
  \href{http://dx.doi.org/10.1088/1126-6708/1998/12/015}{{\em JHEP} {\bfseries
  12} (1998) 015},
\href{http://arxiv.org/abs/hep-th/9806238}{{\ttfamily arXiv:hep-th/9806238
  [hep-th]}}.
%%CITATION = HEP-TH/9806238;%%.

\bibitem{Bhattacharya:2008zy}
J.~Bhattacharya, S.~Bhattacharyya, S.~Minwalla, and S.~Raju, ``{Indices for
  Superconformal Field Theories in 3,5 and 6 Dimensions},''
  \href{http://dx.doi.org/10.1088/1126-6708/2008/02/064}{{\em JHEP} {\bfseries
  02} (2008) 064},
\href{http://arxiv.org/abs/0801.1435}{{\ttfamily arXiv:0801.1435 [hep-th]}}.
%%CITATION = ARXIV:0801.1435;%%.

\bibitem{Narain:1986am}
K.~S. Narain, M.~H. Sarmadi, and E.~Witten, ``{A Note on Toroidal
  Compactification of Heterotic String Theory},''
\href{http://dx.doi.org/10.1016/0550-3213(87)90001-0}{{\em Nucl. Phys.}
  {\bfseries B279} (1987) 369}.
%%CITATION = NUPHA,B279,369;%%.

\bibitem{Feng:2007ur}
B.~Feng, A.~Hanany, and Y.-H. He, ``{Counting gauge invariants: The Plethystic
  program},'' \href{http://dx.doi.org/10.1088/1126-6708/2007/03/090}{{\em JHEP}
  {\bfseries 03} (2007) 090},
\href{http://arxiv.org/abs/hep-th/0701063}{{\ttfamily arXiv:hep-th/0701063
  [hep-th]}}.
%%CITATION = HEP-TH/0701063;%%.

\bibitem{Johnson:1998yw}
C.~V. Johnson, ``{On the (0,4) conformal field theory of the throat},''
  \href{http://dx.doi.org/10.1142/S021773239800262X}{{\em Mod. Phys. Lett.}
  {\bfseries A13} (1998) 2463--2474},
\href{http://arxiv.org/abs/hep-th/9804201}{{\ttfamily arXiv:hep-th/9804201
  [hep-th]}}.
%%CITATION = HEP-TH/9804201;%%.

\bibitem{Gadde:2013ftv}
A.~Gadde and S.~Gukov, ``{2d Index and Surface operators},''
  \href{http://dx.doi.org/10.1007/JHEP03(2014)080}{{\em JHEP} {\bfseries 03}
  (2014) 080},
\href{http://arxiv.org/abs/1305.0266}{{\ttfamily arXiv:1305.0266 [hep-th]}}.
%%CITATION = ARXIV:1305.0266;%%.

\end{thebibliography}
\end{document}